\documentclass[showpacs,amssymb,twocolumn,pre,8pt,nofootinbib,letter]{revtex4-1}

\usepackage[english]{babel}
\usepackage{amsmath,amsfonts,amssymb,amsthm,latexsym,stmaryrd}
\usepackage{mathtools}
\newcommand\underrel[3][]{\mathrel{\mathop{#3}\limits_{%
      \ifx c#1\relax\mathclap{#2}\else#2\fi}}}

\usepackage{verbatim}
\usepackage[utf8]{inputenc}
\usepackage{pst-all}
\usepackage{pstricks}
\usepackage{graphicx}
\usepackage{psfrag}
\usepackage{subfig}
\usepackage{enumitem}
\usepackage{cleveref}
\crefname{equation}{eq.}{eqs.}
\Crefname{equation}{Eq.}{Eqs.}
\usepackage{color}
\usepackage{mathdots}
\usepackage{enumitem}
\usepackage{xcolor}
\usepackage{multirow}
\usepackage{cancel}

\newcommand{\orde}[2]{\ensuremath{\frac{\text{d}{#1}}{\text{d}{#2}}}}
\newcommand{\ordeN}[3]{\ensuremath{\frac{\text{d}^{\,#1}{#2}}{\text{d}{#3}^{#1}}}}

\newcommand{\suchthat}{\mathrel{\ooalign{$\ni$\cr\kern-1pt$-$\kern-6.5pt$-$}}}

\newpsobject{grilla}{psgrid}{subgriddiv=1,griddots=10,gridlabels=6pt}

\def\({\left(}
\def\){\right)}
\def\[{\left[}
\def\]{\right]}
\def\|l{\left|}
\def\|{\right|}
\def\<{\left <}
\def\>{\right>}

\def\emp{\begin{equation}}
\def\emps{\begin{equation*}}
\def\fin{\end{equation}} 
\def\fins{\end{equation*}}
\def\eemp{\begin{equation}\begin{aligned}} 
\def\eemps{\begin{equation*}\begin{aligned}}
\def\ffin{\end{aligned}\end{equation}} 
\def\ffins{\end{aligned}\end{equation*}}

\newcommand{\defeq}{\mathrel{\mathop:}=}

\definecolor{orange}{rgb}{1,0.5,0}
\definecolor{OliveGreen}{rgb}{0,0.6,0}
\definecolor{lightblue}{rgb}{.90,.95,1}
\definecolor{darkblue}{rgb}{0,0,0.5}

\definecolor{airforceblue}{rgb}{0.36, 0.54, 0.66}
\definecolor{aliceblue}{rgb}{0.94, 0.97, 1.0}
\definecolor{alizarin}{rgb}{0.82, 0.1, 0.26}
\definecolor{almond}{rgb}{0.94, 0.87, 0.8}
\definecolor{amaranth}{rgb}{0.9, 0.17, 0.31}
\definecolor{amber}{rgb}{1.0, 0.75, 0.0}
\definecolor{amber(sae/ece)}{rgb}{1.0, 0.49, 0.0}
\definecolor{americanrose}{rgb}{1.0, 0.01, 0.24}
\definecolor{amethyst}{rgb}{0.6, 0.4, 0.8}
\definecolor{anti-flashwhite}{rgb}{0.95, 0.95, 0.96}
\definecolor{antiquebrass}{rgb}{0.8, 0.58, 0.46}
\definecolor{antiquefuchsia}{rgb}{0.57, 0.36, 0.51}
\definecolor{antiquewhite}{rgb}{0.98, 0.92, 0.84}
\definecolor{ao}{rgb}{0.0, 0.0, 1.0}
\definecolor{ao(english)}{rgb}{0.0, 0.5, 0.0}
\definecolor{applegreen}{rgb}{0.55, 0.71, 0.0}
\definecolor{apricot}{rgb}{0.98, 0.81, 0.69}
\definecolor{aqua}{rgb}{0.0, 1.0, 1.0}
\definecolor{aquamarine}{rgb}{0.5, 1.0, 0.83}
\definecolor{armygreen}{rgb}{0.29, 0.33, 0.13}
\definecolor{arsenic}{rgb}{0.23, 0.27, 0.29}
\definecolor{arylideyellow}{rgb}{0.91, 0.84, 0.42}
\definecolor{ashgrey}{rgb}{0.7, 0.75, 0.71}
\definecolor{asparagus}{rgb}{0.53, 0.66, 0.42}
\definecolor{atomictangerine}{rgb}{1.0, 0.6, 0.4}
\definecolor{auburn}{rgb}{0.43, 0.21, 0.1}
\definecolor{aureolin}{rgb}{0.99, 0.93, 0.0}
\definecolor{aurometalsaurus}{rgb}{0.43, 0.5, 0.5}
\definecolor{awesome}{rgb}{1.0, 0.13, 0.32}
\definecolor{azure(colorwheel)}{rgb}{0.0, 0.5, 1.0}
\definecolor{azure(web)(azuremist)}{rgb}{0.94, 1.0, 1.0}
\definecolor{babyblue}{rgb}{0.54, 0.81, 0.94}
\definecolor{babyblueeyes}{rgb}{0.63, 0.79, 0.95}
\definecolor{babypink}{rgb}{0.96, 0.76, 0.76}
\definecolor{ballblue}{rgb}{0.13, 0.67, 0.8}
\definecolor{bananamania}{rgb}{0.98, 0.91, 0.71}
\definecolor{bananayellow}{rgb}{1.0, 0.88, 0.21}
\definecolor{battleshipgrey}{rgb}{0.52, 0.52, 0.51}
\definecolor{bazaar}{rgb}{0.6, 0.47, 0.48}
\definecolor{beaublue}{rgb}{0.74, 0.83, 0.9}
\definecolor{beaver}{rgb}{0.62, 0.51, 0.44}
\definecolor{beige}{rgb}{0.96, 0.96, 0.86}
\definecolor{bisque}{rgb}{1.0, 0.89, 0.77}
\definecolor{bistre}{rgb}{0.24, 0.17, 0.12}
\definecolor{bittersweet}{rgb}{1.0, 0.44, 0.37}
\definecolor{black}{rgb}{0.0, 0.0, 0.0}
\definecolor{blanchedalmond}{rgb}{1.0, 0.92, 0.8}
\definecolor{bleudefrance}{rgb}{0.19, 0.55, 0.91}
\definecolor{blizzardblue}{rgb}{0.67, 0.9, 0.93}
\definecolor{blond}{rgb}{0.98, 0.94, 0.75}
\definecolor{blue}{rgb}{0.0, 0.0, 1.0}
\definecolor{blue(munsell)}{rgb}{0.0, 0.5, 0.69}
\definecolor{blue(ncs)}{rgb}{0.0, 0.53, 0.74}
\definecolor{blue(pigment)}{rgb}{0.2, 0.2, 0.6}
\definecolor{blue(ryb)}{rgb}{0.01, 0.28, 1.0}
\definecolor{bluebell}{rgb}{0.64, 0.64, 0.82}
\definecolor{bluegray}{rgb}{0.4, 0.6, 0.8}
\definecolor{blue-green}{rgb}{0.0, 0.87, 0.87}
\definecolor{blue-violet}{rgb}{0.54, 0.17, 0.89}
\definecolor{blush}{rgb}{0.87, 0.36, 0.51}
\definecolor{bole}{rgb}{0.47, 0.27, 0.23}
\definecolor{bondiblue}{rgb}{0.0, 0.58, 0.71}
\definecolor{bostonuniversityred}{rgb}{0.8, 0.0, 0.0}
\definecolor{brandeisblue}{rgb}{0.0, 0.44, 1.0}
\definecolor{brass}{rgb}{0.71, 0.65, 0.26}
\definecolor{brickred}{rgb}{0.8, 0.25, 0.33}
\definecolor{brightcerulean}{rgb}{0.11, 0.67, 0.84}
\definecolor{brightgreen}{rgb}{0.4, 1.0, 0.0}
\definecolor{brightlavender}{rgb}{0.75, 0.58, 0.89}
\definecolor{brightmaroon}{rgb}{0.76, 0.13, 0.28}
\definecolor{brightpink}{rgb}{1.0, 0.0, 0.5}
\definecolor{brightturquoise}{rgb}{0.03, 0.91, 0.87}
\definecolor{brightube}{rgb}{0.82, 0.62, 0.91}
\definecolor{brilliantlavender}{rgb}{0.96, 0.73, 1.0}
\definecolor{brilliantrose}{rgb}{1.0, 0.33, 0.64}
\definecolor{brinkpink}{rgb}{0.98, 0.38, 0.5}
\definecolor{britishracinggreen}{rgb}{0.0, 0.26, 0.15}
\definecolor{bronze}{rgb}{0.8, 0.5, 0.2}
\definecolor{brown(traditional)}{rgb}{0.59, 0.29, 0.0}
\definecolor{brown(web)}{rgb}{0.65, 0.16, 0.16}
\definecolor{bubblegum}{rgb}{0.99, 0.76, 0.8}
\definecolor{bubbles}{rgb}{0.91, 1.0, 1.0}
\definecolor{buff}{rgb}{0.94, 0.86, 0.51}
\definecolor{bulgarianrose}{rgb}{0.28, 0.02, 0.03}
\definecolor{burgundy}{rgb}{0.5, 0.0, 0.13}
\definecolor{burlywood}{rgb}{0.87, 0.72, 0.53}
\definecolor{burntorange}{rgb}{0.8, 0.33, 0.0}
\definecolor{burntsienna}{rgb}{0.91, 0.45, 0.32}
\definecolor{burntumber}{rgb}{0.54, 0.2, 0.14}
\definecolor{byzantine}{rgb}{0.74, 0.2, 0.64}
\definecolor{byzantium}{rgb}{0.44, 0.16, 0.39}
\definecolor{cadet}{rgb}{0.33, 0.41, 0.47}
\definecolor{cadetblue}{rgb}{0.37, 0.62, 0.63}
\definecolor{cadetgrey}{rgb}{0.57, 0.64, 0.69}
\definecolor{cadmiumgreen}{rgb}{0.0, 0.42, 0.24}
\definecolor{cadmiumorange}{rgb}{0.93, 0.53, 0.18}
\definecolor{cadmiumred}{rgb}{0.89, 0.0, 0.13}
\definecolor{cadmiumyellow}{rgb}{1.0, 0.96, 0.0}
\definecolor{calpolypomonagreen}{rgb}{0.12, 0.3, 0.17}
\definecolor{cambridgeblue}{rgb}{0.64, 0.76, 0.68}
\definecolor{camel}{rgb}{0.76, 0.6, 0.42}
\definecolor{camouflagegreen}{rgb}{0.47, 0.53, 0.42}
\definecolor{canaryyellow}{rgb}{1.0, 0.94, 0.0}
\definecolor{candyapplered}{rgb}{1.0, 0.03, 0.0}
\definecolor{candypink}{rgb}{0.89, 0.44, 0.48}
\definecolor{capri}{rgb}{0.0, 0.75, 1.0}
\definecolor{caputmortuum}{rgb}{0.35, 0.15, 0.13}
\definecolor{cardinal}{rgb}{0.77, 0.12, 0.23}
\definecolor{caribbeangreen}{rgb}{0.0, 0.8, 0.6}
\definecolor{carmine}{rgb}{0.59, 0.0, 0.09}
\definecolor{carminepink}{rgb}{0.92, 0.3, 0.26}
\definecolor{carminered}{rgb}{1.0, 0.0, 0.22}
\definecolor{carnationpink}{rgb}{1.0, 0.65, 0.79}
\definecolor{carnelian}{rgb}{0.7, 0.11, 0.11}
\definecolor{carolinablue}{rgb}{0.6, 0.73, 0.89}
\definecolor{carrotorange}{rgb}{0.93, 0.57, 0.13}
\definecolor{ceil}{rgb}{0.57, 0.63, 0.81}
\definecolor{celadon}{rgb}{0.67, 0.88, 0.69}
\definecolor{celestialblue}{rgb}{0.29, 0.59, 0.82}
\definecolor{cerise}{rgb}{0.87, 0.19, 0.39}
\definecolor{cerisepink}{rgb}{0.93, 0.23, 0.51}
\definecolor{cerulean}{rgb}{0.0, 0.48, 0.65}
\definecolor{ceruleanblue}{rgb}{0.16, 0.32, 0.75}
\definecolor{chamoisee}{rgb}{0.63, 0.47, 0.35}
\definecolor{champagne}{rgb}{0.97, 0.91, 0.81}
\definecolor{charcoal}{rgb}{0.21, 0.27, 0.31}
\definecolor{chartreuse(traditional)}{rgb}{0.87, 1.0, 0.0}
\definecolor{chartreuse(web)}{rgb}{0.5, 1.0, 0.0}
\definecolor{cherryblossompink}{rgb}{1.0, 0.72, 0.77}
\definecolor{chestnut}{rgb}{0.8, 0.36, 0.36}
\definecolor{chocolate(traditional)}{rgb}{0.48, 0.25, 0.0}
\definecolor{chocolate(web)}{rgb}{0.82, 0.41, 0.12}
\definecolor{chromeyellow}{rgb}{1.0, 0.65, 0.0}
\definecolor{cinereous}{rgb}{0.6, 0.51, 0.48}
\definecolor{cinnabar}{rgb}{0.89, 0.26, 0.2}
\definecolor{cinnamon}{rgb}{0.82, 0.41, 0.12}
\definecolor{citrine}{rgb}{0.89, 0.82, 0.04}
\definecolor{classicrose}{rgb}{0.98, 0.8, 0.91}
\definecolor{cobalt}{rgb}{0.0, 0.28, 0.67}
\definecolor{cocoabrown}{rgb}{0.82, 0.41, 0.12}
\definecolor{columbiablue}{rgb}{0.61, 0.87, 1.0}
\definecolor{coolblack}{rgb}{0.0, 0.18, 0.39}
\definecolor{coolgrey}{rgb}{0.55, 0.57, 0.67}
\definecolor{copper}{rgb}{0.72, 0.45, 0.2}
\definecolor{copperrose}{rgb}{0.6, 0.4, 0.4}
\definecolor{coquelicot}{rgb}{1.0, 0.22, 0.0}
\definecolor{coral}{rgb}{1.0, 0.5, 0.31}
\definecolor{coralpink}{rgb}{0.97, 0.51, 0.47}
\definecolor{coralred}{rgb}{1.0, 0.25, 0.25}
\definecolor{cordovan}{rgb}{0.54, 0.25, 0.27}
\definecolor{corn}{rgb}{0.98, 0.93, 0.36}
\definecolor{cornellred}{rgb}{0.7, 0.11, 0.11}
\definecolor{cornflowerblue}{rgb}{0.39, 0.58, 0.93}
\definecolor{cornsilk}{rgb}{1.0, 0.97, 0.86}
\definecolor{cosmiclatte}{rgb}{1.0, 0.97, 0.91}
\definecolor{cottoncandy}{rgb}{1.0, 0.74, 0.85}
\definecolor{cream}{rgb}{1.0, 0.99, 0.82}
\definecolor{crimson}{rgb}{0.86, 0.08, 0.24}
\definecolor{crimsonglory}{rgb}{0.75, 0.0, 0.2}
\definecolor{cyan}{rgb}{0.0, 1.0, 1.0}
\definecolor{cyan(process)}{rgb}{0.0, 0.72, 0.92}
\definecolor{daffodil}{rgb}{1.0, 1.0, 0.19}
\definecolor{dandelion}{rgb}{0.94, 0.88, 0.19}
\definecolor{darkblue}{rgb}{0.0, 0.0, 0.55}
\definecolor{darkbrown}{rgb}{0.4, 0.26, 0.13}
\definecolor{darkbyzantium}{rgb}{0.36, 0.22, 0.33}
\definecolor{darkcandyapplered}{rgb}{0.64, 0.0, 0.0}
\definecolor{darkcerulean}{rgb}{0.03, 0.27, 0.49}
\definecolor{darkchampagne}{rgb}{0.76, 0.7, 0.5}
\definecolor{darkchestnut}{rgb}{0.6, 0.41, 0.38}
\definecolor{darkcoral}{rgb}{0.8, 0.36, 0.27}
\definecolor{darkcyan}{rgb}{0.0, 0.55, 0.55}
\definecolor{darkelectricblue}{rgb}{0.33, 0.41, 0.47}
\definecolor{darkgoldenrod}{rgb}{0.72, 0.53, 0.04}
\definecolor{darkgray}{rgb}{0.66, 0.66, 0.66}
\definecolor{darkgreen}{rgb}{0.0, 0.2, 0.13}
\definecolor{darkjunglegreen}{rgb}{0.1, 0.14, 0.13}
\definecolor{darkkhaki}{rgb}{0.74, 0.72, 0.42}
\definecolor{darklava}{rgb}{0.28, 0.24, 0.2}
\definecolor{darklavender}{rgb}{0.45, 0.31, 0.59}
\definecolor{darkmagenta}{rgb}{0.55, 0.0, 0.55}
\definecolor{darkmidnightblue}{rgb}{0.0, 0.2, 0.4}
\definecolor{darkolivegreen}{rgb}{0.33, 0.42, 0.18}
\definecolor{darkorange}{rgb}{1.0, 0.55, 0.0}
\definecolor{darkorchid}{rgb}{0.6, 0.2, 0.8}
\definecolor{darkpastelblue}{rgb}{0.47, 0.62, 0.8}
\definecolor{darkpastelgreen}{rgb}{0.01, 0.75, 0.24}
\definecolor{darkpastelpurple}{rgb}{0.59, 0.44, 0.84}
\definecolor{darkpastelred}{rgb}{0.76, 0.23, 0.13}
\definecolor{darkpink}{rgb}{0.91, 0.33, 0.5}
\definecolor{darkpowderblue}{rgb}{0.0, 0.2, 0.6}
\definecolor{darkraspberry}{rgb}{0.53, 0.15, 0.34}
\definecolor{darkred}{rgb}{0.55, 0.0, 0.0}
\definecolor{darksalmon}{rgb}{0.91, 0.59, 0.48}
\definecolor{darkscarlet}{rgb}{0.34, 0.01, 0.1}
\definecolor{darkseagreen}{rgb}{0.56, 0.74, 0.56}
\definecolor{darksienna}{rgb}{0.24, 0.08, 0.08}
\definecolor{darkslateblue}{rgb}{0.28, 0.24, 0.55}
\definecolor{darkslategray}{rgb}{0.18, 0.31, 0.31}
\definecolor{darkspringgreen}{rgb}{0.09, 0.45, 0.27}
\definecolor{darktan}{rgb}{0.57, 0.51, 0.32}
\definecolor{darktangerine}{rgb}{1.0, 0.66, 0.07}
\definecolor{darktaupe}{rgb}{0.28, 0.24, 0.2}
\definecolor{darkterracotta}{rgb}{0.8, 0.31, 0.36}
\definecolor{darkturquoise}{rgb}{0.0, 0.81, 0.82}
\definecolor{darkviolet}{rgb}{0.58, 0.0, 0.83}
\definecolor{dartmouthgreen}{rgb}{0.05, 0.5, 0.06}
\definecolor{davysgrey}{rgb}{0.33, 0.33, 0.33}
\definecolor{debianred}{rgb}{0.84, 0.04, 0.33}
\definecolor{deepcarmine}{rgb}{0.66, 0.13, 0.24}
\definecolor{deepcarminepink}{rgb}{0.94, 0.19, 0.22}
\definecolor{deepcarrotorange}{rgb}{0.91, 0.41, 0.17}
\definecolor{deepcerise}{rgb}{0.85, 0.2, 0.53}
\definecolor{deepchampagne}{rgb}{0.98, 0.84, 0.65}
\definecolor{deepchestnut}{rgb}{0.73, 0.31, 0.28}
\definecolor{deepfuchsia}{rgb}{0.76, 0.33, 0.76}
\definecolor{deepjunglegreen}{rgb}{0.0, 0.29, 0.29}
\definecolor{deeplilac}{rgb}{0.6, 0.33, 0.73}
\definecolor{deepmagenta}{rgb}{0.8, 0.0, 0.8}
\definecolor{deeppeach}{rgb}{1.0, 0.8, 0.64}
\definecolor{deeppink}{rgb}{1.0, 0.08, 0.58}
\definecolor{deepsaffron}{rgb}{1.0, 0.6, 0.2}
\definecolor{deepskyblue}{rgb}{0.0, 0.75, 1.0}
\definecolor{denim}{rgb}{0.08, 0.38, 0.74}
\definecolor{desert}{rgb}{0.76, 0.6, 0.42}
\definecolor{desertsand}{rgb}{0.93, 0.79, 0.69}
\definecolor{dimgray}{rgb}{0.41, 0.41, 0.41}
\definecolor{dodgerblue}{rgb}{0.12, 0.56, 1.0}
\definecolor{dogwoodrose}{rgb}{0.84, 0.09, 0.41}
\definecolor{dollarbill}{rgb}{0.52, 0.73, 0.4}
\definecolor{drab}{rgb}{0.59, 0.44, 0.09}
\definecolor{dukeblue}{rgb}{0.0, 0.0, 0.61}
\definecolor{earthyellow}{rgb}{0.88, 0.66, 0.37}
\definecolor{ecru}{rgb}{0.76, 0.7, 0.5}
\definecolor{eggplant}{rgb}{0.38, 0.25, 0.32}
\definecolor{eggshell}{rgb}{0.94, 0.92, 0.84}
\definecolor{egyptianblue}{rgb}{0.06, 0.2, 0.65}
\definecolor{electricblue}{rgb}{0.49, 0.98, 1.0}
\definecolor{electriccrimson}{rgb}{1.0, 0.0, 0.25}
\definecolor{electriccyan}{rgb}{0.0, 1.0, 1.0}
\definecolor{electricgreen}{rgb}{0.0, 1.0, 0.0}
\definecolor{electricindigo}{rgb}{0.44, 0.0, 1.0}
\definecolor{electriclavender}{rgb}{0.96, 0.73, 1.0}
\definecolor{electriclime}{rgb}{0.8, 1.0, 0.0}
\definecolor{electricpurple}{rgb}{0.75, 0.0, 1.0}
\definecolor{electricultramarine}{rgb}{0.25, 0.0, 1.0}
\definecolor{electricviolet}{rgb}{0.56, 0.0, 1.0}
\definecolor{electricyellow}{rgb}{1.0, 1.0, 0.0}
\definecolor{emerald}{rgb}{0.31, 0.78, 0.47}
\definecolor{etonblue}{rgb}{0.59, 0.78, 0.64}
\definecolor{fallow}{rgb}{0.76, 0.6, 0.42}
\definecolor{falured}{rgb}{0.5, 0.09, 0.09}
\definecolor{fandango}{rgb}{0.71, 0.2, 0.54}
\definecolor{fashionfuchsia}{rgb}{0.96, 0.0, 0.63}
\definecolor{fawn}{rgb}{0.9, 0.67, 0.44}
\definecolor{feldgrau}{rgb}{0.3, 0.36, 0.33}
\definecolor{ferngreen}{rgb}{0.31, 0.47, 0.26}
\definecolor{ferrarired}{rgb}{1.0, 0.11, 0.0}
\definecolor{fielddrab}{rgb}{0.42, 0.33, 0.12}
\definecolor{firebrick}{rgb}{0.7, 0.13, 0.13}
\definecolor{fireenginered}{rgb}{0.81, 0.09, 0.13}
\definecolor{flame}{rgb}{0.89, 0.35, 0.13}
\definecolor{flamingopink}{rgb}{0.99, 0.56, 0.67}
\definecolor{flavescent}{rgb}{0.97, 0.91, 0.56}
\definecolor{flax}{rgb}{0.93, 0.86, 0.51}
\definecolor{floralwhite}{rgb}{1.0, 0.98, 0.94}
\definecolor{fluorescentorange}{rgb}{1.0, 0.75, 0.0}
\definecolor{fluorescentpink}{rgb}{1.0, 0.08, 0.58}
\definecolor{fluorescentyellow}{rgb}{0.8, 1.0, 0.0}
\definecolor{folly}{rgb}{1.0, 0.0, 0.31}
\definecolor{forestgreen(traditional)}{rgb}{0.0, 0.27, 0.13}
\definecolor{forestgreen(web)}{rgb}{0.13, 0.55, 0.13}
\definecolor{frenchbeige}{rgb}{0.65, 0.48, 0.36}
\definecolor{frenchblue}{rgb}{0.0, 0.45, 0.73}
\definecolor{frenchlilac}{rgb}{0.53, 0.38, 0.56}
\definecolor{frenchrose}{rgb}{0.96, 0.29, 0.54}
\definecolor{fuchsia}{rgb}{1.0, 0.0, 1.0}
\definecolor{fuchsiapink}{rgb}{1.0, 0.47, 1.0}
\definecolor{fulvous}{rgb}{0.86, 0.52, 0.0}
\definecolor{fuzzywuzzy}{rgb}{0.8, 0.4, 0.4}
\definecolor{gainsboro}{rgb}{0.86, 0.86, 0.86}
\definecolor{gamboge}{rgb}{0.89, 0.61, 0.06}
\definecolor{ghostwhite}{rgb}{0.97, 0.97, 1.0}
\definecolor{ginger}{rgb}{0.69, 0.4, 0.0}
\definecolor{glaucous}{rgb}{0.38, 0.51, 0.71}
\definecolor{gold(metallic)}{rgb}{0.83, 0.69, 0.22}
\definecolor{gold(web)(golden)}{rgb}{1.0, 0.84, 0.0}
\definecolor{goldenbrown}{rgb}{0.6, 0.4, 0.08}
\definecolor{goldenpoppy}{rgb}{0.99, 0.76, 0.0}
\definecolor{goldenyellow}{rgb}{1.0, 0.87, 0.0}
\definecolor{goldenrod}{rgb}{0.85, 0.65, 0.13}
\definecolor{grannysmithapple}{rgb}{0.66, 0.89, 0.63}
\definecolor{gray}{rgb}{0.5, 0.5, 0.5}
\definecolor{gray(html/cssgray)}{rgb}{0.5, 0.5, 0.5}
\definecolor{gray(x11gray)}{rgb}{0.75, 0.75, 0.75}
\definecolor{gray-asparagus}{rgb}{0.27, 0.35, 0.27}
\definecolor{green(colorwheel)(x11green)}{rgb}{0.0, 1.0, 0.0}
\definecolor{green(html/cssgreen)}{rgb}{0.0, 0.5, 0.0}
\definecolor{green(munsell)}{rgb}{0.0, 0.66, 0.47}
\definecolor{green(ncs)}{rgb}{0.0, 0.62, 0.42}
\definecolor{green(pigment)}{rgb}{0.0, 0.65, 0.31}
\definecolor{green(ryb)}{rgb}{0.4, 0.69, 0.2}
\definecolor{green-yellow}{rgb}{0.68, 1.0, 0.18}
\definecolor{grullo}{rgb}{0.66, 0.6, 0.53}
\definecolor{guppiegreen}{rgb}{0.0, 1.0, 0.5}
\definecolor{halayaube}{rgb}{0.4, 0.22, 0.33}
\definecolor{hanblue}{rgb}{0.27, 0.42, 0.81}
\definecolor{hanpurple}{rgb}{0.32, 0.09, 0.98}
\definecolor{hansayellow}{rgb}{0.91, 0.84, 0.42}
\definecolor{harlequin}{rgb}{0.25, 1.0, 0.0}
\definecolor{harvardcrimson}{rgb}{0.79, 0.0, 0.09}
\definecolor{harvestgold}{rgb}{0.85, 0.57, 0.0}
\definecolor{heartgold}{rgb}{0.5, 0.5, 0.0}
\definecolor{heliotrope}{rgb}{0.87, 0.45, 1.0}
\definecolor{hollywoodcerise}{rgb}{0.96, 0.0, 0.63}
\definecolor{honeydew}{rgb}{0.94, 1.0, 0.94}
\definecolor{hookersgreen}{rgb}{0.0, 0.44, 0.0}
\definecolor{hotmagenta}{rgb}{1.0, 0.11, 0.81}
\definecolor{hotpink}{rgb}{1.0, 0.41, 0.71}
\definecolor{huntergreen}{rgb}{0.21, 0.37, 0.23}
\definecolor{iceberg}{rgb}{0.44, 0.65, 0.82}
\definecolor{icterine}{rgb}{0.99, 0.97, 0.37}
\definecolor{inchworm}{rgb}{0.7, 0.93, 0.36}
\definecolor{indiagreen}{rgb}{0.07, 0.53, 0.03}
\definecolor{indianred}{rgb}{0.8, 0.36, 0.36}
\definecolor{indianyellow}{rgb}{0.89, 0.66, 0.34}
\definecolor{indigo(dye)}{rgb}{0.0, 0.25, 0.42}
\definecolor{indigo(web)}{rgb}{0.29, 0.0, 0.51}
\definecolor{internationalkleinblue}{rgb}{0.0, 0.18, 0.65}
\definecolor{internationalorange}{rgb}{1.0, 0.31, 0.0}
\definecolor{iris}{rgb}{0.35, 0.31, 0.81}
\definecolor{isabelline}{rgb}{0.96, 0.94, 0.93}
\definecolor{islamicgreen}{rgb}{0.0, 0.56, 0.0}
\definecolor{ivory}{rgb}{1.0, 1.0, 0.94}
\definecolor{jade}{rgb}{0.0, 0.66, 0.42}
\definecolor{jasper}{rgb}{0.84, 0.23, 0.24}
\definecolor{jazzberryjam}{rgb}{0.65, 0.04, 0.37}
\definecolor{jonquil}{rgb}{0.98, 0.85, 0.37}
\definecolor{junebud}{rgb}{0.74, 0.85, 0.34}
\definecolor{junglegreen}{rgb}{0.16, 0.67, 0.53}
\definecolor{kellygreen}{rgb}{0.3, 0.73, 0.09}
\definecolor{khaki(html/css)(khaki)}{rgb}{0.76, 0.69, 0.57}
\definecolor{khaki(x11)(lightkhaki)}{rgb}{0.94, 0.9, 0.55}
\definecolor{lasallegreen}{rgb}{0.03, 0.47, 0.19}
\definecolor{languidlavender}{rgb}{0.84, 0.79, 0.87}
\definecolor{lapislazuli}{rgb}{0.15, 0.38, 0.61}
\definecolor{laserlemon}{rgb}{1.0, 1.0, 0.13}
\definecolor{lava}{rgb}{0.81, 0.06, 0.13}
\definecolor{lavender(floral)}{rgb}{0.71, 0.49, 0.86}
\definecolor{lavender(web)}{rgb}{0.9, 0.9, 0.98}
\definecolor{lavenderblue}{rgb}{0.8, 0.8, 1.0}
\definecolor{lavenderblush}{rgb}{1.0, 0.94, 0.96}
\definecolor{lavendergray}{rgb}{0.77, 0.76, 0.82}
\definecolor{lavenderindigo}{rgb}{0.58, 0.34, 0.92}
\definecolor{lavendermagenta}{rgb}{0.93, 0.51, 0.93}
\definecolor{lavendermist}{rgb}{0.9, 0.9, 0.98}
\definecolor{lavenderpink}{rgb}{0.98, 0.68, 0.82}
\definecolor{lavenderpurple}{rgb}{0.59, 0.48, 0.71}
\definecolor{lavenderrose}{rgb}{0.98, 0.63, 0.89}
\definecolor{lawngreen}{rgb}{0.49, 0.99, 0.0}
\definecolor{lemon}{rgb}{1.0, 0.97, 0.0}
\definecolor{lemonchiffon}{rgb}{1.0, 0.98, 0.8}
\definecolor{lightapricot}{rgb}{0.99, 0.84, 0.69}
\definecolor{lightblue}{rgb}{0.68, 0.85, 0.9}
\definecolor{lightbrown}{rgb}{0.71, 0.4, 0.11}
\definecolor{lightcarminepink}{rgb}{0.9, 0.4, 0.38}
\definecolor{lightcoral}{rgb}{0.94, 0.5, 0.5}
\definecolor{lightcornflowerblue}{rgb}{0.6, 0.81, 0.93}
\definecolor{lightcyan}{rgb}{0.88, 1.0, 1.0}
\definecolor{lightfuchsiapink}{rgb}{0.98, 0.52, 0.9}
\definecolor{lightgoldenrodyellow}{rgb}{0.98, 0.98, 0.82}
\definecolor{lightgray}{rgb}{0.83, 0.83, 0.83}
\definecolor{lightgreen}{rgb}{0.56, 0.93, 0.56}
\definecolor{lightkhaki}{rgb}{0.94, 0.9, 0.55}
\definecolor{lightmauve}{rgb}{0.86, 0.82, 1.0}
\definecolor{lightpastelpurple}{rgb}{0.69, 0.61, 0.85}
\definecolor{lightpink}{rgb}{1.0, 0.71, 0.76}
\definecolor{lightsalmon}{rgb}{1.0, 0.63, 0.48}
\definecolor{lightsalmonpink}{rgb}{1.0, 0.6, 0.6}
\definecolor{lightseagreen}{rgb}{0.13, 0.7, 0.67}
\definecolor{lightskyblue}{rgb}{0.53, 0.81, 0.98}
\definecolor{lightslategray}{rgb}{0.47, 0.53, 0.6}
\definecolor{lighttaupe}{rgb}{0.7, 0.55, 0.43}
\definecolor{lightthulianpink}{rgb}{0.9, 0.56, 0.67}
\definecolor{lightyellow}{rgb}{1.0, 1.0, 0.88}
\definecolor{lilac}{rgb}{0.78, 0.64, 0.78}
\definecolor{lime(colorwheel)}{rgb}{0.75, 1.0, 0.0}
\definecolor{lime(web)(x11green)}{rgb}{0.0, 1.0, 0.0}
\definecolor{limegreen}{rgb}{0.2, 0.8, 0.2}
\definecolor{lincolngreen}{rgb}{0.11, 0.35, 0.02}
\definecolor{linen}{rgb}{0.98, 0.94, 0.9}
\definecolor{liver}{rgb}{0.33, 0.29, 0.31}
\definecolor{lust}{rgb}{0.9, 0.13, 0.13}
\definecolor{macaroniandcheese}{rgb}{1.0, 0.74, 0.53}
\definecolor{magenta}{rgb}{1.0, 0.0, 1.0}
\definecolor{magenta(dye)}{rgb}{0.79, 0.08, 0.48}
\definecolor{magenta(process)}{rgb}{1.0, 0.0, 0.56}
\definecolor{magicmint}{rgb}{0.67, 0.94, 0.82}
\definecolor{magnolia}{rgb}{0.97, 0.96, 1.0}
\definecolor{mahogany}{rgb}{0.75, 0.25, 0.0}
\definecolor{maize}{rgb}{0.98, 0.93, 0.37}
\definecolor{majorelleblue}{rgb}{0.38, 0.31, 0.86}
\definecolor{malachite}{rgb}{0.04, 0.85, 0.32}
\definecolor{manatee}{rgb}{0.59, 0.6, 0.67}
\definecolor{mangotango}{rgb}{1.0, 0.51, 0.26}
\definecolor{maroon(html/css)}{rgb}{0.5, 0.0, 0.0}
\definecolor{maroon(x11)}{rgb}{0.69, 0.19, 0.38}
\definecolor{mauve}{rgb}{0.88, 0.69, 1.0}
\definecolor{mauvetaupe}{rgb}{0.57, 0.37, 0.43}
\definecolor{mauvelous}{rgb}{0.94, 0.6, 0.67}
\definecolor{mayablue}{rgb}{0.45, 0.76, 0.98}
\definecolor{meatbrown}{rgb}{0.9, 0.72, 0.23}
\definecolor{mediumaquamarine}{rgb}{0.4, 0.8, 0.67}
\definecolor{mediumblue}{rgb}{0.0, 0.0, 0.8}
\definecolor{mediumcandyapplered}{rgb}{0.89, 0.02, 0.17}
\definecolor{mediumcarmine}{rgb}{0.69, 0.25, 0.21}
\definecolor{mediumchampagne}{rgb}{0.95, 0.9, 0.67}
\definecolor{mediumelectricblue}{rgb}{0.01, 0.31, 0.59}
\definecolor{mediumjunglegreen}{rgb}{0.11, 0.21, 0.18}
\definecolor{mediumlavendermagenta}{rgb}{0.8, 0.6, 0.8}
\definecolor{mediumorchid}{rgb}{0.73, 0.33, 0.83}
\definecolor{mediumpersianblue}{rgb}{0.0, 0.4, 0.65}
\definecolor{mediumpurple}{rgb}{0.58, 0.44, 0.86}
\definecolor{mediumred-violet}{rgb}{0.73, 0.2, 0.52}
\definecolor{mediumseagreen}{rgb}{0.24, 0.7, 0.44}
\definecolor{mediumslateblue}{rgb}{0.48, 0.41, 0.93}
\definecolor{mediumspringbud}{rgb}{0.79, 0.86, 0.54}
\definecolor{mediumspringgreen}{rgb}{0.0, 0.98, 0.6}
\definecolor{mediumtaupe}{rgb}{0.4, 0.3, 0.28}
\definecolor{mediumtealblue}{rgb}{0.0, 0.33, 0.71}
\definecolor{mediumturquoise}{rgb}{0.28, 0.82, 0.8}
\definecolor{mediumviolet-red}{rgb}{0.78, 0.08, 0.52}
\definecolor{melon}{rgb}{0.99, 0.74, 0.71}
\definecolor{midnightblue}{rgb}{0.1, 0.1, 0.44}
\definecolor{midnightgreen(eaglegreen)}{rgb}{0.0, 0.29, 0.33}
\definecolor{mikadoyellow}{rgb}{1.0, 0.77, 0.05}
\definecolor{mint}{rgb}{0.24, 0.71, 0.54}
\definecolor{mintcream}{rgb}{0.96, 1.0, 0.98}
\definecolor{mintgreen}{rgb}{0.6, 1.0, 0.6}
\definecolor{mistyrose}{rgb}{1.0, 0.89, 0.88}
\definecolor{moccasin}{rgb}{0.98, 0.92, 0.84}
\definecolor{modebeige}{rgb}{0.59, 0.44, 0.09}
\definecolor{moonstoneblue}{rgb}{0.45, 0.66, 0.76}
\definecolor{mordantred19}{rgb}{0.68, 0.05, 0.0}
\definecolor{mossgreen}{rgb}{0.68, 0.87, 0.68}
\definecolor{mountainmeadow}{rgb}{0.19, 0.73, 0.56}
\definecolor{mountbattenpink}{rgb}{0.6, 0.48, 0.55}
\definecolor{mulberry}{rgb}{0.77, 0.29, 0.55}
\definecolor{mustard}{rgb}{1.0, 0.86, 0.35}
\definecolor{myrtle}{rgb}{0.13, 0.26, 0.12}
\definecolor{msugreen}{rgb}{0.09, 0.27, 0.23}
\definecolor{nadeshikopink}{rgb}{0.96, 0.68, 0.78}
\definecolor{napiergreen}{rgb}{0.16, 0.5, 0.0}
\definecolor{naplesyellow}{rgb}{0.98, 0.85, 0.37}
\definecolor{navajowhite}{rgb}{1.0, 0.87, 0.68}
\definecolor{navyblue}{rgb}{0.0, 0.0, 0.5}
\definecolor{neoncarrot}{rgb}{1.0, 0.64, 0.26}
\definecolor{neonfuchsia}{rgb}{1.0, 0.25, 0.39}
\definecolor{neongreen}{rgb}{0.22, 0.88, 0.08}
\definecolor{non-photoblue}{rgb}{0.64, 0.87, 0.93}
\definecolor{oceanboatblue}{rgb}{0.0, 0.47, 0.75}
\definecolor{ochre}{rgb}{0.8, 0.47, 0.13}
\definecolor{officegreen}{rgb}{0.0, 0.5, 0.0}
\definecolor{oldgold}{rgb}{0.81, 0.71, 0.23}
\definecolor{oldlace}{rgb}{0.99, 0.96, 0.9}
\definecolor{oldlavender}{rgb}{0.47, 0.41, 0.47}
\definecolor{oldmauve}{rgb}{0.4, 0.19, 0.28}
\definecolor{oldrose}{rgb}{0.75, 0.5, 0.51}
\definecolor{olive}{rgb}{0.5, 0.5, 0.0}
\definecolor{olivedrabN3}{rgb}{0.42, 0.56, 0.14}
\definecolor{olivedrabN7}{rgb}{0.24, 0.2, 0.12}
\definecolor{olivine}{rgb}{0.6, 0.73, 0.45}
\definecolor{onyx}{rgb}{0.06, 0.06, 0.06}
\definecolor{operamauve}{rgb}{0.72, 0.52, 0.65}
\definecolor{orange(colorwheel)}{rgb}{1.0, 0.5, 0.0}
\definecolor{orange(ryb)}{rgb}{0.98, 0.6, 0.01}
\definecolor{orange(webcolor)}{rgb}{1.0, 0.65, 0.0}
\definecolor{orangepeel}{rgb}{1.0, 0.62, 0.0}
\definecolor{orange-red}{rgb}{1.0, 0.27, 0.0}
\definecolor{orchid}{rgb}{0.85, 0.44, 0.84}
\definecolor{otterbrown}{rgb}{0.4, 0.26, 0.13}
\definecolor{outerspace}{rgb}{0.25, 0.29, 0.3}
\definecolor{outrageousorange}{rgb}{1.0, 0.43, 0.29}
\definecolor{oxfordblue}{rgb}{0.0, 0.13, 0.28}
\definecolor{oucrimsonred}{rgb}{0.6, 0.0, 0.0}
\definecolor{pakistangreen}{rgb}{0.0, 0.4, 0.0}
\definecolor{palatinateblue}{rgb}{0.15, 0.23, 0.89}
\definecolor{palatinatepurple}{rgb}{0.41, 0.16, 0.38}
\definecolor{paleaqua}{rgb}{0.74, 0.83, 0.9}
\definecolor{paleblue}{rgb}{0.69, 0.93, 0.93}
\definecolor{palebrown}{rgb}{0.6, 0.46, 0.33}
\definecolor{palecarmine}{rgb}{0.69, 0.25, 0.21}
\definecolor{palecerulean}{rgb}{0.61, 0.77, 0.89}
\definecolor{palechestnut}{rgb}{0.87, 0.68, 0.69}
\definecolor{palecopper}{rgb}{0.85, 0.54, 0.4}
\definecolor{palecornflowerblue}{rgb}{0.67, 0.8, 0.94}
\definecolor{palegold}{rgb}{0.9, 0.75, 0.54}
\definecolor{palegoldenrod}{rgb}{0.93, 0.91, 0.67}
\definecolor{palegreen}{rgb}{0.6, 0.98, 0.6}
\definecolor{palemagenta}{rgb}{0.98, 0.52, 0.9}
\definecolor{palepink}{rgb}{0.98, 0.85, 0.87}
\definecolor{paleplum}{rgb}{0.8, 0.6, 0.8}
\definecolor{palered-violet}{rgb}{0.86, 0.44, 0.58}
\definecolor{palerobineggblue}{rgb}{0.59, 0.87, 0.82}
\definecolor{palesilver}{rgb}{0.79, 0.75, 0.73}
\definecolor{palespringbud}{rgb}{0.93, 0.92, 0.74}
\definecolor{paletaupe}{rgb}{0.74, 0.6, 0.49}
\definecolor{paleviolet-red}{rgb}{0.86, 0.44, 0.58}
\definecolor{pansypurple}{rgb}{0.47, 0.09, 0.29}
\definecolor{papayawhip}{rgb}{1.0, 0.94, 0.84}
\definecolor{parisgreen}{rgb}{0.31, 0.78, 0.47}
\definecolor{pastelblue}{rgb}{0.68, 0.78, 0.81}
\definecolor{pastelbrown}{rgb}{0.51, 0.41, 0.33}
\definecolor{pastelgray}{rgb}{0.81, 0.81, 0.77}
\definecolor{pastelgreen}{rgb}{0.47, 0.87, 0.47}
\definecolor{pastelmagenta}{rgb}{0.96, 0.6, 0.76}
\definecolor{pastelorange}{rgb}{1.0, 0.7, 0.28}
\definecolor{pastelpink}{rgb}{1.0, 0.82, 0.86}
\definecolor{pastelpurple}{rgb}{0.7, 0.62, 0.71}
\definecolor{pastelred}{rgb}{1.0, 0.41, 0.38}
\definecolor{pastelviolet}{rgb}{0.8, 0.6, 0.79}
\definecolor{pastelyellow}{rgb}{0.99, 0.99, 0.59}
\definecolor{patriarch}{rgb}{0.5, 0.0, 0.5}
\definecolor{paynesgrey}{rgb}{0.25, 0.25, 0.28}
\definecolor{peach}{rgb}{1.0, 0.9, 0.71}
\definecolor{peach-orange}{rgb}{1.0, 0.8, 0.6}
\definecolor{peachpuff}{rgb}{1.0, 0.85, 0.73}
\definecolor{peach-yellow}{rgb}{0.98, 0.87, 0.68}
\definecolor{pear}{rgb}{0.82, 0.89, 0.19}
\definecolor{pearl}{rgb}{0.94, 0.92, 0.84}
\definecolor{peridot}{rgb}{0.9, 0.89, 0.0}
\definecolor{periwinkle}{rgb}{0.8, 0.8, 1.0}
\definecolor{persianblue}{rgb}{0.11, 0.22, 0.73}
\definecolor{persiangreen}{rgb}{0.0, 0.65, 0.58}
\definecolor{persianindigo}{rgb}{0.2, 0.07, 0.48}
\definecolor{persianorange}{rgb}{0.85, 0.56, 0.35}
\definecolor{peru}{rgb}{0.8, 0.52, 0.25}
\definecolor{persianpink}{rgb}{0.97, 0.5, 0.75}
\definecolor{persianplum}{rgb}{0.44, 0.11, 0.11}
\definecolor{persianred}{rgb}{0.8, 0.2, 0.2}
\definecolor{persianrose}{rgb}{1.0, 0.16, 0.64}
\definecolor{persimmon}{rgb}{0.93, 0.35, 0.0}
\definecolor{phlox}{rgb}{0.87, 0.0, 1.0}
\definecolor{phthaloblue}{rgb}{0.0, 0.06, 0.54}
\definecolor{phthalogreen}{rgb}{0.07, 0.21, 0.14}
\definecolor{piggypink}{rgb}{0.99, 0.87, 0.9}
\definecolor{pinegreen}{rgb}{0.0, 0.47, 0.44}
\definecolor{pink}{rgb}{1.0, 0.75, 0.8}
\definecolor{pink-orange}{rgb}{1.0, 0.6, 0.4}
\definecolor{pinkpearl}{rgb}{0.91, 0.67, 0.81}
\definecolor{pinksherbet}{rgb}{0.97, 0.56, 0.65}
\definecolor{pistachio}{rgb}{0.58, 0.77, 0.45}
\definecolor{platinum}{rgb}{0.9, 0.89, 0.89}
\definecolor{plum(traditional)}{rgb}{0.56, 0.27, 0.52}
\definecolor{plum(web)}{rgb}{0.8, 0.6, 0.8}
\definecolor{portlandorange}{rgb}{1.0, 0.35, 0.21}
\definecolor{powderblue(web)}{rgb}{0.69, 0.88, 0.9}
\definecolor{princetonorange}{rgb}{1.0, 0.56, 0.0}
\definecolor{prune}{rgb}{0.44, 0.11, 0.11}
\definecolor{prussianblue}{rgb}{0.0, 0.19, 0.33}
\definecolor{psychedelicpurple}{rgb}{0.87, 0.0, 1.0}
\definecolor{puce}{rgb}{0.8, 0.53, 0.6}
\definecolor{pumpkin}{rgb}{1.0, 0.46, 0.09}
\definecolor{purple(html/css)}{rgb}{0.5, 0.0, 0.5}
\definecolor{purple(munsell)}{rgb}{0.62, 0.0, 0.77}
\definecolor{purple(x11)}{rgb}{0.63, 0.36, 0.94}
\definecolor{purpleheart}{rgb}{0.41, 0.21, 0.61}
\definecolor{purplemountainmajesty}{rgb}{0.59, 0.47, 0.71}
\definecolor{purplepizzazz}{rgb}{1.0, 0.31, 0.85}
\definecolor{purpletaupe}{rgb}{0.31, 0.25, 0.3}
\definecolor{radicalred}{rgb}{1.0, 0.21, 0.37}
\definecolor{raspberry}{rgb}{0.89, 0.04, 0.36}
\definecolor{raspberryglace}{rgb}{0.57, 0.37, 0.43}
\definecolor{raspberrypink}{rgb}{0.89, 0.31, 0.61}
\definecolor{raspberryrose}{rgb}{0.7, 0.27, 0.42}
\definecolor{rawumber}{rgb}{0.51, 0.4, 0.27}
\definecolor{razzledazzlerose}{rgb}{1.0, 0.2, 0.8}
\definecolor{razzmatazz}{rgb}{0.89, 0.15, 0.42}
\definecolor{red}{rgb}{1.0, 0.0, 0.0}
\definecolor{red(munsell)}{rgb}{0.95, 0.0, 0.24}
\definecolor{red(ncs)}{rgb}{0.77, 0.01, 0.2}
\definecolor{red(pigment)}{rgb}{0.93, 0.11, 0.14}
\definecolor{red(ryb)}{rgb}{1.0, 0.15, 0.07}
\definecolor{red-brown}{rgb}{0.65, 0.16, 0.16}
\definecolor{red-violet}{rgb}{0.78, 0.08, 0.52}
\definecolor{redwood}{rgb}{0.67, 0.31, 0.32}
\definecolor{regalia}{rgb}{0.32, 0.18, 0.5}
\definecolor{richblack}{rgb}{0.0, 0.25, 0.25}
\definecolor{richbrilliantlavender}{rgb}{0.95, 0.65, 1.0}
\definecolor{richcarmine}{rgb}{0.84, 0.0, 0.25}
\definecolor{richelectricblue}{rgb}{0.03, 0.57, 0.82}
\definecolor{richlavender}{rgb}{0.67, 0.38, 0.8}
\definecolor{richlilac}{rgb}{0.71, 0.4, 0.82}
\definecolor{richmaroon}{rgb}{0.69, 0.19, 0.38}
\definecolor{riflegreen}{rgb}{0.25, 0.28, 0.2}
\definecolor{robineggblue}{rgb}{0.0, 0.8, 0.8}
\definecolor{rose}{rgb}{1.0, 0.0, 0.5}
\definecolor{rosebonbon}{rgb}{0.98, 0.26, 0.62}
\definecolor{roseebony}{rgb}{0.4, 0.3, 0.28}
\definecolor{rosegold}{rgb}{0.72, 0.43, 0.47}
\definecolor{rosemadder}{rgb}{0.89, 0.15, 0.21}
\definecolor{rosepink}{rgb}{1.0, 0.4, 0.8}
\definecolor{rosequartz}{rgb}{0.67, 0.6, 0.66}
\definecolor{rosetaupe}{rgb}{0.56, 0.36, 0.36}
\definecolor{rosevale}{rgb}{0.67, 0.31, 0.32}
\definecolor{rosewood}{rgb}{0.4, 0.0, 0.04}
\definecolor{rossocorsa}{rgb}{0.83, 0.0, 0.0}
\definecolor{rosybrown}{rgb}{0.74, 0.56, 0.56}
\definecolor{royalazure}{rgb}{0.0, 0.22, 0.66}
\definecolor{royalblue(traditional)}{rgb}{0.0, 0.14, 0.4}
\definecolor{royalblue(web)}{rgb}{0.25, 0.41, 0.88}
\definecolor{royalfuchsia}{rgb}{0.79, 0.17, 0.57}
\definecolor{royalpurple}{rgb}{0.47, 0.32, 0.66}
\definecolor{ruby}{rgb}{0.88, 0.07, 0.37}
\definecolor{ruddy}{rgb}{1.0, 0.0, 0.16}
\definecolor{ruddybrown}{rgb}{0.73, 0.4, 0.16}
\definecolor{ruddypink}{rgb}{0.88, 0.56, 0.59}
\definecolor{rufous}{rgb}{0.66, 0.11, 0.03}
\definecolor{russet}{rgb}{0.5, 0.27, 0.11}
\definecolor{rust}{rgb}{0.72, 0.25, 0.05}
\definecolor{sacramentostategreen}{rgb}{0.0, 0.34, 0.25}
\definecolor{saddlebrown}{rgb}{0.55, 0.27, 0.07}
\definecolor{safetyorange(blazeorange)}{rgb}{1.0, 0.4, 0.0}
\definecolor{saffron}{rgb}{0.96, 0.77, 0.19}
\definecolor{st.patricksblue}{rgb}{0.14, 0.16, 0.48}
\definecolor{salmon}{rgb}{1.0, 0.55, 0.41}
\definecolor{salmonpink}{rgb}{1.0, 0.57, 0.64}
\definecolor{sand}{rgb}{0.76, 0.7, 0.5}
\definecolor{sanddune}{rgb}{0.59, 0.44, 0.09}
\definecolor{sandstorm}{rgb}{0.93, 0.84, 0.25}
\definecolor{sandybrown}{rgb}{0.96, 0.64, 0.38}
\definecolor{sandytaupe}{rgb}{0.59, 0.44, 0.09}
\definecolor{sangria}{rgb}{0.57, 0.0, 0.04}
\definecolor{sapgreen}{rgb}{0.31, 0.49, 0.16}
\definecolor{sapphire}{rgb}{0.03, 0.15, 0.4}
\definecolor{satinsheengold}{rgb}{0.8, 0.63, 0.21}
\definecolor{scarlet}{rgb}{1.0, 0.13, 0.0}
\definecolor{schoolbusyellow}{rgb}{1.0, 0.85, 0.0}
\definecolor{screamingreen}{rgb}{0.46, 1.0, 0.44}
\definecolor{seagreen}{rgb}{0.18, 0.55, 0.34}
\definecolor{sealbrown}{rgb}{0.2, 0.08, 0.08}
\definecolor{seashell}{rgb}{1.0, 0.96, 0.93}
\definecolor{selectiveyellow}{rgb}{1.0, 0.73, 0.0}
\definecolor{sepia}{rgb}{0.44, 0.26, 0.08}
\definecolor{shadow}{rgb}{0.54, 0.47, 0.36}
\definecolor{shamrockgreen}{rgb}{0.0, 0.62, 0.38}
\definecolor{shockingpink}{rgb}{0.99, 0.06, 0.75}
\definecolor{sienna}{rgb}{0.53, 0.18, 0.09}
\definecolor{silver}{rgb}{0.75, 0.75, 0.75}
\definecolor{sinopia}{rgb}{0.8, 0.25, 0.04}
\definecolor{skobeloff}{rgb}{0.0, 0.48, 0.45}
\definecolor{skyblue}{rgb}{0.53, 0.81, 0.92}
\definecolor{skymagenta}{rgb}{0.81, 0.44, 0.69}
\definecolor{slateblue}{rgb}{0.42, 0.35, 0.8}
\definecolor{slategray}{rgb}{0.44, 0.5, 0.56}
\definecolor{smalt(darkpowderblue)}{rgb}{0.0, 0.2, 0.6}
\definecolor{smokeytopaz}{rgb}{0.58, 0.25, 0.03}
\definecolor{smokyblack}{rgb}{0.06, 0.05, 0.03}
\definecolor{snow}{rgb}{1.0, 0.98, 0.98}
\definecolor{spirodiscoball}{rgb}{0.06, 0.75, 0.99}
\definecolor{splashedwhite}{rgb}{1.0, 0.99, 1.0}
\definecolor{springbud}{rgb}{0.65, 0.99, 0.0}
\definecolor{springgreen}{rgb}{0.0, 1.0, 0.5}
\definecolor{steelblue}{rgb}{0.27, 0.51, 0.71}
\definecolor{stildegrainyellow}{rgb}{0.98, 0.85, 0.37}
\definecolor{straw}{rgb}{0.89, 0.85, 0.44}
\definecolor{sunglow}{rgb}{1.0, 0.8, 0.2}
\definecolor{sunset}{rgb}{0.98, 0.84, 0.65}
\definecolor{tan}{rgb}{0.82, 0.71, 0.55}
\definecolor{tangelo}{rgb}{0.98, 0.3, 0.0}
\definecolor{tangerine}{rgb}{0.95, 0.52, 0.0}
\definecolor{tangerineyellow}{rgb}{1.0, 0.8, 0.0}
\definecolor{taupe}{rgb}{0.28, 0.24, 0.2}
\definecolor{taupegray}{rgb}{0.55, 0.52, 0.54}
\definecolor{teagreen}{rgb}{0.82, 0.94, 0.75}
\definecolor{tearose(orange)}{rgb}{0.97, 0.51, 0.47}
\definecolor{tearose(rose)}{rgb}{0.96, 0.76, 0.76}
\definecolor{teal}{rgb}{0.0, 0.5, 0.5}
\definecolor{tealblue}{rgb}{0.21, 0.46, 0.53}
\definecolor{tealgreen}{rgb}{0.0, 0.51, 0.5}
\definecolor{tawny}{rgb}{0.8, 0.34, 0.0}
\definecolor{terracotta}{rgb}{0.89, 0.45, 0.36}
\definecolor{thistle}{rgb}{0.85, 0.75, 0.85}
\definecolor{thulianpink}{rgb}{0.87, 0.44, 0.63}
\definecolor{ticklemepink}{rgb}{0.99, 0.54, 0.67}
\definecolor{tiffanyblue}{rgb}{0.04, 0.73, 0.71}
\definecolor{tigerseye}{rgb}{0.88, 0.55, 0.24}
\definecolor{timberwolf}{rgb}{0.86, 0.84, 0.82}
\definecolor{titaniumyellow}{rgb}{0.93, 0.9, 0.0}
\definecolor{tomato}{rgb}{1.0, 0.39, 0.28}
\definecolor{toolbox}{rgb}{0.45, 0.42, 0.75}
\definecolor{tractorred}{rgb}{0.99, 0.05, 0.21}
\definecolor{trolleygrey}{rgb}{0.5, 0.5, 0.5}
\definecolor{tropicalrainforest}{rgb}{0.0, 0.46, 0.37}
\definecolor{trueblue}{rgb}{0.0, 0.45, 0.81}
\definecolor{tuftsblue}{rgb}{0.28, 0.57, 0.81}
\definecolor{tumbleweed}{rgb}{0.87, 0.67, 0.53}
\definecolor{turkishrose}{rgb}{0.71, 0.45, 0.51}
\definecolor{turquoise}{rgb}{0.19, 0.84, 0.78}
\definecolor{turquoiseblue}{rgb}{0.0, 1.0, 0.94}
\definecolor{turquoisegreen}{rgb}{0.63, 0.84, 0.71}
\definecolor{tuscanred}{rgb}{0.51, 0.21, 0.21}
\definecolor{twilightlavender}{rgb}{0.54, 0.29, 0.42}
\definecolor{tyrianpurple}{rgb}{0.4, 0.01, 0.24}
\definecolor{uablue}{rgb}{0.0, 0.2, 0.67}
\definecolor{uared}{rgb}{0.85, 0.0, 0.3}
\definecolor{ube}{rgb}{0.53, 0.47, 0.76}
\definecolor{uclablue}{rgb}{0.33, 0.41, 0.58}
\definecolor{uclagold}{rgb}{1.0, 0.7, 0.0}
\definecolor{ufogreen}{rgb}{0.24, 0.82, 0.44}
\definecolor{ultramarine}{rgb}{0.07, 0.04, 0.56}
\definecolor{ultramarineblue}{rgb}{0.25, 0.4, 0.96}
\definecolor{ultrapink}{rgb}{1.0, 0.44, 1.0}
\definecolor{umber}{rgb}{0.39, 0.32, 0.28}
\definecolor{unitednationsblue}{rgb}{0.36, 0.57, 0.9}
\definecolor{unmellowyellow}{rgb}{1.0, 1.0, 0.4}
\definecolor{upforestgreen}{rgb}{0.0, 0.27, 0.13}
\definecolor{upmaroon}{rgb}{0.48, 0.07, 0.07}
\definecolor{upsdellred}{rgb}{0.68, 0.09, 0.13}
\definecolor{urobilin}{rgb}{0.88, 0.68, 0.13}
\definecolor{usccardinal}{rgb}{0.6, 0.0, 0.0}
\definecolor{uscgold}{rgb}{1.0, 0.8, 0.0}
\definecolor{utahcrimson}{rgb}{0.83, 0.0, 0.25}
\definecolor{vanilla}{rgb}{0.95, 0.9, 0.67}
\definecolor{vegasgold}{rgb}{0.77, 0.7, 0.35}
\definecolor{venetianred}{rgb}{0.78, 0.03, 0.08}
\definecolor{verdigris}{rgb}{0.26, 0.7, 0.68}
\definecolor{vermilion}{rgb}{0.89, 0.26, 0.2}
\definecolor{veronica}{rgb}{0.63, 0.36, 0.94}
\definecolor{violet}{rgb}{0.56, 0.0, 1.0}
\definecolor{violet(colorwheel)}{rgb}{0.5, 0.0, 1.0}
\definecolor{violet(ryb)}{rgb}{0.53, 0.0, 0.69}
\definecolor{violet(web)}{rgb}{0.93, 0.51, 0.93}
\definecolor{viridian}{rgb}{0.25, 0.51, 0.43}
\definecolor{vividauburn}{rgb}{0.58, 0.15, 0.14}
\definecolor{vividburgundy}{rgb}{0.62, 0.11, 0.21}
\definecolor{vividcerise}{rgb}{0.85, 0.11, 0.51}
\definecolor{vividtangerine}{rgb}{1.0, 0.63, 0.54}
\definecolor{vividviolet}{rgb}{0.62, 0.0, 1.0}
\definecolor{warmblack}{rgb}{0.0, 0.26, 0.26}
\definecolor{wenge}{rgb}{0.39, 0.33, 0.32}
\definecolor{wheat}{rgb}{0.96, 0.87, 0.7}
\definecolor{white}{rgb}{1.0, 1.0, 1.0}
\definecolor{whitesmoke}{rgb}{0.96, 0.96, 0.96}
\definecolor{wildblueyonder}{rgb}{0.64, 0.68, 0.82}
\definecolor{wildstrawberry}{rgb}{1.0, 0.26, 0.64}
\definecolor{wildwatermelon}{rgb}{0.99, 0.42, 0.52}
\definecolor{wisteria}{rgb}{0.79, 0.63, 0.86}
\definecolor{xanadu}{rgb}{0.45, 0.53, 0.47}
\definecolor{yaleblue}{rgb}{0.06, 0.3, 0.57}
\definecolor{yellow}{rgb}{1.0, 1.0, 0.0}
\definecolor{yellow(munsell)}{rgb}{0.94, 0.8, 0.0}
\definecolor{yellow(ncs)}{rgb}{1.0, 0.83, 0.0}
\definecolor{yellow(process)}{rgb}{1.0, 0.94, 0.0}
\definecolor{yellow(ryb)}{rgb}{1.0, 1.0, 0.2}
\definecolor{yellow-green}{rgb}{0.6, 0.8, 0.2}
\definecolor{zaffre}{rgb}{0.0, 0.08, 0.66}
\definecolor{zinnwalditebrown}{rgb}{0.17, 0.09, 0.03}

\usepackage{bm}

\newcommand{\highlight}[1]{\colorbox{gray!13}{$\displaystyle#1$}}

\newcommand{\FDMele}[3]{a_{#1|#2}^{(#3)}}
\newcommand{\FDMset}[1]{\mathcal{A}_{|#1}}
\newcommand{\FDMmat}[1]{\mathbf{A}^{(#1)}}

\newcommand{\GREEN}[3]{G_{#1|\theta'}^{#2,#3}}
\newcommand{\GREENM}[1]{\mathbf{G}_{#1|\theta'}}
\newcommand{\GREENMV}{\mathbb{G}_{\theta'}}

\newcommand{\PMATM}{\mathbb{U}}
\newcommand{\PMATME}[4]{\mathbb{U}_{#1,#2}^{#3,#4}}
\newcommand{\PMATMINV}{\mathbb{U}^{-1}}
\newcommand{\PMATEINV}[4]{[\mathbb{U}^{-1}]_{#1,#2}^{\,#3,#4}}
\newcommand{\PMATEINVD}[2]{[\mathbb{U}^{-1}]^{\,#1,#2}}
\newcommand{\PMATED}[2]{\mathbb{U}^{\,#1,#2}}

\newcommand{\QFUNC}[1]{P\left(#1\right)}
\newcommand{\QMAT}[1]{P^{#1}}
\newcommand{\QMATM}{\mathbf{P}}

\newcommand{\FFUNC}[2]{Q_{#1}\left(#2\right)}
\newcommand{\FMAT}[2]{Q_{#1}^{#2}}
\newcommand{\FMATM}[1]{\mathbf{Q}_{#1}}

\newcommand{\GFUNC}[3]{R_{#1,#2}\left(#3\right)}
\newcommand{\GMAT}[3]{R_{#1,#2}^{#3}}
\newcommand{\GMATM}[2]{\mathbf{R}_{#1,#2}}

\newcommand{\VMAT}{\mathbb{V}}
\newcommand{\EMATT}{\mathbb{E}_{\theta'}}

\newcommand{\VMATE}[4]{\mathbb{V}_{#1,#2}^{#3,#4}}
\newcommand{\EMATE}[3]{\mathbb{E}_{#1|\theta'}^{#2,#3}}

\newcommand{\AMAT}{\mathbb A}

\newcommand{\AMATE}[4]{\mathbb{A}_{#1,#2}^{#3,#4}}
\newcommand{\AMATED}[2]{\mathbb{A}^{#1,#2}}
\newcommand{\BMATED}[2]{\mathbb{B}^{#1,#2}}

\newcommand{\BMATE}[4]{\mathbb{B}_{#1,#2}^{#3,#4}}
\newcommand{\SMATM}[1]{\mathbb{S}^{#1}}
\newcommand{\SMATE}[3]{{\mathbb S}_{#1,#2}^{#3}}

\newcommand{\SMATEINV}[3]{\bigl[\mathbb{S}^{#3}\bigr]_{#1,#2}^{-1}}

\makeatletter
\def\@affil@script#1#2#3#4{
 \@ifnum{#1=\z@}{}{

  \begingroup
   \frontmatter@affiliationfont
   \@ifnum{\c@affil<\affil@cutoff}{}{
    \def\@thefnmark{#1}\@makefnmark
   }
   \ignorespaces#3
   \@if@empty{#4}{}{\frontmatter@footnote{#4}}
   
  \endgroup
 }
}
\makeatother

\begin{document}

\author{Alejandro Ferrero Botero}
\email{aferrero@ucatolica.edu.co}
\affiliation{Departamento de Ciencias B\'asicas, Universidad Cat\'olica de Colombia - Bogot\'a, Colombia}
\author{Juan Pablo Mallarino}
\email{jp.mallarino50@uniandes.edu.co}
\affiliation{Facultad de Ciencias --- Laboratorio Computacional HPC, Universidad de los Andes - Bogot\'a, Colombia}

\title{Approximate solution of two dimensional disc-like systems by one dimensional reduction: an approach through the Green function formalism using the Finite Elements Method}

\date{\today}

\begin{abstract}
We present a comprehensive study for common second order PDE's in two dimensional \emph{disc-like} systems and show how their solution can be approximated by finding the Green function of an effective one dimensional system. After elaborating on the formalism, we propose to secure an exact solution via a Fourier expansion of the Green function, which entails to solve an infinitely countable system of differential equations for the Green-Fourier modes that in the simplest case yields the source-free Green distribution. We present results on non separable systems---or such whose solution cannot be obtained by the usual variable separation technique---on both annulus and disc geometries, and show how the resulting one dimensional Fourier modes potentially generate a near-exact solution. Numerical solutions will be obtained via finite differentiation using FDM or FEM with the three-point stencil approximation to derivatives. Comparing to known exact solutions, our results achieve an estimated numerical relative error below $10^{-6}$.
\end{abstract}

\maketitle

\section{Introduction}
\label{sec:intro}

In the present work we elaborate on the FEM for solving complex two-dimensional partial differential equations (DE) using a Green {\it function} construction. Green's method has been employed extensively in Physics for solving Laplace's equation and associates in a cornucopia of areas, such as Quantum and Statistical Mechanics. In quantum mechanics, for example, the method of nonequilibrium Green’s functions (NEGF) has been used to study the Brownian motion of a quantum oscillator \citep{Schwinger1961}, quantum thermal transport \citep{Wang2014,Foster2019}, derive quantum kinetic equations \citep{zbMATH03187565}, study hadronic physics \citep{ALKOFER2001281}, among others. In statistical mechanics, some of the applications of the Green functions include the predictions of some observables \citep{Lucarini2018}, help to describe 1D hydrodynamic models \citep{PhysRevLett.120.240601}, finding electrical properties of some physical systems \citep{PhysRevA.98.032509,PhysRevB.89.245430}, study nonextensive statistical mechanics with new normalized $q$-expectation values \citep{LENZI2000503}, and so much more. Even the Green functions are used in quantum field theory to describe the propagators of quantum fields in the perturbative regime. 

Not only are Green functions useful to solve systems described by inhomogeneous differential equations, but they can also be used to describe thermodynamic properties. For instance, the density and correlations of particles immersed in two dimensional two component plasmas at certain temperatures can be described by sets of Green functions \citep{cornu:2444,Ferrero2007,Ferrero2014}. 

In order to study how an inhomogenous partial DE can be solved by the method we propose, we start defining a differential operator
\eemp\label{eq:operator}
\hat{\mathcal{L}}_{\{\mathbf{r}\}}\square=(\vec{\nabla}_{\{\mathbf{r}\}}+\vec{f}(\mathbf{r}))\cdot(\vec{\nabla}_{\{\mathbf{r}\}}\square) +
g({\mathbf{r}})\square,
\ffin
acting on a scalar field in $\Re^d$, with $d$ the dimension of the system---\emph{i.e.} $\mathbf{r}\in\Re^d$. This operator is known in other contexts as the Liouville operator; via this definition, we often describe the evolution of a relevant quantity $\psi$ by means of the equation $\partial_t\psi(\mathbf{r},t)-\hat{\mathcal{L}}\psi(\mathbf{r},t) = 0$ as it is the case of the wave function in quantum mechanics. For example, in the diffusion phenomenon the functions take the form $\vec{f}(\mathbf{r},t) = \vec{\nabla}D(\textbf{r},t)$ and $g(\textbf{r}) = 0$, and for the Helmholtz equation $\vec{f}(\mathbf{r},t) = 0$ and $g(\textbf{r}) = m^2$, with $m$ a constant.

Finding solutions to the latter has motivated the development of numerical methods that grow in number and complexity. For instance, using Restricted Boltzmann Machines we can engineer an artificial neural network that is able to accurately sample the probability distribution for quantum statistical systems \citep{PhysRevB.96.205152,PhysRevE.96.022131}. However, some effort can be made from a mathematical point of view prior to implementing a full scale numerical calculation.

Green's function---or more precisely, distribution---is perhaps the most interesting artifact of a huge bag of tricks that we have when facing differential equations. Its power relies on the possibility of \emph{inverting} the differential operator $\hat{\mathcal{L}}$ to solve the inhomogeneous equation
\eemp
\hat{\mathcal{L}}\psi(\mathbf{r}) = \phi(\mathbf{r}),
\label{eq:original-problem}
\ffin
with $\psi(\mathbf{r})$ and $\phi(\mathbf{r})$ two scalar functions. Hinting that its existence, the Green Distribution, is conditioned by some properties of $\hat{\mathcal{L}}$.

A disadvantage of the Green methodology is the duplication of degrees of freedom, encouraging researchers to find $\psi(\mathbf r)$ directly. Our aim is not to develop a generalized theory for an arbitrary problem and number of dimensions. Despite this, we can look into the consequences of \emph{breaking down} one dimension by focusing on the simple two-dimensional case.

Two dimensional systems are of great interest in statistical mechanics \citep{cornu:2444,Ferrero2007,Ferrero2014}, material sciences \citep{Novoselov666,articleFG,C5NR01052G}, quantum computing \citep{articleSR,doi:10.1021/nl0518472}, high energy physics \citep{articleSF,GAMBOASARAVI1981239}, ionic fluids \citep{articlePA}, theoretical mathematics \citep{doi:10.1063/1.533014,doi:10.1063/1.1897183}, and many others.

The outline of the paper is as follows. We first remind some relevant known results for the Green's function construction in \cref{sec:framework} prior to presenting the strategy to move from 2D to 1D in \cref{subsec:2d-to-1d}. We lay out a clever geometric interpretation of the result in \cref{w-and-inverse-of-L} followed by a connection to a relevant theory for Hilbert Space functions in \cref{subsec:sturm-liouville}. Consequently, we next discuss its implications towards finding the Green function using FDM in \cref{sec:method}. We present some mathematical results that include the solution of some known results for testing purposes, the implementation of the method in a non-separable 2D system, and a discussion of how the algorithm can be adapted to solve the heat diffusion problem in thermal equilibrium in   \cref{sec:numerical_results}. Finally, we wrap up the conclusions in \cref{sec:conclusions}. Intermediate calculations and numerical details are left for further inspection in appendices. 

\section{Framework}
\label{sec:framework}

We start studying the Green function formalism by postulating the convolution identity from the Dirac distribution,
\eemp
\psi(\mathbf{r}) = \int_\mathbf{r^\prime}\psi(\mathbf{r^\prime})\,\delta(\mathbf{r^\prime}-\mathbf{r})\,w(\mathbf{r^\prime},\mathbf{r})\,\text{d}\mathbf{r^\prime},
\label{eq:dirac-identity}
\ffin
with $w(\mathbf{r^\prime},\mathbf{r})$ a weight function properly defined by two conditions; the first of which $w(\mathbf{r},\mathbf{r})=1$. Now by defining $G(\mathbf{r^\prime},\mathbf{r})$ as,
\eemp
\hat{\mathcal{L}}_{\{\mathbf r^\prime\}}G(\mathbf{r^\prime},\mathbf{r}) = \delta(\mathbf{r^\prime}-\mathbf{r}),
\label{eq:green-definition}
\ffin
with $\delta(\mathbf{r^\prime}-\mathbf{r})=\delta(\mathbf{r}-\mathbf{r^\prime})$ the $\Re^d$ Dirac delta distribution, then,
\eemp
\psi(\mathbf{r}) = \int_\mathbf{r^\prime}\psi(\mathbf{r^\prime})\,\[\hat{\mathcal{L}}_{\{\mathbf{r^\prime}\}}G(\mathbf{r^\prime},\mathbf{r})\]\,w(\mathbf{r^\prime},\mathbf{r})\,\text{d}\mathbf{r^\prime}.
\label{eq:green-demo1}
\ffin

The second condition over $w(\mathbf{r},\mathbf{r^\prime})$ will be determined in such a way that $\hat{\mathcal{L}}$ is \emph{self-adjoint} (Hermitian), or equivalently
\eemps
\int_\mathbf{r^\prime}\psi(\mathbf{r^\prime})\,&\[\hat{\mathcal{L}}_{\{\mathbf{r^\prime}\}}G(\mathbf{r^\prime},\mathbf{r})\]\,w(\mathbf{r^\prime},\mathbf{r})\,\text{d}\mathbf{r^\prime} = \\
&\int_\mathbf{r^\prime}\[\hat{\mathcal{L}}_{\{\mathbf{r^\prime}\}}\psi(\mathbf{r^\prime})\]\,G(\mathbf{r^\prime},\mathbf{r})\,w(\mathbf{r^\prime},\mathbf{r})\,\text{d}\mathbf{r^\prime} +\,\text{b.c.},
\ffins
with added Dirichlet or Neumann boundary conditions (b.c.). Direct substitution into \cref{eq:green-demo1}, using \cref{eq:original-problem}, yields
\eemp
\psi(\mathbf{r}) = \int_\mathbf{r^\prime}G(\mathbf{r^\prime},\mathbf{r})\phi(\mathbf{r^\prime})\,w(\mathbf{r^\prime},\mathbf{r})\,\text{d}\mathbf{r^\prime} +\,\text{b.c.}\,.
\label{eq:green-ansatz}
\ffin

\subsection{On the nature of $w(\mathbf{r},\mathbf{r^\prime})$ and $\hat{\mathcal{L}}^{-1}$}
\label{w-and-inverse-of-L}
This former known result deserves a more delicate look, particularly, on the existence of the weight function, and how previous solution relates with the usual convolution theorem $\psi(\mathbf{r}) = \int_\mathbf{r^\prime}G(\mathbf{r},\mathbf{r'})\phi(\mathbf{r^\prime})\,\text{d}\mathbf{r^\prime} +\,\text{b.c.}$. As mentioned, an appropriate choice for the weight function ensures that \cref{eq:green-demo1} reproduces \cref{eq:green-ansatz}. This is done by using the Green's and Divergence theorem in \cref{eq:green-demo1} to perform an integration by parts. After simplifications, we realize that by choosing the weight function such that $\vec{\nabla}_{\{\mathbf{r^\prime}\}}w(\mathbf{r^\prime},\mathbf{r})-w(\mathbf{r^\prime},\mathbf{r})\vec{f}(\mathbf{r^\prime}) = 0$ (See \cref{app:weight} for further details) we ensure that the operator is self-adjoint! This is essential to Green's method. Hence, if no weight function exists, we might be forced to use other analytical and/or numerical procedures in order to find $\psi$. For that matter, the range of problems that we aim to analyze is narrowed down to the few ones satisfying the aforementioned condition; despite this, a great many of this subset are of special interest for Mathematics and Physics. 

Assuming $w(\mathbf{r^\prime},\mathbf{r})$ exists we are able to incorporate the premise for \cref{eq:green-ansatz} yielding exactly,
\eemp\label{eq:sol-tot}
\psi(\mathbf{r}) =& \int_\mathbf{r^\prime}G(\mathbf{r^\prime},\mathbf{r})\phi(\mathbf{r^\prime})\,w(\mathbf{r^\prime},\mathbf{r})\,\text{d}\mathbf{r^\prime}+\oint_{\partial\mathbf{r^\prime}}w(\mathbf{r^\prime},\mathbf{r})\times\\
&\,\Big[\psi(\mathbf{r^\prime})\vec{\nabla}_{\{\mathbf{r^\prime}\}}G(\mathbf{r^\prime},\mathbf{r})-G(\mathbf{r^\prime},\mathbf{r})\vec{\nabla}_{\{\mathbf{r^\prime}\}}\psi(\mathbf{r^\prime})\Big]\cdot\mathbf{n}\text{d}S'.
\ffin
Note how the second term depends on $\psi(\mathbf{r})$'s boundary conditions. By choosing identical conditions and trivial values for the Green distribution function at the boundaries ($G=0$ for Dirichlet or $G^\prime=0$ for Neumann) we are capable of solving an infinite number of alike boundary value problems.

The discussion for the existence of the weight function can be answered mathematically. Given the relationship required, the weight is defined as $w(\mathbf{r^\prime},\mathbf{r})\defeq\exp[-\gamma(\mathbf{r^\prime},\mathbf{r})]$, yielding $\vec{\nabla}_{\{\mathbf{r^\prime}\}}\gamma(\mathbf{r^\prime},\mathbf{r})=-\vec{f}(\mathbf{r^\prime})$. Considering that the curl of the gradient of any scalar function is trivial then $\gamma(\mathbf{r^\prime},\mathbf{r})$ exists if and only if $\vec{\nabla}\times\vec{f} = 0$, which means that $\vec{f}$ must be a conservative vector field! Within this view, $\gamma(\mathbf r',\mathbf r)$ represents the scalar potential associated with a force. Anticipating this last restriction, the solution for $\gamma(\mathbf{r^\prime},\mathbf{r})$ is independent of a path that simply connects $\mathbf{r}$ to $\mathbf{r^\prime}$ yielding,
\eemp\label{eq:weight-function}
w(\mathbf{r^\prime},\mathbf{r})\equiv\frac{e^{-\gamma(\mathbf{r^\prime})}}{e^{-\gamma(\mathbf{r})}}=\frac{1}{w(\mathbf{r},\mathbf{r^\prime})},
\ffin
reflecting on the symmetry of the distribution as it will be shown later. Finally, we are ready to define the inverse operator of $\hat{\mathcal{L}}$ as
\eemp\label{eq:inv-op}
\hat{\mathcal{L}}_{\{\mathbf{r}\}}^{-1}\square=\int_{\mathbf{r^\prime}}\,G(\mathbf{r^\prime},\mathbf{r})\,\square(\mathbf{r^\prime})\,w(\mathbf{r^\prime},\mathbf{r})\text{d}\mathbf{r^\prime},
\ffin
conditioned by the boundary-value problem, which in turn defines $G(\mathbf{r^\prime},\mathbf{r})$ from \cref{eq:green-definition}.

There is one last piece of the puzzle to be resolved and it is related to the symmetry of the Green function distribution. Let us evaluate $\hat{\mathcal{L}}_{\{\mathbf r\}}G(\mathbf{r^\prime},\mathbf{r})$---i.e., the operator acting on the second variable. Direct application of $\hat{\mathcal{L}}_{\{\mathbf r\}}$ on \cref{eq:sol-tot}, using \cref{eq:original-problem},
\eemp
\phi(\mathbf{r}) =& \int_\mathbf{r^\prime}\hat{\mathcal{L}}_{\{\mathbf r\}}\{G(\mathbf{r^\prime},\mathbf{r})\,w(\mathbf{r^\prime},\mathbf{r})\}\phi(\mathbf{r^\prime})\,\text{d}\mathbf{r^\prime}+\hat{\mathcal{L}}_{\{\mathbf r\}}\{\text{b.c.}\}\,,
\label{eq:symmetry-identity}
\ffin
hints how this operator appears to work and leads us to anticipate the convolution of a Dirac distribution. Indeed this is true. To clarify, here we exchanged integral and Liouville operators because they are acting on separate variables, and the weight and Green functions (except at $\mathbf{r^\prime}=\mathbf{r}$) are differentiable.

This conjecture can be proved from the following statement: two separate problems with different boundary values and identical inhomogeneous differential equation ---\cref{eq:original-problem}---share the same Green function distribution and satisfy \cref{eq:symmetry-identity}; therefore, by comparing equations for any two cases leads to $\hat{\mathcal{L}}_{\{\mathbf r\}}\{\text{b.c.}\}=0$ because we can always choose convenient trivial boundary values (\emph{i.e.} $\text{b.c.}=0$) in one case. Consequently,
\emp
\hat{\mathcal{L}}_{\{\mathbf r\}}\{G(\mathbf{r^\prime},\mathbf{r})\,w(\mathbf{r^\prime},\mathbf{r})\}\equiv\delta(\mathbf{r^\prime}-\mathbf{r})\,,
\label{eq:symmetry-delta}
\fin
an equality that bears meaning in the sense of the distributions. These final result unravels the symmetry of the Green distribution function via the weight function, \emph{i.e.}
\emp
\label{eq:symmetry-green}
G(\mathbf{r^\prime},\mathbf{r})=G(\mathbf{r},\mathbf{r^\prime})\,w(\mathbf{r},\mathbf{r^\prime}).
\fin

An interesting question now arises, and it is related to the possibility of using \cref{eq:symmetry-green} to drop the weight function out of the equation. This operation, with the addition of the relation $\vec\nabla_{\{\mathbf r'\}}w(\mathbf r,\mathbf r')=-w(\mathbf r,\mathbf r')\vec f(\mathbf r')$, leads to,
\eemp\label{eq:sol-tot1}
\psi(\mathbf{r}) =& \int_\mathbf{r^\prime}G(\mathbf r ,\mathbf r')\phi(\mathbf{r^\prime})\,\text{d}\mathbf{r^\prime}+\oint_{\partial\mathbf{r^\prime}}
\Big[\psi(\mathbf r')\vec{\nabla}_{\{\mathbf{r^\prime}\}}G(\mathbf r, \mathbf r')
\\
&-G(\mathbf r,\mathbf r')\bigl[\psi(\mathbf r')\vec f(\mathbf r')+\vec{\nabla}_{\{\mathbf r'\}} \psi(\mathbf{r^\prime})\bigr]\Big]\cdot\mathbf{n}\text{d}S'.
\ffin
Notice that  for Neumann boundary conditions (NBC), unlike Dirichlet (DBC), both $\psi(\mathbf r)$ and its derivative ---at the boundaries--- are necessary. Ergo, \cref{eq:sol-tot1} is inconvenient for NBC unless either $\psi$ vanishes or $\vec f(\mathbf r)=0$. In such a case, it deems necessary to use the version that incorporates weight function.

Actually, the vector field $\vec f$ does not appear in some of the Liouville operators used in physics. For instance, the Green function associated with the electrostatic field satisfies the relation $\vec\nabla^2G(\mathbf r,\mathbf r')=-4\pi\delta(\mathbf r-\mathbf r')$---the irrelevant factor of $-4\pi$ appears by convenience. The static regime of the Klein Gordon equation---which also leads to the Yukawa potential---also follows a similar behavior, as its associated Green function in 2D is $K_0(\mu r)$, satisfying the DE $(\vec\nabla^2-\mu^2)G(\mathbf r)=-2\pi\delta(\mathbf r)$ \citep{PhysRevD.55.3830}. Clearly, $\vec f$ is  absent in both systems.

Yet the DEs describing the behavior of other physical systems such as the driven damped harmonic oscillator,\footnote{The 1D driven damped harmonic oscillator is modeled by the DE $m\ddot x+b \dot x+kx=F(t)$. The damping constant $b$ plays the role of $\vec f$ in this one dimensional system. Although the time $t$ is the relevant variable describing this system (instead of the position $x$), the one dimensional formalism we describe is analog to this model.} the diffusion equation at thermal equilibrium with an anisotropic diffusion coefficient, and the electrostatic potential in the presence of anisotropic media, include the existence of a vector field $\vec f$---see section \ref{sec:numerical_results} for more details about the first system.

Surprisingly, any dependence on the weight function in \cref{eq:sol-tot1} has vanished. As previously stated, the weight function can only be defined when $\vec f$ is a conservative field. Then, an important question now arises: is \cref{eq:sol-tot1} still valid for non-conservative vector fields? This in a fundamental question that can be addressed in a future work. Since the main purpose is to present a compact and rigorous algorithm to solve the Green function in 2D space, we will restrict our analysis to only the supported cases.

\subsection{Boundary conditions}
\label{subsec:G-boundaries}

As previously stated, the Green function conveniently inherits identical types of conditions as the target function $\psi$ at the boundaries. These can be summarized as,
\eemp
\text{DBC}:\,\,\to\,\,&G(\mathbf{r},\mathbf{r'})\vert_{\mathbf{r}\text{ at }R_{\text{ext}/\text{int}}}=0,\\
\text{NBC}:\,\,\to\,\,&\partial_{\mathbf{r}}G(\mathbf{r},\mathbf{r'})\vert_{\mathbf{r}\text{ at }R_{\text{ext}/\text{int}}}=0.
\ffin
However, there are two hidden additional conditions that must be satisfied enforced by the presence of Dirac's distribution. The rationale behind is that without them $G=0$ will be a solution to the Green function for the simple boundary value problem. While this is directly visible for Dirichlet, notice that it also applies for Neumann's case. The added restrictions appear at the artificial boundary $\mathbf{r}=\mathbf{r'}$ implying continuity of $G$ and discontinuity of the local derivative. Both are essential to secure a non--zero solution. Continuity is often regarded considering that the Green distribution is still a function and its derivatives up to second order exist in the classical sense of the DE everywhere except at $\mathbf{r}=\mathbf{r'}$. Though there is a stronger argument that stems from the fact that the annulus and the disc are Lipschitz domains \citep{doi:10.1080/03605302.2010.489629}, in DE it always results convenient to decide what do we take as an acceptable solution to any problem, which is our particular case here.

\begin{figure}
\centering\includegraphics[scale=0.68]{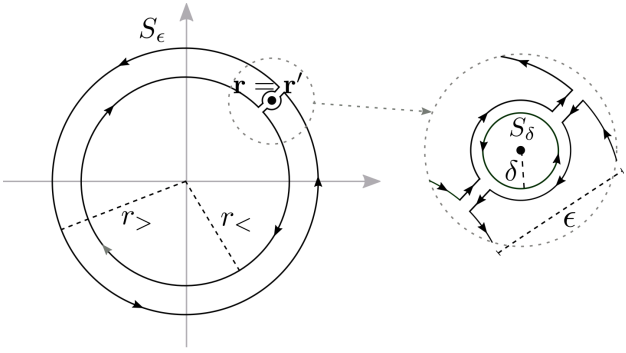}\caption{The contours used in path integration: to the left $S_\epsilon$ and $S_\delta$ to the right. The radius and thickness are chosen purposely as $\delta<\epsilon/2$ to take the $\epsilon\to0$ limit.}\label{fig:contour}
\end{figure}

Turning to the plane $\mathbf{r}=\mathbf{r'}$, conditions are derived directly from \cref{eq:green-definition} by integrating over $\mathbf{r}$ inside the volume delimited by the surface $S_\delta$ enclosing $\mathbf{r'}$ such that it is contained inside a vecinity---$\mathcal{V}_\delta$---of $\mathbf{r'}$ (see right of \cref{fig:contour} for an artistic view). Using the divergence theorem, the condition simplifies to,
\eemp\label{eq:cond-discon}
\lim_{\delta\to0}\oint_{S_{\delta}}\vec{\nabla}_{\{\mathbf{r}\}} G(\mathbf r,\mathbf r')\cdot\hat n\text{d}S=1\,,
\ffin
where we have kept the leading contributing term while taking the limit. For the one dimensional case it yields the relation $G'(r_{>}^{\prime},r')-G'(r_{<}^{\prime},r')=1$.

\subsection{Connection with the Sturm--Liouville problem}
\label{subsec:sturm-liouville}

The weight function, if existent, is able to transform the Liouville operator into a self--adjoint differential operator. Notice that action of $w(\mathbf r',\mathbf r)$ on \cref{eq:operator},
\begin{align}
    w(&\mathbf r',\mathbf r)\hat{\mathcal{L}}_{\{\mathbf{r'}\}}\square=\nonumber\\
    &\vec\nabla_{\{\mathbf{r'}\}}\cdot\bigl[w(\mathbf r',\mathbf r)\vec\nabla_{\{\mathbf{r'}\}} \square\bigr]+w(\mathbf r',\mathbf r)g(\mathbf r')\square\,,
    \label{eq:operator-sturm-liouville}
\end{align}
yields the otherwise known Sturm--Liouville form for PDE's. Namely, the Sturm--Liouville differential operator reads then as,
\begin{align}
    \hat{\mathcal{L}}_{\{\mathbf{r'}\}}^{\text{SL}}\square=w(\mathbf r',\mathbf r)\hat{\mathcal{L}}_{\{\mathbf{r'}\}}\square\,.
\end{align}
Consequently, operating onto the Green distribution function gives equal results for both operators, \emph{i.e.} $\hat{\mathcal{L}}_{\{\mathbf{r'}\}}^{\text{SL}}G(\mathbf r',\mathbf r )=\delta(\mathbf r-\mathbf r')$.

This last result connects the Sturm-Liouville problem with null eigenvalues and the Green function distribution problem where the former is the solution to the first strictly when $\mathbf r\neq \mathbf r'$ under either Dirithlet or Neumann boundary conditions. Resulting from this, $G(\mathbf r',\mathbf r )$ is continuous everywhere and differentiable at $\mathbf r\neq \mathbf r'$; the behavior of its  derivative at $\mathbf r=\mathbf{r'}$ is dictaminated by the Liouville operator in the domain of the problem and  specified by the Dirac Delta distribution.

\subsection{Moving from 2D to 1D: the infinite coupling}
\label{subsec:2d-to-1d}

As noted before, let us elaborate on the simplest scenario where a reduction of the dimension of the problem significantly improves our chances of procuring a general solution. Assume we would like to find the two-dimensional Green function in accordance with \cref{eq:green-definition}. Taking advantage of the completeness of the Fourier infinite expansion of any periodic function, we propose to solve the two-dimensional DE in polar coordinates; the dimensional reduction occurs due to the periodicity in $\theta$ that does not take place in Cartesian coordinates.

Although the Laplace operator in polar coordinates is known to be separable in the variables $r$ and $\theta$, the introduction of the additional terms in \cref{eq:operator}, as already mentioned, may lead to a DE that cannot be split conveniently. \Cref{eq:green-definition} is then given by,
\begin{align}\label{eq:dif-eq-gen}
&\Bigl[\frac{\partial^2}{\partial r^2}+\frac{1}{r}\frac{\partial}{\partial r}+\frac{1}{r^2}\frac{\partial^2}{\partial\theta^2}+f_r(r,\theta)\frac{\partial}{\partial r}+\nonumber\\
&\,\,f_\theta(r,\theta)\frac{1}{r}\frac{\partial}{\partial \theta}+g(r,\theta)\Bigr]G(\mathbf r,\mathbf r')=\delta(\mathbf r-\mathbf r')\,,
\end{align}
where convenient periodic conditions that must be satisfied suggests we should expand the Green function distribution in Fourier modes. Tentatively, we can resort to an expansion\footnote{The Fourier expansion,
\begin{align*}
    f(r,\theta)=\sum_{\mu\in\mathbb{Z}}f_\mu(r)e^{i\mu\theta}\,\,;\,\,f_\mu(r)=\frac{1}{2\pi}\int_{0}^{2\pi}f(r,\theta)e^{-i\mu\theta}d\theta\,.
\end{align*}
} of the form $\sum_{\lambda} e^{i\lambda(\theta-\theta')}G_\lambda$ to match the delta distribution expansion---\emph{i.e.} $\delta(\mathbf r-\mathbf r')=\frac{1}{2\pi r}\delta(r-r')\sum_\lambda e^{i\lambda(\theta-\theta')}$---but since we cannot guarantee that the $G_\lambda$ coefficients are $\theta'$-independent (only under \emph{proper} angular symmetry conditions) then we will assume $G(\mathbf r,\mathbf r')$ expands as,
\emp
\label{eq:green-fourier-exp}
G(\mathbf r,\mathbf r')=\frac{1}{2\pi}\sum_{\lambda\in\mathbb{Z}} e^{i\lambda\theta}G_\lambda(r,r',\theta').
\fin
With that in mind and multiplying \cref{eq:dif-eq-gen} by $r^2$ to avoid divergences at $r=0$,
\begin{align}\label{eq:dif-eq-gen1}
\sum_{\lambda\in\mathbb{Z}}e^{i\lambda\theta}\Bigl[&r^2\,\ordeN{2}{}{r}+r(1+rf_r)\orde{}{r}+(-\lambda^2+i\lambda r\,f_\theta+\nonumber\\
&r^2g)\Bigr]G_\lambda=\sum_{\lambda\in\mathbb{Z}}e^{i(\lambda-\lambda')\theta}r\delta(r-r')\,.
\end{align}

Notice how we cannot obtain a solution because there remains a residual dependence of $\theta$ in functions $\vec{f}$ and $g$. Despite this, a simplification can be manufactured when they are replaced by their Fourier series form before integrating on $\theta$ over a full period. This step yields our master equation where we deduce that the $G_\lambda(r,r',\theta')$ modes satisfy the DE\footnote{We have dropped out the dependencies of all functions on $r$, $\theta$, $r^\prime$, and $\theta^\prime$ facilitating a comprehensible reading.},
\begin{align}\label{eq:dif-eq-gen5}
&r^2 G_{\lambda}''+r\, G'_{\lambda}-\lambda^2G_{\lambda}\nonumber\\
&+\sum_{\mu\in\mathbb{Z}}r^2f_{r\,\mu} G'_{\lambda-\mu}+\sum_{\mu\in\mathbb{Z}}\bigl[ir(\lambda-\mu)f_{\theta\,\mu}+r^2g_\mu\bigr] G_{\lambda-\mu}
\nonumber\\
&\,\,=r\delta(r-r')e^{-i\lambda\theta'}\,.
\end{align}
This final result shows we have accomplished to reduce the rank of the effective Green function to solve at the expense of requiring a countable large number of these Green modes. It is remarkable how the dependence on $\theta'$ is delegated to a quasi-negligible term at the right hand side of the equation. We will develop this argument further in the following sections.

This formulation represents an infinitely coupled system of linear differential equations that unsurprisingly contains the solution to the Green function for the classical source-free wave function; the structure of functions $\vec{f}$ and $g$ defines the strength of the entanglement of Green's free wave modes appearing in the rate at which the Fourier coefficients---functions---go to zero with increasing mode frequency. For simplicity, we opt to recall Green's Fourier modes as $\lambda$-modes, and $\vec{f}$ and $g$'s modes as $\mu$-modes suggested by the indexes employed in the equation above.

\subsection{More on boundary conditions of the $\lambda$-modes}
\label{subsec:bc-lambda-modes}

One last effort must be done to explain how boundary conditions are inherited along the free-wave modes. The key to this understanding depends on the geometry of the problem and the originating expansion from \cref{eq:green-fourier-exp}; we can identify two cases for disc--like systems: the \emph{annulus} and the \emph{disc}. Other geometries will be studied in a future work. For the annulus, either under Dirichlet or Neumann boundary conditions, the function or its derivative must vanish at the boundaries. This can be met if all modes preserve the vanishing values at both inner and outer boundaries ---under uniform convergence. In doing so, we guarantee to meet all requisites for the Green function and a solution is obtained. Conversely, preserving boundary conditions for the disc is not trivial because we do not have one but two boundaries (the second at $r\to0^+$). Due to the oscillating behavior of $e^{i\lambda\theta}$ with $\theta$ at $r\to0^+$, all $\lambda\neq0$ modes must vanish at the origin to ensure continuity of the Green distribution function. This can be enforced examining \cref{eq:dif-eq-gen5} as $r\to0^+$. Discontinuity due to the source at $r=r'$ may be neglected for now to realize that we can, while approaching the origin, consider the behavior of each $r^0$, $r^1$, and $r^2$ terms \emph{independently}. We draw then conveniently,
\eemps
\label{eq:bc-triplet}
r^0\left[-\lambda^2G_{\lambda}\right] &\underrel[c]{r\shortrightarrow0}{=} 0\,,\\
r^1\Big[G_{\lambda}^{\prime}+i\sum_{\mu\in\mathbb{Z}}(\lambda-\mu)f_{\theta\mu}G_{\lambda-\mu}\Big] &\underrel[c]{r\shortrightarrow0}{=} 0\,,\\
r^2\Big[G_{\lambda}^{\prime\prime}+\sum_{\mu\in\mathbb{Z}}f_{r\mu}G_{\lambda-\mu}^{\prime}+\sum_{\mu\in\mathbb{Z}}g_{\mu}G_{\lambda-\mu}\Big] &\underrel[c]{r\shortrightarrow0}{=} 0,,
\ffins
where we can choose, via $r^0$ terms, that $G_\lambda=0\,\,\forall\,\,\lambda\neq0$ and $G_0\neq0$. Plugging this sequentially into $r^1$ and $r^2$ terms hints $G_{\lambda}^{\prime}=0$ and $G_{\lambda}^{\prime\prime}=0$ ($\forall\lambda\neq0$) assuming that $\lim_{r\shortrightarrow0}f_{\theta\mu}G_{\lambda-\mu}=0$ and $\lim_{r\shortrightarrow0}f_{r\mu}G_{\lambda-\mu}^{\prime}=0\,\,\wedge\,\,\lim_{r\shortrightarrow0}g_{\mu}G_{\lambda-\mu}=0$ (with the exception of $\lambda=0$ where the $\mu=0$ term remains, thus we will choose $G_{0}^{\prime\prime}=-g_0G_0$) respectively.

This is supported from continuity of $g(\mathbf{r})$ everywhere in the disc and from the definition of $\vec{f}(\mathbf{r})$, where limited by the existence of $w(\mathbf{r'},\mathbf{r})$, as the gradient of a scalar function. If such function, $\gamma(\mathbf{r})$, where to be free of pathologies and differentiable everywhere in the disc (including the origin) then $\lim_{r\shortrightarrow0}f_{r\lambda}=0$ and $\lim_{r\shortrightarrow0}f_{\theta\lambda}=0$ for $\lambda\neq0$. It remains to say, that in order to fulfill all above conditions we will require that $f_{\theta0}$ and $f_{r0}$ are finite as $r\to0^+$. Looking under the hood of these assumptions, note that consequently the $0$-mode has a logarithmic divergence when $r'=0$, \emph{i.e.} $G_0\underrel[c]{r\shortrightarrow0}{\propto}\,-\log r$.

In summary, the conditions for the disc at the origin are the following two only for $r'>0$ (see section \ref{sec:method} for numerical details)
\eemp
\label{eq:boundary-modes}
G_{\lambda}(0^+,r',\theta')&=0\,\,\,\forall\,\lambda\neq0\,,\\
G_{\lambda}^{\prime}(0^+,r',\theta')&=0\,\,\,\forall\,\lambda\,.
\ffin
Exceptions and particularities emerging from the specific form of functions $\vec{f}$ and $g$ must be taken into account when detailing the boundary conditions and may alter the relationships obtained above.

This relationship has to be completed with the resulting relationship at the artificial boundary $r=r'$ obtained when using a complementary surface $S_\epsilon$ corresponding to an open ring of $\epsilon>0$ thickness ---see left \cref{fig:contour} and \cref{eq:cond-discon}. This gives,
\eemp\label{eq:cond-discon1}
r'\lim_{\epsilon\to0}\int_0^{2\pi}\bigl[\partial_r G(\mathbf r,\mathbf r')_>-\partial_r G(\mathbf r,\mathbf r')_<\bigr]\text{d}\theta=1\,.
\ffin
which entails the radial averaged contribution. We have eliminated angular contributions by selecting the convenient contour $S_\epsilon$ suggesting a pathway to extend it to the $\lambda$-modes. Direct substitution of \cref{eq:green-fourier-exp} along with a convenient choice of unity---inspired by \cref{eq:dif-eq-gen5}---gives us ultimately,
\eemp\label{eq:cond-discon2-p}
r'\bigl[G\,'_{\!\lambda}(r'_>,r',\theta')-G\,'_{\!\lambda}(r'_<,r',\theta')\bigr]=e^{-i\lambda\theta'}\,,
\ffin
where primes denote partial derivatives with respect to the first argument at both left ($<$) and right ($>$) hand sides of $r'$. Substituting this relationship in the differential equation, we obtain a similar relationship for the second derivatives, essential to the numerical method, as follows,
\eemp\label{eq:cond-discon2-pp}
(r')^2\bigl[G\,''_{\!\lambda}(r'_>,r',\theta')-&G\,''_{\!\lambda}(r'_<,r',\theta')\bigr]=\\
&-e^{-i\lambda\theta'}\left[1+r'\,f_r(r',\theta')\right]\,.
\ffin

For the disc, the $r'=0$ case must be clarified. In polar coordinates, Dirac's distribution is best described as absent of angular dependence, which entails that for all $\lambda$-modes except $\lambda=0$ it is exactly zero. Therefore, the boundary at the origin for each $\lambda$-mode is dictated by symmetry except for $\lambda=0$. This last, carries the logarithmic divergence. This means that conditions for non-zero modes are unchanged. For the zero mode and due to symmetry $G_{0}^{\prime}|_{r\to0^+} = -G_{0}^{\prime}|_{r\to0^-}$ and $G_{0}^{\prime\prime}|_{r\to0^+} = G_{0}^{\prime\prime}|_{r\to0^-}$; however, due to the divergence a cutoff must be set in place. Such a choice of cutoff will be discussed later.

\section{Finite Differences Method, FDM or FEM on a regular grid}
\label{sec:method}

The FDM, or uniform mesh FEM, has been used extensively in the literature to find approximate solutions for many physical systems and its stability makes it a suitable candidate to obtain a numerical Green distribution function. Some examples include the one-dimensional Schr\"odinger equation \citep{TRUHLAR1972123}, the Poisson equation for Electrodynamics \citep{articleJomaa}, the Euler equations of inviscid fluid flow \citep{STEGER1978175}, solutions to 1D and 2D Burgers' equation \citep{articleBurger} and the time-fractional diffusion equation \citep{LIN20071533}. From the mathematical perspective, the same method has been implemented to solve elliptic, hyperbolic and parabolic partial DEs on irregular meshes \citep{articleIzaian}, with interfaces \citep{articleJo}, or in finding optimal algorithms on nontrivial meshes \citep{KWAK199977}.

Orchestrating an exact solution to \cref{eq:dif-eq-gen5} is virtually not possible. There are four cases where an analytical approach can be attempted: two cases where either $\vec{f}$ or $g$ are zero, requiring to find a base of $G_\lambda$'s that can decouple the system---hence, a diagonalization---, the unique case where the same base applies to both coupling matrices accompanying $G_{\lambda}^{\prime}$ and $G_\lambda$, and the trivial \emph{free}-wave ($\vec{f}$ and $g$ zero). Excluding the latter, finding this diagonalizing operator for the first three cases will be addressed in a future study.

Therefore, we will compute a numerical solution where we approximate the operator with finite differentiation (the finite difference method ---FDM or FEM for a regular grid) and bind expansions to include all relevant Green and function modes up to a calculated cutoff; maximum and minimum modes will be chosen respectively for $\lambda$- and $\mu$-modes symmetrically as $|\lambda| \leq L$ and $|\mu| \leq M$ considering that $M\leq L$ for reasons that will be clarified afterwards.

Since the Green function is twice differentiable, when $\mathbf{r}\neq\mathbf{r}'$, its Fourier series converges uniformly and its coefficients decay at least as $\lambda^{-2}$, conditioned by equally well behaved functions $\vec{f}$ and $g$. Then, a possible educated choice of $L$ is the minimum integer such that the sum of $1/k^2$ up to $L$ exceeds $\frac{\pi^2}{6}p$, with $p$ a percentage of accuracy; for example, to achieve at most $1\%$ of estimation error we require $L>60$.

Numerical details and calculations performed henceforth are presented solely for the 3--point--stencil. The strategy for the implementation of more accurate approximations will only be mentioned and briefly discussed; their details will be left for the reader to carry them out. Other minor and mayor details regarding the procedure will be addressed in a future work.

\subsection{A {\it large} matrix equation}

\label{subsec:methodA}

The Finite Differences Method (FDM or FEM---finite elements method---with uniform grid) is a simple approach to computing derivatives of functions at a point by using Taylor expansions on a discretized mesh.\footnote{The choice of whether dissecting uniformly or non-uniformly is highly dependable on the problem. For example, if we were interested in fracture dynamics we would prefer a non-uniform grid to model complex material topologies.} In doing so, a derivative will rely on knowledge of the values of the function in neighboring sites. Such is the art of computing derivatives. The number of neighboring sites to be taken into consideration determines the degree of which the function approaches to the point value. For instance, in the so called three-point stencil (the site in question and its two adjacent neighbors), the first and second derivatives are accurate up to order square of the mesh size.

\begin{center}
\footnotesize
    \begin{table}
    \setlength{\tabcolsep}{4pt}
    \renewcommand{\arraystretch}{1.5}
    \begin{tabular}{|c|c|}
    \hline
    Variable or function & Equivalent array \\ \hline
    \multirow{2}{*}{$r$ and $r'$} & $r^{j}=r^0+h\,j$ with $r^0=R_{\text{int}}$ \\
     & $j\in\{0,1,2,\dots,N\}$ \\ \hline
    $G_\lambda(r,r',\theta')$ & $\GREEN{\lambda}{j}{k}=G_\lambda(r^j,{r^{\prime}}^{k},\theta')$ \\ \hline
    $\QFUNC{r}\defeq r^2$ & $\QMAT{j}=\QFUNC{r^j}$ \\ \hline
    $\FFUNC{\mu}{r}\defeq r^2f_{r\,\mu}(r)+r\delta_{0\mu}$ & $\FMAT{\mu}{j}=\FFUNC{\mu}{r^j}$ \\ \hline
    $\GFUNC{\lambda}{\mu}{r}\defeq r^2g_\mu(r)-\lambda^2\delta_{0\mu}$ & \multirow{2}{*}{$\GMAT{\lambda}{\mu}{j}=\GFUNC{\lambda}{\mu}{r^j}$} \\
    $+ir(\lambda-\mu)f_{\theta\,\mu}(r)$ & \\ \hline
    \end{tabular}
    \caption{A summary on the change of notation from continuous to discrete form}\label{tab:discretization}
    \end{table}
\normalsize
\end{center}

Going back to our problem in \cref{eq:dif-eq-gen5}, we turn to a simply redefined one dimensional DE for a sketch of the forthcoming operations. The left hand side reads rewritten as,
\begin{gather*}
\QFUNC{r}\ordeN{2}{G_{\lambda}(r,r',\theta')}{r}+\sum_{\mu\in\mathbb{Z}}\FFUNC{\mu}{r}\orde{G_{\lambda-\mu}(r,r',\theta')}{r}\\
+\sum_{\mu\in\mathbb{Z}}\GFUNC{\lambda}{\mu}{r}G_{\lambda-\mu}(r,r',\theta').
\end{gather*}
In transforming the continuous variables $r$ and $r'$ into a discrete equally--spaced mesh of size $h$, we will adopt matrix notation for variables and functions; ergo, for $N$ partitions defining $N+1$ points $h=(R_{\text{ext}}-R_{\text{int}})/N$ in a disc--like geometry. For clarity, we summarize notation changes in \cref{tab:discretization}. This procedure applied over the aforementioned equation gives for $r\neq r'$,
\begin{gather*}
\QMAT{j}\bigg[\frac{1}{h^2}\sum_{\eta\in\FDMset{j}}\FDMele{\eta}{j}{2}\GREEN{\lambda}{j+\eta}{k}\bigg]+\sum_{\mu\in\mathbb{Z}}\FMAT{\mu}{j}\bigg[\frac{1}{h}\sum_{\eta\in\FDMset{j}}\FDMele{\eta}{j}{1}\GREEN{\lambda-\mu}{j+\eta}{k}\bigg]\\
+\sum_{\mu\in\mathbb{Z}}\GMAT{\lambda}{\mu}{j}\GREEN{\lambda-\mu}{j}{k}+\mathcal{O}(h^\xi),
\end{gather*}
with $\xi$ the order of approximation, $\FDMset{j}$ the set of neighbor site indices, and $\FDMele{\eta}{j}{n}$ the respective coefficient (namely the {\it finite difference coefficient} included into a matrix representation $\FDMmat{n}$---see \cref{app:fem-method}) of the $\eta$-th neighbor required to compute the $n$-th derivative up to a predetermined order of accuracy \citep{Fornberg1988}; in the three-point stencil case, $\xi=2$. Both sets of neighbor indices and coefficients depend on the information of the site $j$ under inspection; if for example we are at or near an interface, boundary or discontinuity then the strategy for choosing neighbors may differ; we might be interested in computing derivatives using only points in regions where it makes sense.

With some reorganization, the generated discrete DE can be regarded as a matrix multiplication. To see this, first we realize that by understanding $\GREEN{\lambda}{j}{k}$ as the $(j,k)$-th element of a constructed matrix $\GREENM{\lambda}$---of size $N+1\times N+1$---we can envision a column matrix vector $\GREENMV$ that contains all $\lambda$-modes, or all of $\{\GREENM{\lambda}\,\forall\,\lambda\in\mathbb{Z}\}$, where all operations from the previous complex array equation are condensed into an equally conceived matrix $\PMATM$ multiplying $\GREENMV$. The following is a view of $\GREENMV$,
\begin{widetext}
\begin{align}
    \GREENMV=\left(\begin{array}{c}
         \vdots  \\
         \GREENM{-L}  \\
         \vdots  \\
         \GREENM{-1}  \\
         \GREENM{0}  \\
         \GREENM{1}  \\
         \vdots  \\
         \GREENM{L}  \\
         \vdots
    \end{array}\right)\,\,,\,\,\text{with}\,\,\GREENM{\lambda}=\left(\begin{array}{ccccc}
        \GREEN{\lambda}{0}{0} & \GREEN{\lambda}{0}{1} & \cdots & \GREEN{\lambda}{0}{N-1} & \GREEN{\lambda}{0}{N} \\
        \GREEN{\lambda}{1}{0} & \GREEN{\lambda}{1}{1} & \cdots & \GREEN{\lambda}{1}{N-1} & \GREEN{\lambda}{1}{N} \\
        \vdots & \vdots & \ddots &\vdots & \vdots\\
        \GREEN{\lambda}{N-1}{0} & \GREEN{\lambda}{N-1}{1} & \cdots & \GREEN{\lambda}{N-1}{N-1} & \GREEN{\lambda}{N-1}{N} \\
        \GREEN{\lambda}{N}{0} & \GREEN{\lambda}{N}{1} & \cdots & \GREEN{\lambda}{N}{N-1} & \GREEN{\lambda}{N}{N}
        \end{array}\right).
\end{align}
\end{widetext}

In principle, both matrices are infinitely large but for practical terms they will be truncated on both $\lambda$- and $\mu$-modes as mentioned in the previous section. Despite this numerical simplification that will be carried out in the numerical analysis, the infinite matrix $\PMATM$ has a well defined structure as will be detailed in \cref{subsec:P-matrix}.

Finally, the terms to the right of \cref{eq:dif-eq-gen5} vanish for all $r\neq r'$ leading us to believe that if $\PMATM$ is invertible then the solution to the discrete Green function $\GREENMV$ is identically zero. However, attention should be paid at $r=r'$ for its effect discards the trivial solution. Along with the other geometrical boundary conditions the problem will now have a \emph{unique} solution. These boundary conditions will be addressed in \cref{subsec:discrete-bc}.

\subsection{Infinite matrix $\PMATM$}
\label{subsec:P-matrix}

To understand the structure of $\PMATM$ we turn to the set of operations for a particular $\lambda$-mode. Seeing as $\PMATM$ is infinite we may encode rows by the integer value of the mode being solved and columns by the value of the mode being correlated. Thus taking row $\lambda$ from $\PMATM$,
\begin{widetext}

\begin{align}
    \PMATM_\lambda\GREENMV=
    \left(
    \underset{\cdots\,,\,M\,,\,\cdots,\,3\,,\, 2\,,\,1\,\longleftarrow\,\mu\,|}
    {\cdots,\overbrace{\frac{1}{h}\FMATM{\mu}+\GMATM{\lambda}{\mu}}^{\text{column}\,\lambda-\mu},\cdots,}
    \underset{\mu=0}{\overbrace{\frac{1}{h^2}\QMATM+\frac{1}{h}\FMATM{0}+\GMATM{\lambda}{0}}^{\text{column}\,\lambda}}
    \underset{|\,\mu\longrightarrow\, -1\,,\,-2\,,\,\cdots\,,\,-M\,,\,\cdots}{\,,\cdots,\,\overbrace{\frac{1}{h}\FMATM{\mu}+\GMATM{\lambda}{\mu}}^{\text{column}\,\lambda-\mu},\cdots}
    \right)
    \left(\begin{array}{cl}
     \vdots & \\
     \GREENM{\lambda-\mu} & \Bigr\}\,{\scriptstyle\text{row}\, \lambda-\mu} \\
     \vdots & \\
     \GREENM{\lambda} & \Bigr\}\,{\scriptstyle\text{row}\, \lambda} \\
     \vdots & \\
     \GREENM{\lambda-\mu} & \Bigr\}\,{\scriptstyle\text{row}\, \lambda-\mu} \\
     \vdots &
    \end{array}\right),
\end{align}
\end{widetext}
with the following definitions for matrices $\QMATM$, $\FMATM{\mu}$, $\GMATM{\lambda}{\mu}$,
\begin{align}
    \QMATM=\left(\begin{array}{cccc}
        \QMAT{0} & 0 & \cdots & 0 \\
        0 & \QMAT{1} & \cdots & 0 \\
        \vdots & \vdots & \ddots & \vdots\\
        0 & 0 & \cdots & \QMAT{N}
        \end{array}\right)\times\FDMmat{2},
\end{align}
\begin{align}
    \FMATM{\mu}=\left(\begin{array}{cccc}
        \FMAT{\mu}{0} & 0 & \cdots & 0 \\
        0 & \FMAT{\mu}{1} & \cdots & 0 \\
        \vdots & \vdots & \ddots & \vdots\\
        0 & 0 & \cdots & \FMAT{\mu}{N}
        \end{array}\right)\times\FDMmat{1},
\end{align}

\begin{align}
    \GMATM{\lambda}{\mu}=\left(\begin{array}{cccc}
        \GMAT{\lambda}{\mu}{0} & 0 & \cdots & 0 \\
        0 & \GMAT{\lambda}{\mu}{1} & \cdots & 0 \\
        \vdots & \vdots & \ddots & \vdots\\
        0 & 0 & \cdots & \GMAT{\lambda}{\mu}{N}
        \end{array}\right).
\end{align}

One last remark on matrix $\PMATM$ is that the density of non--zero entries is at most $3/N$ for the three-point stencil. For $N$ sufficiently large, it will become essential to find a way to manage such sparsity for all speedups, data-compression and efficiency in memory footprint.

\subsection{Discrete Boundary Conditions}

\label{subsec:discrete-bc}

Retaking conditions detailed thoroughly in \cref{subsec:G-boundaries} and at end of \cref{subsec:2d-to-1d} we are now in capacity of parameterizing the values of $\GREEN{\lambda}{j}{k}$. This parametrization should further reflect the behavior of the $\delta$--function. The following are the conditions for the 3--point--stencil: (\emph{i.}) at $r=R_{\text{ext}}$,
\eemp\label{eq:condN}
\GREEN{\lambda}{N}{k}&=0\,\,\,\,\,\,\textrm{for DBC}\,;
\\
\highlight{\GREEN{\lambda}{N+1}{k}}-\GREEN{\lambda}{N-1}{k}&=0\,\,\,\,\,\,\textrm{for NBC}\,,
\ffin
(\emph{ii.}) at $R_{\text{int}}$ for the annulus,
\eemp\label{eq:cond0}
\GREEN{\lambda}{0}{k}&=0\,\,\,\,\,\,\textrm{for DBC}\,;
\\
\GREEN{\lambda}{1}{k}-\highlight{\GREEN{\lambda}{-1}{k}}&=0\,\,\,\,\,\,\textrm{for NBC}\,,
\ffin
(\emph{iii.}) for the disc at $R_{\text{int}}$ (disregarding $G_{\lambda}^{\prime}=0$ for now),
\eemp\label{eq:cond0D01}
\GREEN{0}{1}{k}-\highlight{\GREEN{0}{-1}{k}}=0&\,\,\,\,\,\,\lambda=0,\\
\GREEN{\lambda}{0}{k}=0&\,\,\,\,\,\,\lambda\neq 0,
\ffin
and, finally, (\emph{iv.}) at the interface $r=r'$ the condition reads,
\footnotesize
\eemp\label{eq:disc-r}
\bigl(\GREEN{\lambda}{k_>+1}{k}-\highlight{\GREEN{\lambda}{k_>-1}{k}}\bigr)-\bigl(\highlight{\GREEN{\lambda}{k_<+1}{k}} -\GREEN{\lambda}{k_<-1}{k}\bigr)=\frac{2h}{r^{k}}e^{-i\lambda\theta'}.
\ffin
\normalsize
Here we have adopted the subscript convention of $<,>$ to refer to points to the left and right of the site of derivative evaluation. Note how all equations above reference and highlight a few fictitious points. The mesh points that lay outside or \emph{beyond} the valid grid are $\GREEN{\lambda}{N+1}{k}$, $\GREEN{\lambda}{-1}{k}$, $\GREEN{\lambda}{k_<+1}{k}$, and $\GREEN{\lambda}{k_>-1}{k}$. These spurious terms must be dealt with and simplified in order to be able to incorporate readily all conditions.

\subsection{A Non--Trivial Matrix Equation and a Solution}
\label{sec:mat-ele}

We will now show the explicit matrix equation associated with the conditions described above. As mentioned, they depend on the degree of accuracy that we choose, or equivalently, the stencil. We will describe the procedure for the three-point stencil and further discuss how to generalize for higher orders of approximation.

With the boundary relationships in mind, here in \cref{eq:condN,eq:cond0,eq:cond0D01,eq:disc-r},  \cref{eq:dif-eq-gen5} (multiplied by $-h^2$) equates partially to zero (when $r\neq r'$) as,
\begin{align*}
&2P^j\GREEN{\lambda}{j}{k}\!-P^j(\GREEN{\lambda}{j+1}{k}+\GREEN{\lambda}{j-1}{k})-h^2\sum_{\mu}\GMAT{\lambda}{\mu}{j} \GREEN{\lambda-\mu}{j}{k}\nonumber\\
&-\frac{h}{2}\sum_{\mu}\FMAT{\mu}{j}( \GREEN{\lambda-\mu}{j+1}{k}-\GREEN{\lambda-\mu}{j-1}{k})=0\,,
\end{align*}
where via \cref{eq:disc-r} the latter can be used to simplify both spurious terms (appearing at $r=r'$) $\GREEN{\lambda}{k_<+1}{k}$ and $\GREEN{\lambda}{k_>-1}{k}$. After crossing out these terms by iterative substitution we obtain a generalized expression for the above valid for \emph{almost} every point in the grid.
The general discrete equation yields for $r'>0$,
\begin{align}\label{eq:dif-disc1}
&2P^j\GREEN{\lambda}{j}{k}\!-P^j(\GREEN{\lambda}{j+1}{k}+\GREEN{\lambda}{j-1}{k})-h^2\sum_{\mu}\GMAT{\lambda}{\mu}{j} \GREEN{\lambda-\mu}{j}{k}\nonumber\\
&-\frac{h}{2}\sum_{\mu}\FMAT{\mu}{j}( \GREEN{\lambda-\mu}{j+1}{k}-\GREEN{\lambda-\mu}{j-1}{k})=-hr^{j}\delta^{j,k}e^{-i\lambda\theta'} \times \nonumber\\
&\left\{1-\frac{h^2}{4(r^k)^2}[1+r^kf_r(r^k,\theta')]^2\bigl[1-\delta_{\lambda,0}\bigr]\right\},
\end{align}
where the new term that accounts for the boundary condition at $r=r'$ has appeared. Due to the absence of a left-hand limit  as $r'=0$, according to \cref{eq:disc-r}, this term is exactly $-hr^{j}\delta^{j,k}e^{-i\lambda\theta'}$ at the origin. Actually, this condition holds for the mode $\lambda=0$ in general due to translational invariance---this invariance is clearly absent for the other modes. The terms composing the right hand side of last equation can be viewed as of order of mesh--size or order of radial distance from the origin as follows,
\begin{enumerate}
    \item $-\frac{h^3}{4}f_r(r^k,\theta')$, a surprising third order correction due to the vector field appearing after substituting the interface difference in derivatives.
    \item $h\,r^k=h\,R_{\textrm{int}}+h^2k$, the leading order that substituted yields a first order constant term and a second order increasing term.
    \item $-\frac{h^3}{4}\frac{1}{r^k}=-\frac{h^3}{4}\frac{1}{R_{\textrm{int}}+h\,k}$, a negative term significant closer to the origin. As expected, the behavior of the discrete version near zero validates our previous choice of boundary condition for the disc.
    \item For $r'=0$, we must implement a cutoff such that $r^0=\epsilon>0$ instead of zero to avoid numerical divergences. The choice for $\epsilon$ will be discussed below.
\end{enumerate}
Because the error in the differential equation is of $O(h^4)$, we should incorporate all terms to the calculation. However, we will neglect the higher order term---first term---since this will simplify our calculations of $\psi(\mathbf r)$.

This final expression is valid everywhere including the controversial $j=0,N$ points, where either Dirichlet or Neumann conditions complete \cref{eq:dif-disc1} at the borders. In those two cases, substitutions must take place following \cref{eq:condN,eq:cond0,eq:cond0D01}. After replacements, and due to the nature of derivative calculation in the three-point stencil, rows from $\PMATM$ corresponding to exterior and interior borders are modified. See the substitution rules in \cref{tab:rules1a,tab:rules2a}.

We now define our complete matrix system as $\PMATM\cdot\GREENMV=-h\VMAT\cdot \EMATT$. Among other things, the right hand side accounts for the contribution of Dirac's distribution. The two additional definitions appearing correspond to first a distance parameter generalized into $\alpha^j_\lambda$, a new object that incorporates the boundary conditions at both $j=0$ and $j=N$. Notice, for instance, that keeping the term $r^j$ at every point does not explain the vanishing of the Green function at the boundaries when DBC are considered, neither does it describe the correct behavior at $r=0$ for a disk. Actually, when the last condition is considered, an ultraviolet cutoff $\epsilon$---such that $\epsilon\to0$---must be introduced to avoid divergences, as seen in previous works \citep{cornu:2444,Ferrero2014,Ferrero2007}. Such cutoff is not surprising, as the 2D Green distribution has a natural divergence at $\mathbf r=\mathbf r'$ and a logarithmic behavior near the origin when $\mathbf{r'}=0$. Although the appearance of this divergence can easily be visualized after studying the behaviour of \cref{eq:dif-disc1} at $j=0$ for the mode $\lambda=0$ in a disk, its existence at any point---also for an annulus---is guaranteed by the infinite number of $\lambda$-modes that must be summed up to obtain an exact solution. Therefore, it is not surprising that $\epsilon$ and $h$ are related---see section \ref{sec:numerical_results} for more details.

Matrix terms are written as,
\begin{subequations}
\begin{align}\label{eq:eleV}
    \VMATE{\lambda}{\mu}{j}{k}&=\alpha_{\lambda}^{j}\delta_{\lambda,\mu}\delta^{j,k}\,,\\\label{eq:eleE}
    \EMATE{\lambda}{j}{k}&=e^{-i\lambda\theta'}\delta^{j,k},
\end{align}
\end{subequations}
and the solution to the $\lambda$--modes matrix $\GREENMV$ is,
\emp
\GREENMV=-h\AMAT\cdot\EMATT,
\fin
where we have defined $\AMAT=\PMATMINV\cdot\VMAT$ assuming that $\PMATM$ is invertible.

Matrix $\AMAT$ was declared because it has interesting symmetry properties that will be discussed in the next section.
\Cref{tab:rules1a,tab:rules2a,tab:rules2b,tab:rules3a} outline how to fill the matrix elements of the objects we have described.

\begin{center}
    \begin{table}
    \setlength{\tabcolsep}{3pt}
    \renewcommand{\arraystretch}{1.45}
    \begin{tabular}{|c|c|c|c|c|c|}
    \hline
    Mode & $\PMATME{\lambda}{\lambda}{j}{j}$ & $\PMATME{\lambda}{\lambda}{j}{j\pm1}$ & $\PMATME{\lambda}{\lambda-\mu}{j}{j}$ & $\PMATME{\lambda}{\lambda-\mu}{j}{j\pm1}$\\
    \hline
    $\forall\,\lambda$ & $2\QMAT{j}-h^2\GMAT{\lambda}{0}{j}$ & $-\QMAT{j}\mp \frac{h}{2}\FMAT{0}{j}$ & $-h^2\GMAT{\lambda}{\mu}{j}$ & $\mp \frac{h}{2}\FMAT{\mu}{j}$ \\
    \hline
    \end{tabular}
    \caption{Nonvanishing matrix elements of $\PMATM$ for $1\leq j\leq N-1$ for an annulus and a disc.}\label{tab:rules1a}
    \end{table}
\end{center}
\begin{center}
    \begin{table}
    \setlength{\tabcolsep}{2pt}
    \renewcommand{\arraystretch}{1.35}
    \begin{tabular}{|c|c|c|c|c|c|c|}
    \hline
   & Mode & $ j$ & $\PMATME{\lambda}{\lambda}{j}{j}$ & $\PMATME{\lambda}{\lambda}{j}{j+1}$ & $\PMATME{\lambda}{\lambda}{j}{j-1}$ & $\PMATME{\lambda}{\lambda-\mu}{j}{j}$ \\
    \hline
    (D) & $\forall \,\lambda$ & $0,N$ & 1 & 0, N/A & N/A, 0 & 0 \\ 
    \hline
    (N) & $\forall \,\lambda$ & $0$ & $2\QMAT{0}\!-h^2\GMAT{\lambda}{0}{0}$ & $-2\QMAT{1}$ & N/A & $-h^2\GMAT{\lambda}{\mu}{0}$ \\ 
    \hline
    (N) & $\forall\, \lambda$ & $N$ & $2\QMAT{N}\!-h^2\GMAT{\lambda}{0}{N}$ & N/A & $-2\QMAT{N-1}$ & $-h^2\GMAT{\lambda}{\mu}{N}$ \\ 
    \hline
    \end{tabular}
    \caption{Nonvanishing matrix elements of matrix $\PMATM$ that define DBC (D) and NBC (N) for an annulus. N/A specifies those elements that lay outside $\PMATM$. }\label{tab:rules2a}
    \end{table}
\end{center}
\begin{center}
    \begin{table}
    \setlength{\tabcolsep}{2pt}
    \renewcommand{\arraystretch}{1.35}
    \begin{tabular}{|c|c|c|c|c|c|c|}
    \hline
    Mode & $ j$ & $\PMATME{\lambda}{\lambda}{j}{j}$ & $\PMATME{\lambda}{\lambda}{j}{j+1}$ & $\PMATME{\lambda}{\lambda}{j}{j-1}$ & $\PMATME{\lambda}{\lambda-\mu}{j}{j}$ \\ 
    \hline
    \hline
    $\lambda=0$ & $0$ & $2-h^2g_0^{0}$ & $-2$ & N/A &  $0$ \\
    \hline
    $\lambda=0$ & $N$ & $1$ & N/A & $0$ & $0$ \\
    \hline
    $\lambda\neq 0$ & $0,N$ & $1$ & $0$, N/A & N/A, $0$ & $0$\\
    \hline
    \hline
    $\lambda=0$ & $0$ & $2-h^2g_0^{0}$ & $-2$ & N/A & $0$ \\ 
    \hline
    $\forall\,\lambda$ & $N$ & $2\QMAT{N}\!-h^2\GMAT{\lambda}{0}{N}$ & N/A & $-2\QMAT{N-1}$ & $-h^2\GMAT{\lambda}{\mu}{N}$ \\ \hline
    $\lambda\neq0$ & $0$ & $1$ & $0$ & N/A & $0$ \\
    \hline
    \end{tabular}
    \caption{Nonvanishing matrix elements of matrix $\PMATM$ that define DBC (shown above) and NBC (shown below) for a disc. N/A specifies those elements that lay outside $\PMATM$.}\label{tab:rules2b}
    \end{table}
\end{center}

\begin{center}
    \begin{table}
    \setlength{\tabcolsep}{3pt}
    \renewcommand{\arraystretch}{1.40}
    \begin{tabular}{|c|c|c|c|c|c|c|}
    \hline
    Mode & $ j$ & $\alpha^{j}_{\lambda}$ (DBC) & $\alpha^{j}_{\lambda}$ (NBC) & Geom.\\
    \hline
    $\lambda=0$ & $0$ & $\epsilon^{-1}$ & $\epsilon^{-1}$ & (D) \\ 
    \hline
    $\lambda=0$ & $0$ & $0$ & $r^0$ & (A) \\ 
    \hline
    $\lambda=0$ & $1\leq j\leq N-1$ & $r^j$ & $r^j$ & (A, D) \\
    \hline
    $\lambda=0$ & $N$ & $0$ & $r^N$ & (A, D) \\
    \hline
    $\lambda\neq 0$ & $0$ & $0$ & $0$ & (D) \\ 
    \hline
    $\lambda\neq 0$ & $0$ & $0$ & $r^0-\frac{h^2}{4r^0}$ & (A) \\ 
    \hline
    $\lambda\neq 0$ & $1\leq j\leq N-1$ & $r^j-\frac{h^2}{4r^j}$ & $r^j-\frac{h^2}{4r^j}$ & (A,\,D) \\ 
    \hline
    $\lambda\neq 0$ & $N$ & $0$ & $r^N-\frac{h^2}{4r^N}$ & (A,\,D) \\ 
    \hline
    \end{tabular}
    \caption{Elements $\alpha_\lambda^j$ that define DBC and NBC for an annulus (A) and a disc (D).}\label{tab:rules3a}
    \end{table}
\end{center}

\subsection{The parameter $\alpha$ and the symmetry of $\AMAT$}
\label{subsec:distance-param-alpha-and-A}

A closed relation can be found for the matrix describing the entire Green function. Using the results from previous section it is
\begin{align}\label{eq:mult2}
    G^{j,k}_{\theta,\theta'}=-\frac{h}{2\pi}\sum_{\lambda,\,\mu}\,e^{i\lambda\theta}e^{-i\mu\theta'}\AMATE{\lambda}{\mu}{j}{k}\,,
\end{align}
where $\AMATE{\lambda}{\mu}{j}{k}=\PMATEINV{\lambda}{\mu}{j}{k}\,\alpha_{\mu}^{k}$. As previously mentioned, the parameter $\alpha_\lambda^k$ generalizes the radial parameter $r^k$, including the boundary conditions. On the other hand, it is worthwhile to state the symmetry conditions that $\AMAT$ satisfies\footnote{A more detailed derivation can be found in appendix \ref{app:properties}; $\overline{z}$ denotes the complex conjugate of $z$.}
\eemp\label{eq:properties}
&\operatorname{Im}\(\AMATE{0}{0}{j}{k}\)=0\,\,\,,\,\,\,\\ &\AMATE{-\lambda}{0}{j}{k}=\overline{\AMATE{\lambda}{0}{j}{k}}\,\,\,,\,\,\,\AMATE{0}{-\mu}{j}{k}=\overline{\AMATE{0}{\mu}{\!j}{k}}\,\,\,,\,\,\,\\
&\AMATE{-\lambda}{\mu}{j}{k}=\overline{\AMATE{\lambda}{-\mu}{j}{k}}\,\,\,,\,\,\,\AMATE{-\lambda}{-\mu}{j}{k}=\overline{\AMATE{\lambda}{\mu}{j}{k}}\,.
\ffin
Using \cref{tab:rules3a}, it is easy to see that the matrix elements $\PMATEINV{\lambda}{\mu}{j}{k}$ satisfy the same symmetry properties.

\subsection{The algorithm}
\label{subsec:the-algorithm}

The algorithm for a numerical solution can be summarized as follows:
\begin{enumerate}[noitemsep]\label{steps}
    \item The values of $L$ and $M$ are determined according to the required level of approximation.
    \item We fill all elements described in table \ref{tab:discretization}; the matrix elements $\PMATME{\lambda}{\mu}{j}{k}$ are filled by blocks using the rules shown in tables \ref{tab:rules1a}, \ref{tab:rules2a}, \ref{tab:rules2b}. Matrix elements $\alpha_\lambda^j$ are also filled according to table \ref{tab:rules3a}.
    \item Matrix ${\PMATM}$ is inverted and so matrix $\AMAT$ is computed.   
    \item The Green function is computed according to \cref{eq:mult2}.
\end{enumerate}
Using previous results, we can deduce a closed form for $\psi(\mathbf r)$ for both DBC and NBC using the conventions stated in \cref{eq:sol-tot} and \cref{eq:sol-tot1}. Although there are many ways to perform the integrals stated in previous equations, and the reader can choose the method that he or she prefers, a sketch of these solutions, using the trapezoid rule, is shown in appendices \ref{app:psi-exp} and \ref{app:function-sc}.

\begin{center}
{\bf Particular cases and properties}
\end{center}

We will analyze some particular cases that can be deduced from the procedure explained above. We will start focusing on the one-dimensional case.

\begin{center}
    {\bf One dimensional case}
\end{center}

The analysis of a Green function in one dimension requires an appropriate definition of a general DE obeyed by the Green function $G(x,x')$. Unfortunately, a direct analysis of the results by studying \cref{eq:dif-eq-gen5} is not straightforward due to the clear differences between the Laplacians in cartesian and polar coordinates. Let us imagine a general second order DE of the form $\mathcal L_x\psi(x)=\phi(x)$, where the  Green function satisfies the relation
\eemp\label{eq:DE-gen}
    \mathcal L_x G(x,x')&=P(x)\frac{d^2G}{dx^2}+Q(x)\frac{dG}{dx}+R(x)G\\
    &=\beta\delta(x-x')\,.
\ffin
The function $P(x)$ might not be necessary, as it can be eliminated by division, but its inclusion allows us to have a more general analysis. The constant therm $\beta$ seems clumsily placed, as its value is usually 1. Nonetheless, some formalisms define the Green function by means of the operator $\mathcal L_{x}G(x,x')=-\delta(x-x')$, thus introducing a change of sign that can be contemplated in our study. 

By following a similar analysis as that shown above, we can deduce an appropriate recurrence relation for \cref{eq:DE-gen}, which is
\eemp\label{eq:DE-gen2}
&(2P^j-h^2R^j)G^{j,k}\!-(P^j+\frac{h}{2}Q^j)G^{j+1,k}\\
&\!-(P^j-\frac{h}{2}Q^j)G^{j-1,k}
=-h\beta\,\delta^{j,k}.
\ffin
Having confined the system within the domain $x\in[x^0,x^N]$, our step size is now $h=(x^N-x^0)/N$.

From this point on, we can apply the results obtained for the two dimensional problem in this study. Notice that, in the absence of modes that account for the angular dependence, we can always say that $\AMATE{\lambda}{\mu}{j}{k}=\AMATE{\lambda}{\mu}{j}{k}\delta_{\lambda0}\delta_{\mu0}$. Therefore, \cref{eq:mult2} becomes
\begin{align}\label{eq:G1D}
    G^{jk}=-h\AMATED{j}{k}\,,\,\,\,\textrm{where}\,\,\,\AMATED{j}{k}=\PMATEINVD{j}{k}\alpha^{k}
\end{align}
and $\alpha^k$ accounts for the boundary conditions. By making the association $x^j=x^0+hj$, the elements described in \cref{eq:G1D}, for $1\leq j\leq N-1$, are now filled using the following rules: 
\begin{enumerate}[noitemsep]
\item $\PMATED{j}{j}=2P^j-h^2R^j$. 
\item $\PMATED{j}{j\pm 1}=-P^j\mp \frac{h}{2}Q^j$.
\item $\PMATED{0}{0}=\PMATED{N}{N}=1$ for DBC. For NBC: $\PMATED{0}{0}=2q^0-h^2b^0$, $\PMATED{N}{N}=2P^N-h^2R^N$, $\PMATED{0}{1}=-2P^{1}$ and $\PMATED{N}{N-1}=-2P^{N-1}$.
\item $\alpha^j=\beta$.
\item $\alpha^0=\alpha^N=0$ for DBC. For NBC: $\alpha^0=\beta$ and $\alpha^N=\beta$. 
\end{enumerate}

The weight function, which now guarantees the symmetry condition $G^{kj}=w^{jk}G^{jk}$ takes the form
\begin{align}\label{eq:weight-1D}
    w(x',x)=\frac{e^{U(x')}}{e^{U(x)}}\,\,,\,\,\,
    U(z)=\frac{1}{P(z)}
    \exp\bigg[\int_{z_0}^{z}\frac{Q(y)}{P(y)}dy\bigg]\,,
\end{align}
where $z_0$ is an irrelevant constant. Solutions for $\psi(x)$ with both DBC and NBC using the trapezoid rule as method of integration are shown in appendix \ref{app:function-sc}.

\begin{center}
{\bf Monopole--like case}
\end{center}

This takes place when both $\vec f(\mathbf r)$ and $g(\mathbf r)$ have no significant angular dependence, so the mode $\mu=0$ is their only relevant contribution; this implies that $\FMAT{\mu}{j}=\GMAT{\lambda}{\mu}{j}=0$ for $\mu\neq 0$. The off-diagonal matrices $\PMATME{\lambda}{\lambda-\mu}{j}{k}$ thus vanish---this leads to a block diagonal ${\PMATM}$ matrix---and so the system becomes separable in the radial and angular variables. Having now the relation $\AMATE{\lambda}{\mu}{j}{k}=\AMATE{\lambda}{\mu}{j}{k}\delta_{\lambda,\mu}$, \cref{eq:mult2} reduces to
 \begin{align}\label{eq:mult2a}
    G^{\,jk}_{\theta\theta'}=-\frac{h}{2\pi}\sum_{\lambda}e^{i\lambda(\theta-\theta')}\AMATE{\lambda}{\lambda}{j}{k}\,.
\end{align}
Notice that each mode can now be solved independently.




\subsection{Beyond the three-point stencil}
\label{subsec:beyond}

As mentioned, we only showed an explicit analysis for a three-point stencil approximation. This method can be generalized to include the contribution of more neighbors in the derivative terms, i.e., higher order stencils that provide more accurate degrees of approximation in $h$. In spite of its simplicity, the three-point stencil has the great advantage that the spurious terms that arise from the boundary conditions can be eliminated in a simple fashion. 

The description of the system with a five-point stencil, for instance, will increase the amount of terms different from zero in $\PMATM$---for example, terms of form $\PMATME{\lambda}{\lambda-\mu}{j}{j\pm2}$ will provide non-zero contributions. Having a higher degree of approximation, that demands the inclusion of more non-trivial matrix terms, the grid size can be reduced. Although there is no guarantee that the inversion process is optimized in time when the contributions of more neighbors are included, as the matrices are highly sparse, there is a clear optimization of memory storage.

Yet a great disadvantage that higher stencils inherit is the elimination of the spurious terms that come from the boundary conditions. For instance, when we deal with the condition at $j=k$ ($r=r'$), \cref{eq:disc-r} will include more coefficients outside the grid, so the recurrence relation that is obtained will not be able to eliminate all of them---at least, using the same procedure we implemented. Therefore, a different approach must be performed. A possible solution could be expanding the derivatives around a point different from the center, so avoiding the spurious terms; this process is studied in detail in \citep{Fornberg1988}. Nonetheless, this could be discussed in a future study.

\section{Numerical Results}
\label{sec:numerical_results}

We will use the formalism described above to solve some particular examples.

\begin{center}
    {\bf Example 1: A one dimensional case}
\end{center}

As a first example, let us study a one dimensional system with a known analytical solution, useful to test the formalism we have described. Let us suppose we want to solve the DE in the domain $[0,4]$
\begin{align}\label{eq:DE-exam1}
x^2\psi''(x)+x\psi'(x)+(x^2-4)\psi(x)=J_4(x)\,,   
\end{align}
where $J_n(x)$ and $Y_n(x)$ are the Bessel functions of first and second kind of order $n$. Using \cref{eq:weight-1D}, we can easily deduce that $w(x',x)=x/x'$.

\begin{figure}
    \centering
    \includegraphics[scale=0.6]{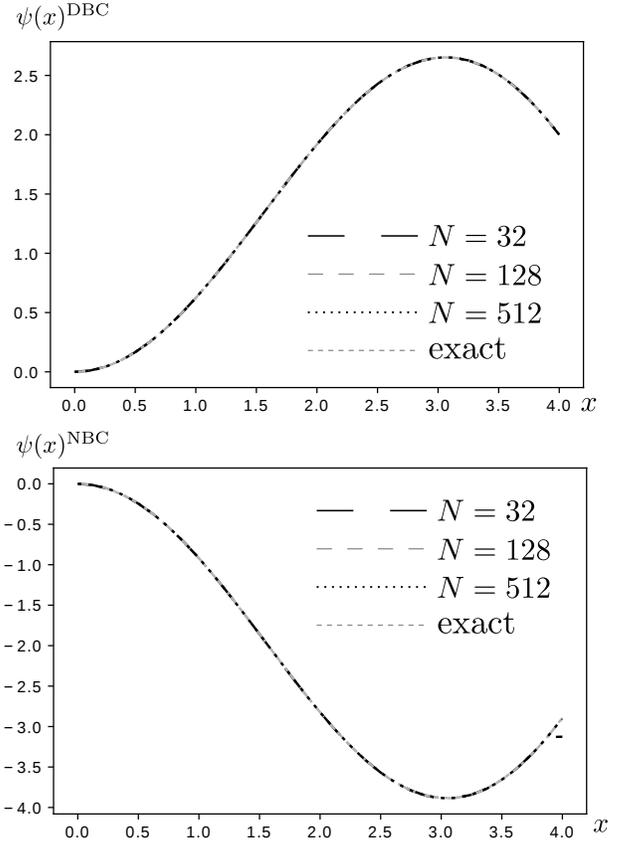}
    \caption{Above: solution to the DE given by \cref{eq:DE-exam1} with the initial conditions $\psi(0)=0$ and $\psi(4)=2$. Below: solution to the DE given by \cref{eq:DE-exam1} with the initial conditions $\psi'(0)=0$ and $\psi'(4)=2$. Table \ref{tab:ex1} analyzes the accuracy of the numerical solutions. Note: we made $x_0=10^{-6}$ to avoid numerical divergences at $x=0$.}
    \label{fig:example1}
\end{figure}

The conditions $\psi^0=0$ and $\psi^N=2$ (Dirichlet), lead to the analytical solution,
\eemp
\psi(x)^{\textrm{DBC}}=&\frac{1}{24 J_2(4)}[J_2(x)(48-2J_4(4)+\\
&\,\,\,+\pi x J_2(4)J_4(x)Y_1(x))\\
&\,\,\,-\pi x J_2(4)J_1(x)J_4(x)Y_2(x)]\,.
\ffin
Conversely, with conditions $f'^0=0$ and $f'^N=2$ (Neumann), the analytical solution yields,
\eemp
\psi(x)^{\textrm{NBC}}=&\frac{1}{24x^3(J_1(4)-J_3(4))}[(2J_0(4)+3J_1(4))\times\\
&\,\,\,(x(x^2-24)J_0(x)-8(x^2-6)J_1(x))\\
&\,\,\,+x^3J_2(x)(96+2J_0(4)-3J_1(4))]\,.
\ffin
Analytic and numerical results are compared for both cases in \cref{fig:example1} and table \ref{tab:ex1}---see \cref{eq:psi1D-1} and \cref{eq:psi1D-2} for explicit expressions using a numerical approach.

It is interesting to contrast the numerical solutions shown in \cref{fig:example1} using the weight function and the one that arises without the weight function formalism---performing the replacement $w^{jk}G^{jk}\to G^{kj}$. Interestingly, from \cref{tab:ex1} we conclude that the introduction of the weight function leads to a more accurate result.

\begin{center}
    \begin{table}
    \setlength{\tabcolsep}{3pt}
    \renewcommand{\arraystretch}{1.35}
    \begin{tabular}{|c|c|c|c|c|}
    \hline
    Func. & $\psi_{N=32}$ & $\psi_{N=128}$ & $\psi_{N=512}$ \\
    \hline
    \hline
    \hline
    $\psi_{max}^{\textrm{DBC}}$ & $2.6510$ & $2.6526$ & $2.6525$ \\
    \hline
    $\textrm{PE}$ & $5.6550\!\times\!10^{-2}$ & $3.7700\!\times\!10^{-3}$ & $1.9221\!\times\!10^{-4}$ \\
    \hline
    $\textrm{MSE}$ & $3.2824$ & $2.6156\!\times\! 10^{-9}$ & $1.0287\!\times\! 10^{-11}$ \\
    \hline
    \hline
    $\psi_{max}^{\textrm{DBC}}$ & $2.7372$ & $2.6736$ & $2.6577$ \\
    \hline
    $\textrm{PE}$ & $3.1939$ & $7.9615\!\times\!10^{-1}$ & $1.9671\!\times\!10^{-1}$ \\
    \hline
    $\textrm{MSE}$ & $3.4935$ & $2.1177\!\times\! 10^{-4}$ & $1.3148\!\times\! 10^{-5}$ \\
    \hline
    \hline
    \hline
    $f_{min}^{\textrm{NBC}}$ & $-3.8852$ & $-3.8853$ & $-3.8856$ \\
    \hline
    $\textrm{PE}$ & $1.4438\!\times\!10^{-2}$ & $1.2060\!\times\!10^{-2}$ & $3.1806\!\times\!10^{-3}$ \\
    \hline
    $\textrm{MSE}$ & $1.6316\!\times\!10^{-3}$ & $2.3096\!\times\! 10^{-5}$ & $3.5993\!\times\! 10^{-7}$ \\
    \hline
    \end{tabular}
    \caption{Analysis of the accuracy of the two numerical solutions to eq. (\ref{eq:DE-exam1}). The first two blocks use the conditions $\psi(0)=0$ and $\psi(4)=2$ and show the value of the maximum---exact value $\psi^{\textrm{DBC}}_{max}=2.6524822\dots$, its percentage error (PE), and mean square error (MSE) of the function along the domain $[0,4]$; the first block uses the weight function, the second one does not. The block below shows something similar for the conditions $\psi'(0)=0$ and $\psi'(4)=4$ and focus on the global minimum ---exact value $\psi ^{\textrm{NBC}}_{min}=-3.88574186\dots$ using the weight function---remember that for NBC the formalism with the weight function is required.}
    \label{tab:ex1}
    \end{table}
\end{center}

\begin{center}
{\bf Example 2: The two-dimensional Helmholtz equation with imaginary wave number}
\end{center}

We now shift our attention to solve a two dimensional system. Let us consider the DE
\begin{align}\label{eq:Green-ex2}
    (\vec\nabla^2-m^2)G(\mathbf r,\mathbf r')=\delta(\mathbf r-\mathbf r')\,.
\end{align}
In the absence of the vector field $\vec f(\mathbf r)$, we conclude that $w(\mathbf r,\mathbf r')=1$; besides, the system is separable in the radial and angular coordinates. The solution to last equation confined in a large disc of radius $r^N=R_{\textrm{ext}}$ with Dirichlet boundary conditions can be found analytically. Adapting the result found in \citep{Ferrero2014}, we deduce that the Green function associated with \cref{eq:Green-ex2} is
\begin{align}\label{eq:solG2}
    G(\mathbf r,\mathbf r')=-\frac{m^2}{2\pi}\sum_\lambda &e^{i\lambda(\theta-\theta')}\Big[I_\lambda(mr_<)K_\lambda(mr_>)\nonumber\\
    &-t_\lambda(R)I_\lambda(mr)I_\lambda(mr')\Big]\,,
\end{align}
where $r_>$ and $r_<$ are the maximum and minimum between $r$ and $r'$, $I_\lambda(x)$ and $K_\lambda(x)$ are the well-known modified Bessel function of the second kind and $t_\lambda(R)=\frac{K_\lambda(mR)}{I_\lambda(mR)}$. Taking a look to \cref{eq:solG2} we deduce that the Green function diverges---many distributions are formally infinite. Actually, the first term of previous sum can be reduced to $-\frac{m^2}{2\pi}K_0(m|\mathbf r-\mathbf r'|)$.

Notice that the Green function diverges logarithmically as as $\mathbf r=\mathbf r'$, as $K_0(x)_{x\to0}\sim \ln(2/x)-\gamma$, with $\gamma$ the Euler Mascheroni constant. A cutoff $s$, which represents a minimum separation distance between $\mathbf r$ and $\mathbf r'$ \citep{cornu:2444}, is usually introduced to address this divergence. In turn, $s$ might be related to $L$ and $\epsilon$.

We now implement the numerical analysis to verify last solution noticing that $\QMAT{j}=(r^j)^2$, $\FMAT{\mu}{j}(r)=\alpha^j=r^j\delta_{\mu,0}$, and $\GMAT{\lambda}{\mu}{j}=-(\lambda^2+m^2\QMAT{j})\delta_{\mu,0}$.

\begin{center}
\begin{figure}
     \includegraphics[scale=0.65]{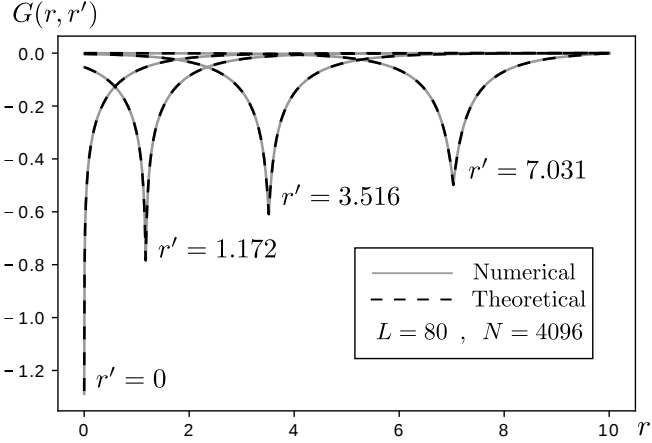}
    \caption{Solution to \cref{eq:Green-ex2} using the numerical solution explained in section \ref{sec:method} (continuous gray lines) and using \cref{eq:solG2} (black dashed lines) for $\theta=\theta'$ and different values of $r'$. We have chosen in all cases units such that $ m=1$. We also used $L=80$ for both cases. In the numerical solutions $N=4096$ and $\epsilon=0.25h$, in the analytical solution $s=0.15h$. The values of the minima and the Mean square error (MSE) are shown in table \ref{tab:ex2}.} \label{fig:ex2}
\end{figure}
\end{center}

The following step is determining an appropriate value for $\epsilon$. From \cref{fig:ex2} we see that the solution for a fixed value of $r'$ shows a peak at $\mathbf r=\mathbf r'$. As we sum up all modes the magnitude of the height of the peaks must be infinite. However, the introduction of $L$ guarantees the peaks to be finite. The height of the peak at $r=r'=0$ is associated with $\epsilon$, as the functions $K_\lambda(r)$ diverge at $r=0$. The cutoff $\epsilon$ is chosen in such way that the height of the peaks in the neighborhood of $r=r'=0$---i.e., $|r-r'|=O(h)$---are close enough. We found that for $N\geq 2^7$, $\epsilon\simeq 0.25 h$. Surprisingly, we found that $\epsilon$ does not depend on $L$ for large enough $L$.

We recall that this is an approximation; the exact solution for the distribution is found in the limits $\epsilon\to0$ and $L\to\infty$. Comparisons between the numerical and analytical results are shown in \cref{fig:ex2} and \cref{tab:ex2}.

\begin{center}
    \begin{table}
    \setlength{\tabcolsep}{3pt}
    \renewcommand{\arraystretch}{1.35}
    \begin{tabular}{|c|c|c|c|c|}
    \hline
    & \multicolumn{4}{|c|}{Values of the minima}\\
    \hline
    $G(r,r')$ & $r'=0$ & $r'=1.172$ & $r'=3.516$ & $r'=7.031$ \\
    \hline
    NS & $-1.288$ & $-0.783$ &  $-0.609$ & $-0.498$ \\
    \hline
    AS & $-1.2778$ & $-0.7835$ & $-0.6087$ & $-0.4984$ \\
    \hline
    $\textrm{PE}$ & $0.8263$ & $3.566^{(-2)}$ & $5.148^{(-3)}$ & $1.603^{(-3)}$ \\
    \hline
    $\textrm{MSE}$ & $3.629^{(-8)}$ & $1.343^{(-10)}$ & $5.053^{(-12)}$ & $6.379^{(-13)}$ \\
    \hline
    \end{tabular}
    \caption{Values of the minima shown in \cref{fig:ex2} for the Numerical solution (NS) and analytical solution (AS). PE means percentage error (from $0$ to $100\%$) and MSE is the mean square error. $x^{(y)}$ stands for $x\times 10^{y}$.}\label{tab:ex2}
    \end{table}
\end{center}
Last result will now be used to solve a inhomogeneous equation of the form $(\vec\nabla^2-m^2)\psi(\mathbf r)=\phi(\mathbf r)$, whose general solution is provided in appendix \ref{app:function-sc} with $r^0=R_{\textrm{int}}=0$.

\begin{center}
\begin{figure}
     \includegraphics[scale=0.60]{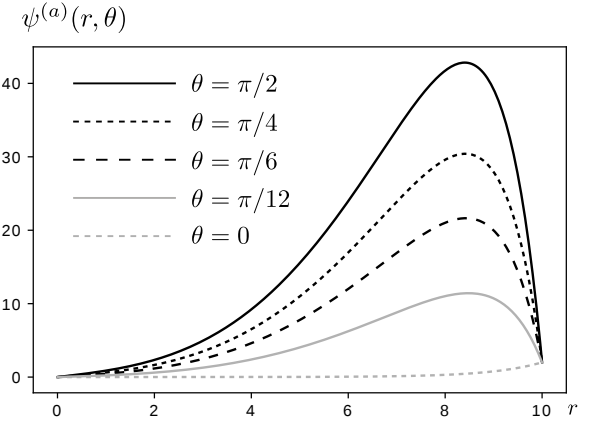}
    \caption{Solution to $\psi^{(a)}(r,\theta)$ for different values of $\theta$ in units in which $m=1$. We used $N=256$.} \label{fig:exam2a}
\end{figure}
\end{center}

\begin{center}
\begin{figure}
     \includegraphics[scale=0.60]{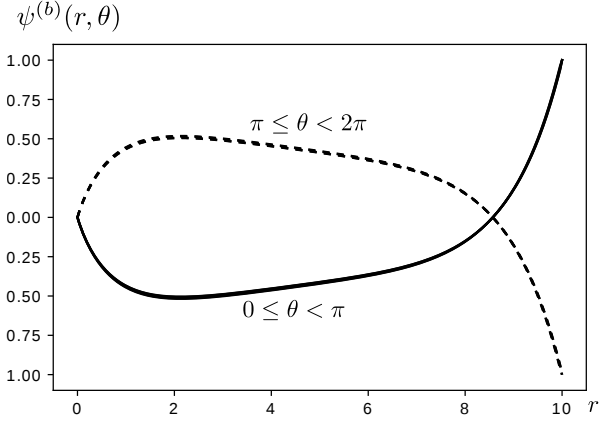}
    \caption{Solution to $\psi^{(b)}(r,\theta)$ for different values of $\theta$ in units in which $m=1$. The continuous line shows the solution for $\theta=\{\frac{\pi}{12},\frac{\pi}{6},\frac{\pi}{4},\frac{\pi}{2}\}$, the dotted line for $\theta=\{-\frac{\pi}{12},-\frac{\pi}{6},-\frac{\pi}{4},-\frac{\pi}{2}\}$. We used $N=256$ and $L=80$.} \label{fig:exam2b}
\end{figure}
\end{center}

Now, let us consider $\phi(\mathbf r)$ to be a function defined over a disc of radius $R=10$ and study the two following cases:
\begin{enumerate}
    \item[(a)] $\psi^{(a)}(r,\theta)$, with $\phi(\mathbf r)=-\frac{1}{10}r\sin\theta$ and $\psi(R,\theta)=2$.
    \item[(b)] $\psi^{(b)}(r,\theta)$, with $\phi(\mathbf r)=\left\{\begin{array}{ccc}r^{-1/2}&,&0\leq\theta< \pi\\-r^{-1/2}&,&\pi\leq\theta<2\pi\end{array}\right.$ and $\psi(R,\theta)=\left\{\begin{array}{ccc}1&,&0\leq\theta< \pi\\-1&,&\pi\leq\theta<2\pi\end{array}\right.$.
\end{enumerate}
Solutions to $\psi^{(a)}(r,\theta)$ and $\psi^{(b)}(r,\theta)$ for some angles are shown in \cref{fig:exam2a} and \cref{fig:exam2b}, respectively.

\begin{center}{\bf Example 3: A pedagogical example}\end{center}

\begin{center}
\begin{figure}
     \includegraphics[scale=0.63]{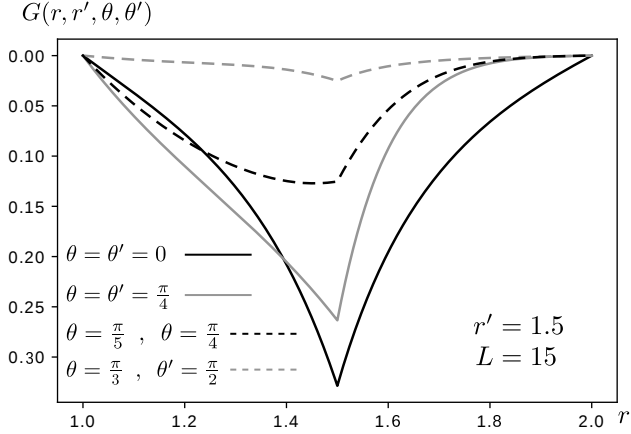}
    \caption{Solution to $G(\mathbf r,\mathbf r')$ with DBC, as given in example 3 for different values. We set $h=\frac{1}{256}$ and made $L=15$.} \label{fig:exam3a1}
\end{figure}
\end{center}

\begin{center}
\begin{figure}
     \includegraphics[scale=0.63]{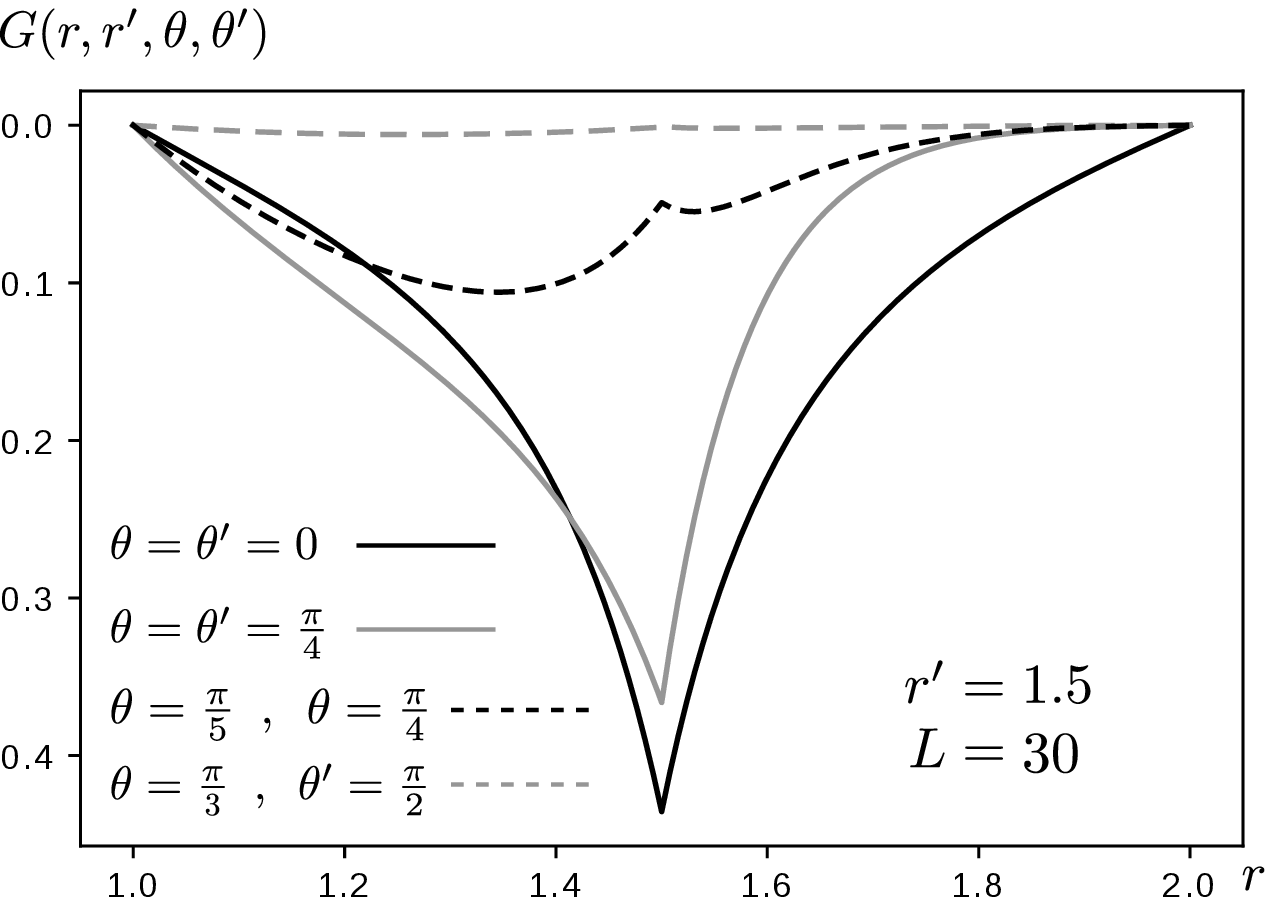}
    \caption{Solution to $G(\mathbf r,\mathbf r')$ with DBC, as given in example 3 for different values. We set $h=\frac{1}{256}$ and made $L=30$.} \label{fig:exam3a2}
\end{figure}
\end{center}

\begin{center}
\begin{figure}
     \includegraphics[scale=0.63]{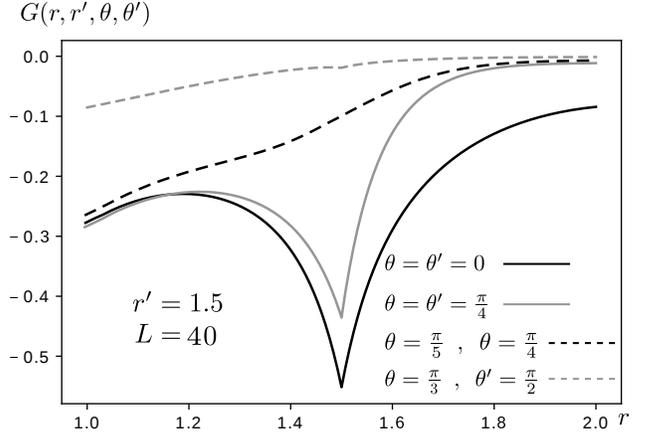}
    \caption{Solution to $G(\mathbf r,\mathbf r')$ with NBC, as given in example 3 for different values. We set $h=\frac{1}{256}$ and made $L=40$.} \label{fig:exam3b}
\end{figure}
\end{center}

Now let us apply the same formalism to solve another two-dimensional problem. Let us suppose that we want to find the Green function associated with the two-dimensional DE $[\vec\nabla+\vec\nabla Z(\mathbf r)]\cdot\vec\nabla\psi(\mathbf r)=\phi(\mathbf r)$, where $Z(x,y)=2x^2y^2$. After transforming the system to polar coordinates, we can see that  $f_r=r^3(1-\cos4\theta)$ and $f_\theta=r^3\sin4\theta$. The elements defined in table \ref{tab:discretization} now become
\begin{subequations}
\begin{align}
\label{eq:def1a}
\FMAT{\mu}{j}&=r^j\delta_{0,\mu}+(r^j)^5\bigl[\delta_{0,\mu}-\frac{1}{2}(\delta_{4,\mu}+\delta_{-4,\mu})\bigr]\,,\\
\label{eq:def3a}
\GMAT{\lambda}{\mu}{j}&=-\lambda^2\delta_{0,\mu}+\frac{1}{2}(r^j)^4(\lambda-\mu)(\delta_{4,\mu}-\delta_{-4,\mu})\,.
\end{align}
\end{subequations}
The system will be confined in an annulus or internal radius $R_{\textrm{int}}=1$ and external radius $R_{\textrm{ext}}=2$. Fig. \ref{fig:exam3a1} and \cref{fig:exam3a2} show the Green function for DBC and some particular parameters but different values of $L$. Fig. \ref{fig:exam3b} shows the results for different parameters under NBC.

Notice from \cref{fig:exam3a1} and \cref{fig:exam3a2} how the Green functions become zero at the borders and present a discontinuity at $r=r'$. As expected, the larger $L$, the larger the magnitude of the value at that point; however, this value decreases as $|\theta -\theta'|$ increases. For NBC, as illustrated in \cref{fig:exam3b}, something similar happens. However, the derivatives at the borders are now the ones which tend no be zero. While this is clear as $r\to 2$, the asymptotic behavior toward zero close to the inner border can be appreciated.

When using the Green function to solve a particular inhomogeneous equation, it is clear that $\vec\nabla\times f=0$, so the weight function exists. It is 
\begin{align}\label{eq:wEx3}
w(\mathbf r',\mathbf r)=\frac{e^{\frac{1}{4}r'^4(1-\cos4\theta')}}{e^{\frac{1}{4}r^4(1-\cos4\theta)}}=\frac{\sum_\lambda w_\lambda(r')e^{i\lambda\theta'}}{w(\mathbf r)}\,,
\end{align}
whose only nonvanishing modes in discrete coordinates are given by $w_{4\lambda}^j=(-1)^\lambda e^{\frac{1}{4}(r^ j)^4}I_\lambda(\frac{1}{4}(r^j)^4)$. Although we are not interested in finding $\psi(\mathbf r)$ for a particular boundary problem, all the steps are carried out to accomplish this goal.

\begin{center}{\bf Example 4: The Stationary Diffusion Equation}\end{center}

It is worthwhile to describe how our formalism can be adapted to solve the diffusion equation at ``thermal'' equilibrium. Let $\psi(\mathbf r)$ and $D(\mathbf r)$ represent the density of the diffusion material and the anisotropic diffusion coefficient, respectively. In the stationary regime, $\psi(\mathbf r)$ satisfies the DE 
\begin{align}\label{eq:diffusion}
    D(\mathbf r)\vec\nabla^2\psi(\mathbf r)+\vec\nabla D(\mathbf r)\cdot\vec\nabla\psi(\mathbf r)=0\,.
\end{align}
Although \cref{eq:diffusion} does not have the standard form shown in \cref{eq:original-problem}, after dividing \cref{eq:diffusion} by $D(\mathbf r)$ and defining $\vec f(\mathbf r)$ as $\vec f(\mathbf r)=\frac{1}{D(\mathbf r)}\vec\nabla D(\mathbf r)$, the standard form can be obtained.\footnote{The diffusion coefficient is assumed to be well--behaved within the annulus or disc. However, $\vec f$ can have poles within the same domain.} The weight function is now guaranteed to exist, as $\vec\nabla\times\vec f=-\frac{1}{D^2}\vec\nabla D\times\vec\nabla D+\frac{1}{D}\vec\nabla\times\vec\nabla D=0$. Actually, $w(\mathbf r,\mathbf r')=\frac{D(\mathbf r)}{D(\mathbf r')}$.

The viability of our method to solve \cref{eq:diffusion} depends on the particular form of the diffusion coefficient. The following possibilities may arise: (a) $\vec f$ has no poles in the two dimensional domain; (b) $\vec f$ has a divergence in $r=0$ that can be eliminated once $\vec f$ is multiplied by $r$; (c) the divergence at $r=0$---or any other radial divergence---previously discussed still persists after multiplication by $r$; and (d) $D^{-1}$ has poles for some $\theta\in[0,2\pi)$.

The cases (a) and (b) can be solved with the regular  procedure we have described; the modes $f_{r\,\mu}$ and $f_{\theta\,\mu}$ are well--behaved and so the elements $\FMAT{\mu}{j}$ and $\GMAT{\lambda}{\mu}{j}$ defined in table \ref{tab:discretization} exist. The possibility stated in (c) demands a redefinition of $\vec f$ to eliminate any possible radial divergence; however, this redefinition does not guarantee the existence of the weight function. The situation described in (d) is problematic, as some of the modes $f_{r\,\mu}$ and $f_{\theta\,\mu}$ are divergent. Last situation is alleviated by working with the original DE, \cref{eq:diffusion}; nonetheless, the process that we must follow to solve a system whose mathematical form differs from \cref{eq:original-problem} has not been described in this work.

Similar analysis can be performed as we deal with the Poisson's equation associated with electrostatic potential in an anisotropic media, among others.

\section{Conclusions}
\label{sec:conclusions}

In this paper we have analyzed the Green function formalism and studied under which conditions such mechanism can be used to obtain the solution of an inhomogeneous DE. Particularly, we found that there exists a function, which we called the weight function, that makes the Liouville operator self-adjoint. This function also defines the symmetry properties of the Green function (how it is transformed under the exchange of $\mathbf r$ and $\mathbf r'$).

After decomposing the Green function as a sum of Fourier modes, an infinite set of coupled second order differential for the radial variable is found. While such set decouples when the initial DE is separable in the radial and angular variables, the coupling in the modes arises as the vector field $f(\mathbf r)$ and the scalar function $g(\mathbf r)$ are expressed as a sum of Fourier modes. 

An algorithm to solve the Green function associated with a general class of Liouville operator was solved using a FEM. We used a simple three-point stencil approach to approximate the solution and focused on both Dirichlet and Neumann boundary conditions. A set of approximations was made, which included a truncation of the infinite number of modes, a minimum distance when the system is confined in a disc, and the discard of the term $-\frac{h^3}{4}f_r(r^k,\theta')$. While the first two approximations are well--justified because the Green function has a natural divergence, the last one was performed by convenience (anyway, it provides a very small contribution).

The algorithm was verified by comparing with known results and obtaining very small percentage errors. An additional example whose solution cannot be found by means of the regular algorithms was shown. 

We consider that the presented method is a useful attempt to solve Green functions of operators whose radial and angular variables cannot be separated. However, we expect this algorithm can be improved by other authors in the future to obtain more accuracy without the need of creating huge matrix systems, which demand large storage memory and computational time. Some of the improvements may include the implementation of the method for higher order stencils, an optimized calculation either mathematically or numerically of $\epsilon$, simplified formulas for the calculation of $\psi(\mathbf r)$ or ``on the go'' algorithms that do not require the inversion of the matrix or the storage of temporal information.

\begin{acknowledgements}
This work was partially funded by Universidad Cat\'olica de Colombia.
\end{acknowledgements}


\bibliography{biblio}

\onecolumngrid
\newpage
\appendix
\section{Deduction of the weight function}
\label{app:weight}

The relation obeyed by the weight function that makes the Liouville operator self-adjoint can be deduced by performing a direct substitution of \cref{eq:operator} into \cref{eq:green-demo1} and using Green's: $\int_\mathcal{V}\phi(\nabla^2\psi)\,\text{d}\mathbf{r}=\int_\mathcal{V}\psi(\nabla^2\phi)\,\text{d}\mathbf{r}+\oint_{\partial\mathcal{V}}\Big[\phi(\vec{\nabla}\psi)-\psi(\vec{\nabla}\phi)\Big]\cdot\mathbf{n}\,\text{d}S$ and the Divergence: $\int_\mathcal{V}\textbf{a}\cdot(\vec{\nabla}\psi)\,\text{d}\mathbf{r}=\oint_{\partial\mathcal{V}}\psi\,\textbf{a}\cdot\mathbf{n}\,\text{d}S - \int_\mathcal{V}\psi(\vec{\nabla}\cdot\textbf{a})\,\text{d}\mathbf{r}$ theorems. After writing it conveniently, the result of this operation is
\eemps\label{eq:convo-w}
\psi(\mathbf{r}) = &\int_\mathbf{r^\prime}G(\mathbf{r^\prime},\mathbf{r})\Big[w(\mathbf{r^\prime},\mathbf{r})\bigl[\nabla_{\{\mathbf{r^\prime}\}}^{2}\psi(\mathbf{r^\prime})\bigr]\,+\bigl[\vec\nabla_{\{\mathbf r'\}}w(\mathbf r,\mathbf r')\bigr] \cdot\bigl[\vec\nabla_{\{\mathbf r'\}}\psi(\mathbf r')\bigr]+w(\mathbf{r^\prime},\mathbf{r})g(\mathbf{r^\prime})\psi(\mathbf{r^\prime})\Big]\text{d}\mathbf{r^\prime}\,+\\
&\int_{\mathbf r'}\Big\{\bigl[\vec{\nabla}_{\{\mathbf{r^\prime}\}}w(\mathbf{r^\prime},\mathbf{r})-w(\mathbf{r^\prime},\mathbf{r})\vec{f}(\mathbf{r^\prime})\bigr]\cdot\vec{\nabla}_{\{\mathbf{r^\prime}\}}\psi(\mathbf{r^\prime})\,+
\vec{\nabla}_{\{\mathbf{r^\prime}\}}\cdot\bigl[\vec{\nabla}_{\{\mathbf{r^\prime}\}}w(\mathbf{r^\prime},\mathbf{r})-w(\mathbf{r^\prime},\mathbf{r})\vec{f}(\mathbf{r^\prime})\bigr]\psi(\mathbf{r^\prime})\Big\}\text{d}\mathbf{r^\prime}\,+\\
&\oint_{\partial\mathbf{r^\prime}}\Big[w(\mathbf{r^\prime},\mathbf{r})\bigl[\psi(\mathbf{r^\prime})\vec{\nabla}_{\{\mathbf{r^\prime}\}}G(\mathbf{r^\prime},\mathbf{r})-G(\mathbf{r^\prime},\mathbf{r})\vec{\nabla}_{\{\mathbf{r^\prime}\}}\psi(\mathbf{r^\prime})\bigr]-G(\mathbf{r^\prime},\mathbf{r})\psi(\mathbf{r^\prime})\big[\vec{\nabla}_{\{\mathbf{r^\prime}\}}w(\mathbf{r^\prime},\mathbf{r})-w(\mathbf{r^\prime},\mathbf{r})\vec{f}(\mathbf{r^\prime})\big]\Big]\cdot\mathbf{n}\text{d}S'.
\ffins
Notice that under the choice $\vec{\nabla}_{\{\mathbf{r^\prime}\}}w(\mathbf{r^\prime},\mathbf{r})-w(\mathbf{r^\prime},\mathbf{r})\vec{f}(\mathbf{r^\prime})=0$, last equation transforms into \cref{eq:sol-tot}.

\section{Finite elements method, matrix elements}
\label{app:fem-method}
The elements of matrices $\FDMmat{n}$ introduced in section \ref{subsec:methodA} depend on the required level of accuracy and the central site $\eta$ that we choose; a general algorithm to deduce such elements is shown in \citep{Fornberg1988}. In the simplest case, as we choose $\eta=0$ as the central point in conjunction with the two closets neighbors---three-point stencil approximation---we have the following relations \citep{Fornberg1988,forsythe2013finite}
\begin{align}\label{eq:derivatives}
\FDMmat{1}=\left(\begin{array}{rrrrrrr}
     0 & 1 & 0 & \cdots & 0 & 0 & 0 \\
     -1 & 0 & 1 & \cdots & 0 & 0 & 0 \\
     0 & -1 & 0 &  \cdots & 0 & 0 & 0 \\
     \vdots & \vdots & \vdots & \ddots & \vdots & \vdots & \vdots \\
     0 & 0 & 0 & \cdots & 0 & 1 & 0 \\
     0 & 0 & 0 & \cdots & -1 & 0 & 1 \\
     0 & 0 & 0 &  \cdots & 0 & -1 & 0 \\
\end{array}\right)_{\!\!N+1\times N+1},\,\,\,\,\,\,
\FDMmat{2}=\left(\begin{array}{rrrrrrr}
     -2 & 1 & 0 & \cdots & 0 & 0 & 0 \\
     1 & -2 & 1 & \cdots & 0 & 0 & 0 \\
     0 & 1 & -2 &  \cdots & 0 & 0 & 0 \\
     \vdots & \vdots & \vdots & \ddots & \vdots & \vdots & \vdots \\
     0 & 0 & 0 & \cdots & -2 & 1 & 0 \\
     0 & 0 & 0 & \cdots & 1 & -2 & 1 \\
     0 & 0 & 0 &  \cdots & 0 & 1 & -2 \\
\end{array}\right)_{\!\!N+1\times N+1}.
\end{align}
Notice how the matrices $\FDMmat{n}$ must be truncated at the boundaries; this is a natural consequence of the FEM, coming from the boundary conditions. 

\section{Derivatives at the boundaries for DBC}

If the Green function is used to find a nonhomgeneous function with DBC, the derivatives of the Green function at the borders are needed---see \cref{eq:sol-tot} and \cref{eq:sol-tot1}. Combining \cref{eq:dif-disc1} with the boundary conditions stated in \cref{eq:condN,eq:cond0,eq:cond0D01,eq:disc-r}, we deduce the following two relations ($j=0$ and $j=N$ refer to the two possible boundaries)
\begin{align}\label{eq:derivNBC}
    G\,'^{\,(0,N),k}_{\theta,\theta'}&=\frac{1}{2\pi}\sum_{\lambda,\mu}e^{i\lambda\theta}e^{-i\mu\theta'}\BMATE{\lambda}{\mu}{(0,N)}{k}\,,\,\textrm{where}\nonumber\\
    \BMATE{\lambda}{\mu}{(0,N)}{k}&=\mp \QMAT{0,N}\sum_{\nu}\SMATEINV{\lambda}{\nu}{(0,N)}\AMATE{\nu}{\mu}{(1,N-1)}{k}\,.
\end{align}
The matrix elements associated with $\SMATM{(0,N)}$ are $\SMATE{\lambda}{\lambda}{(0,N)}=\QMAT{0,N}\mp\frac{h}{2}\FMAT{0}{0,N}$ and  $\SMATE{\lambda}{\lambda-\mu}{(0,N)}=\mp\frac{h}{2}\FMAT{\mu}{0,N}$. Since $\psi$ is known at the boundaries, the derivatives $G\,'^{\,0,0}_{\theta,\theta'}$ and $G\,'^{\,N,N}_{\theta,\theta'}$ are irrelevant; additionally, a disk only requires the calculation of $G\,'^{\,N,k}_{\theta,\theta'}$. The symmetric elements $G_{\theta,\theta'}'^{j(0,N)}$ can be found similarly, in terms of the transpose elements $\BMATE{\lambda}{\mu}{j}{(0,N)}$, which are defined according to last expression by performing the index change and transposition $\AMATE{\nu}{\mu}{(1,N-1)}{k}\to \AMATE{\nu}{\mu}{j}{(1,N-1)}$. In one dimension the derivatives are  $G\,'^{\,(0,N)k}=\mp\frac{P^{0,N}\AMATED{(1,N-1)}{k}}{P^{0,N}\mp\frac{h}{2}Q^{0,N}}$.

\section{Symmetry properties of some matrix elements}
\label{app:properties}

Expanding \cref{eq:mult2} to eliminate the negative modes, we can write the Green function as
\begin{align}\label{Green-exp}
G_{\theta,\theta'}^{j,k}&=-\frac{h}{2\pi}\bigg[\AMATE{0}{0}{j}{k}
+\sum_{\lambda\geq1}\Big\{\bigl[\AMATE{-\lambda}{0}{j}{k}+\AMATE{\lambda}{0}{j}{k}\bigr]\cos(\lambda\theta)+i\bigl[-\AMATE{-\lambda}{0}{j}{k}+\AMATE{\lambda}{0}{j}{k}\bigr]\sin(\lambda\theta)\Big\}
+\sum_{\mu\geq1}\Big\{\bigl[\AMATE{0}{-\mu}{j}{k}+\AMATE{0}{\mu}{j}{k}\bigr]\cos(\mu\theta')
\nonumber\\
&\phantom{=}+i\bigl[\AMATE{0}{-\mu}{j}{k}-\AMATE{0}{\mu}{j}{k}\bigr]\sin(\mu\theta)\Big\}
+\sum_{\lambda,\,\mu\geq1}\Big\{
\bigl[\AMATE{-\lambda}{-\mu}{j}{k}+\AMATE{-\lambda}{\mu}{j}{k}+\AMATE{\lambda}{-\mu}{j}{k}+\AMATE{\lambda}{\mu}{j}{k}\bigr]\cos(\lambda\theta)\cos(\mu\theta')
\nonumber\\
&\phantom{=}+i\bigl[\AMATE{\lambda}{-\mu}{j}{k}+\AMATE{\lambda}{\mu}{j}{k}-\AMATE{-\lambda}{-\mu}{j}{k}-\AMATE{-\lambda}{\mu}{j}{k}\bigr]\sin(\lambda\theta)\cos(\mu\theta')+i\bigl[\AMATE{-\lambda}{-\mu}{j}{k}-\AMATE{-\lambda}{\mu}{j}{k}+\AMATE{\lambda}{-\mu}{j}{k}-\AMATE{\lambda}{\mu}{j}{k}\bigr]\cos(\lambda\theta)\sin(\mu\theta')\nonumber\\
&\phantom{=}+\bigl[\AMATE{-\lambda}{-\mu}{j}{k}-\AMATE{-\lambda}{\mu}{j}{k}-\AMATE{\lambda}{-\mu}{j}{k}+\AMATE{\lambda}{\mu}{j}{k}\bigr]\sin(\lambda\theta)\sin(\mu\theta')\Big\}\bigg]\,.
\end{align}
Since the Green function must be real for real Liouville operators, we demand that the imaginary contributions of last expression must vanish. Hence, we have the restrictions stated in \cref{eq:properties}.

\section{Expansion of the Green function as sines and cosines}
\label{app:green-sc}

This expansion allows us to write the Green function as a sum of real elements, explicitly showing that the Green function is real. Using the properties stated in \cref{eq:properties} into \cref{Green-exp}, we find that

\eemp\label{eq:green-sc}
G_{\theta\theta'}^{jk}=&-\frac{h}{2\pi}\textrm{Re}\bigl(\AMATE{0}{0}{j}{k}\bigr)
-\frac{h}{\pi}\sum_{\lambda\geq1}\Big[\textrm{Re}\bigl(\AMATE{\lambda}{0}{j}{k}\bigr)\cos(\lambda\theta)+\textrm{Re}\bigl(\AMATE{0}{\lambda}{j}{k}\bigr)\cos(\lambda\theta')
-\textrm{Im}\bigl(\AMATE{\lambda}{0}{j}{k}\bigr)\sin(\lambda\theta)+\textrm{Im}\bigl(\AMATE{0}{\lambda}{j}{k}\bigr)\sin(\lambda\theta')\Big]
\\
&-\frac{h}{\pi}\sum_{\lambda,\,\mu\geq1}\Big[
\textrm{Re}\bigl(\AMATE{\lambda}{\mu}{j}{k}\bigr)\cos(\lambda\theta-\mu\theta')
+\textrm{Re}\bigl(\AMATE{\lambda}{-\mu}{j}{k}\bigr)\cos(\lambda\theta+\mu\theta')\Big]
\\
&+\frac{h}{\pi}\sum_{\lambda,\,\mu\geq1}\Big[
\textrm{Im}\bigl(\AMATE{\lambda}{\mu}{j}{k}\bigr)\sin(\lambda\theta-\mu\theta')
+\textrm{Im}\bigl(\AMATE{\lambda}{-\mu}{j}{k}\bigr)\sin(\lambda\theta+\mu\theta')\Big]\,.
\ffin
In the presence of angular symmetry, $A_{\lambda\mu}^{jk}=A_{\lambda\mu}^{jk}\delta_{\lambda\mu}$, so last equation reduces to
\eemp
G_{\theta\theta'}^{jk}=&-\frac{h}{2\pi}\textrm{Re}\bigl(\AMATE{0}{0}{j}{k}\bigr)-\frac{h}{\pi}\sum_{\lambda\geq1}\Big[
\textrm{Re}\bigl(\AMATE{\lambda}{\lambda}{j}{k}\bigr)\cos[\lambda(\theta-\theta')]-\textrm{Im}\bigl(\AMATE{\lambda}{\lambda}{j}{k}\bigr)\sin[\lambda(\theta-\theta')]\Big]\,.
\ffin

\section{Computation of $\psi(\mathbf r)$ as an exponential expansion}
\label{app:psi-exp}

In this section we will derive expressions for \cref{eq:sol-tot} and \cref{eq:sol-tot1} for both DBC and NBC. Although both approaches must lead to the same results, it is worthwhile to show how both relations can be found through the formalism we have described.

For convenience, we will split $\psi(r,\theta)$ into a \textit{volume} ($V$) and \textit{surface} ($S$) contribution ---the \textit{volume} contribution is the term containing the integral over $\mathbf r'$ in \cref{eq:sol-tot} and \cref{eq:sol-tot1}; the \textit{surface} contribution is the one containing the integral over the closed surface $\partial\mathbf r'$ in the same equations. For DBC and NBC, $\psi$ can be written in discrete coordinates as
\begin{subequations}
\begin{align}\label{eq:psiDBC}
(\psi_\theta^j)^{\textrm{DBC}}&=(\psi_\theta^j)_{V}+(\psi_\theta^j)_{S}^{\textrm{DBC}}\,,
\\\label{eq:psiNBC}   
(\psi_\theta^j)^{\textrm{NBC}}&=(\psi_\theta^j)_{V}+(\psi_\theta^j)_{S}^{\textrm{NBC}}\,.
\end{align}
\end{subequations}
There are many ways to evaluate numerically an integral. We will use one of the simplest, however, very efficient, ways to do so, the so called trapezoid rule. Due to the discretization we have used, this rule will be applied to evaluate the the radial integrals, appearing in the {\it volume} contributions; the integrals over angular coordinates will be evaluated  directly using the Fourier expansions of the functions involved.

\begin{center}
    {\bf Using the weight function}
\end{center}

Having adopted the convention described in \cref{eq:sol-tot}, we start performing a Fourier expansions of the external field: $\phi(\mathbf r')\to \phi^{k}_{\theta'} =\sum_{\lambda}\phi^{k}_\lambda e^{i\lambda\theta'}$, the weight function: $w(\mathbf r',\mathbf r)\to\frac{w^k_{\theta'} }{w(r^j,\theta)}= \frac{1}{w(r^j,\theta)}\sum_{\lambda}w^{k}_\lambda e^{i\lambda\theta'}$---and something similar for the boundary conditions $\psi_{\theta'}^{(0,N)}$ and $\psi\,'^{\,(0,N)}_{\theta'}$. Now, we will define the function
\begin{align}\label{eq:xiExp}
    \xi_w(k,M^{k,j},\eta^k,\theta)&=\frac{1}{2\pi}
    \int_{0}^{2\pi}d\theta'\sum_{\lambda,\,\mu}e^{i\lambda\theta'}M_{\lambda,\mu}^{k,j}\,e^{-i\mu\theta}\sum_{\nu}\eta_\nu^{k}e^{i\nu\theta'}\sum_{\rho}w_\rho^{k}e^{i\rho\theta'}
    =\sum_{\lambda,\,\mu,\,\nu}e^{-i\mu\theta}M^{k,j}_{\lambda,\mu}\,\eta_\nu^{k}\,w_{-\lambda-\nu}^k\,.
\end{align}
This definition will be used to define the {\it volume}- and {\it surface}-terms.

Since the \textit{volume}-term can be written as $\int_{r^0}^{r^N}r'dr'\int_0^{2\pi}w(\mathbf r',\mathbf r)G(\mathbf r',\mathbf r)\phi(\mathbf r')d\theta'$ ($r^0=R_{\textrm{int}},\,r^N=R_{\textrm{ext}}$), we can say that
\eemp\label{eq:volume-term}
(\psi^{j}_{\theta})_V=&-\frac{h^2}{w(r^j,\theta)}\Big[\sum_{k=1}^{N-1}r^k\xi_w(k,\AMATED{k}{j},\phi^k,\theta)+\frac{1}{2}r^0\xi_w(0,\AMATED{0}{j},\phi^0,\theta)+\frac{1}{2}r^N\xi_w(N,\AMATED{N}{j},\phi^0,\theta)\Big]\,.
\ffin
The \textit{surface}--term that arises in DBC can be expanded as $r'\int_0^{2\pi}w(\mathbf r',\mathbf r)\psi(\mathbf r')\partial_{r'}G(\mathbf r',\mathbf r)  d\theta'\big\vert_{r^0}^{r^N}$. Similarly as shown above, in discrete coordinates it is given by
 \begin{align}\label{eq:surface-term-DBC}
    (\psi^{j}_\theta)^{\textrm{DBC}}_S&=\frac{1}{w(r^j,\theta)}\Big[\,r^N\xi_w(N,\BMATED{N}{j},\psi^N,\theta)-r^0\xi_w(0,\BMATED{0}{j},\psi^0,\theta)\Big]\,.
\end{align}
Finally, the {\it surface}--term $-\oint_{\partial \mathbf r'}w(\mathbf r',\mathbf r)G(\mathbf r',\mathbf r)\vec\nabla_{\{\mathbf r'\}}\psi(\mathbf r')\cdot\mathbf n dS' $ that appears in NBC is now expanded as $-r'\int_0^{2\pi}w(\mathbf r',\mathbf r)G(\mathbf r',\mathbf r)\partial_{r'}\psi(\mathbf r')d\theta'\big\vert_{r^0}^{r^N}$, it now becomes
\eemp\label{eq:surface-term-NBC}
(\psi^{j}_\theta)^{\textrm{NBC}}_S&=\frac{h}{w(r^j,\theta)}\Big[\,r^N\xi_w(N,\AMATED{N}{j},\psi^N,\theta)-r^0\xi_w(0,\AMATED{0}{j},\psi^0,\theta)\Big]\,.
\ffin
Remarks: the matrix elements of matrices $\AMAT$ and $\BMATED{(0}{N)}$ are given by \cref{eq:mult2} and \cref{eq:derivNBC}, respectively---the indices associated to the position in the blocks have been omitted by convenience. The function $\psi_\theta^j$ is defined in the interval $1\leq j\leq N-1$; in DBC the terms $\psi_\theta^0$ and $\psi_\theta^ N$ are given, in NBC the function at the borders is not accurate enough due to the discontinuity of the Green function at the borders.

\begin{center}
    {\bf Using no weight function}
\end{center}

When we adapt the convention stated in \cref{eq:sol-tot1}, eqs. (\ref{eq:volume-term})--(\ref{eq:surface-term-NBC}) are slightly modified. We now define the function $\xi$ as
\eemp\label{eq:xiExp1}
\xi(k,M^{j,k},\eta^k,\theta)=&\int_0^{2\pi}\frac{d\theta'}{2\pi}
\sum_{\lambda,\,\mu}e^{i\lambda\theta}M_{\lambda,\mu}^{j,k}\,e^{-i\mu\theta'}\sum_{\nu}\eta^{k}_\nu e^{i\nu\theta'}
=\sum_{\lambda,\,\mu}e^{i\lambda\theta}M_{\lambda,\mu}^{j,k}\,\eta^{k}_\mu\,.
\ffin
We can now conclude that
\begin{align}\label{eq:volume-term1}
    (\psi_\theta^j)_V&=-h^2\sum_{k=1}^{N-1} \Big[r^k\xi(k,\AMATED{j}{k},\phi^k,\theta)+\frac{1}{2}r^0\xi(0,\AMATED{j}{0},\phi^0,\theta)+\frac{1}{2}r^N\xi(N,\AMATED{j}{N},\phi^N,\theta)
    \Big]\,\\\label{eq:surface-term1-DBC}
    (\psi_\theta^j)_S^{\textrm{DBC}}&=r^N\xi(N,\BMATED{j}{N},\psi^N,\theta)-r^0\xi(0,\BMATED{j}{0},\psi^0,\theta)\,\\\label{eq:surface-term1-NBC}
    (\psi_\theta^j)_S^{\textrm{NBC}}&=hr^N\bigl[\xi(N,\AMATED{j}{N},f_r^N,\theta)+\xi(N,\AMATED{j}{N},\psi'^{N},\theta)\bigr]-hr^0\bigl[\xi(0,\AMATED{j}{0},f_r^0,\theta)+\xi(0,\AMATED{j}{0},\psi'^{0},\theta)\bigr]\,.
    \end{align}

\section{Computation of the inhomogeneous function as expansion of trigonometric functions}
\label{app:function-sc}

It is now useful to expand the relations shown in previous section as trigonometric functions. Although the expressions found are much longer, this allows us to use the symmetry properties, \cref{eq:properties}, to get rid of irrelevant terms and explicitly express $\psi(\mathbf r)$ as a real function. Besides, the exponential expansion defined in appendix \ref{app:psi-exp} might introduce some spurious imaginary contributions, which may arise by as a consequence of the truncating process of matrix $\PMATM$---the complex conjugate counterparts of some modes may be discarded in this process. Taking advantage of the definitions used in appendix \ref{app:psi-exp}, \cref{eq:volume-term} to \cref{eq:surface-term-NBC} are still valid when we adopt the convention stated in \cref{eq:sol-tot}; similarly, when the convention \cref{eq:sol-tot1} is adopted, \cref{eq:volume-term1} to \cref{eq:surface-term1-NBC} are also valid. Now, we only need to expand $\xi_w$ and $\xi$ eliminating the negative complex modes to express them as sum of real modes. By doing so, \cref{eq:xiExp} becomes

\eemp\label{eq:xi}
&\xi_w(k,{M^{k,j},\eta^k},\theta)=\textrm{Re}\bigl(M_{0,0}^{k,j}\bigr)\textrm{Re}\bigl(\eta_0^{k}\bigr)\textrm{Re}\bigl(w_0^{k}\bigr)     \\
&+2\sum_{\lambda\geq1}\textrm{Re}\bigl(\eta_0^{k}\bigr)\Big[\textrm{Re}\bigl(M_{0,0}^{k,j}\bigr)X_\lambda^k+Y_\lambda^{k,j}
+\textrm{Re}\bigl(w_0^{k}\bigr)
\bigl[\textrm{Re}\bigl(M_{0,\lambda}^{k,j}\bigr)\cos(\lambda\theta)+\textrm{Im}\bigl(M_{0,\lambda}^{k,j}\bigr)\sin(\lambda\theta)\bigr]\Big]
\\
&+4\sum_{\lambda,\,\mu\geq1}X_\mu^k
\bigl[\textrm{Re}\bigl(M_{0,\lambda}^{k,j}\bigr)\cos(\lambda\theta)+M_{0,\lambda}^{\mathbb{I}\,k,j}\sin(\lambda\theta)\bigr]
\\
&+2\!\sum_{\lambda,\,\mu\geq1}\!\textrm{Re}\bigl(\eta_0^{k}\bigr)\Big[\bigl[\textrm{Re}\bigl(w_\lambda^{k}\bigr)M_{(+)\lambda,\mu}^{(1)k,j}
+\textrm{Im}\bigl(w_\lambda^{k}\bigr)M_{(+)\lambda,\mu}^{(2)k,j}\bigr]\cos(\mu\theta)
-\bigl[\textrm{Im}\bigl(w_\lambda^{k}\bigr)M_{(-)\lambda,\mu}^{(1)k,j}
-\textrm{Re}\bigl(w_\lambda^{k}\bigr)M_{(-)\lambda,\mu}^{(2)k,j}\bigr]\sin(\mu\theta)\Big]
\\
&+2\!\sum_{\lambda,\mu\geq1}\!\Big[
\textrm{Re}\bigl(\eta_{\mu}^{k}\bigr)\bigl[\textrm{Re}\bigl(M_{\lambda,0}^{k,j}\bigr)w_{+(\lambda,\mu)}^{(1)k}
+\textrm{Im}\bigl(M_{\lambda,0}^{k,j}\bigr)w_{+(\lambda,\mu)}^{(2)k}\bigr]
+\textrm{Im}\bigl(\eta_{\mu}^{k}\bigr)\bigl[\textrm{Re}\bigl(M_{\lambda,0}^{k,j}\bigr)w_{-(\lambda,\mu)}^{(2)k}
-\textrm{Im}\bigl(M_{\lambda,0}^{k,j}\bigr)w_{-(\lambda,\mu)}^{(1)k}\bigr]\Big]
\\
&+2\!\sum_{\lambda,\mu,\nu\geq1}
\Big[M_{(+)\lambda,\mu}^{(1)k,j}
\bigl[Z^{(1)k}_{(\lambda,\nu)}
+Z^{(2)k}_{(\lambda,\nu)}\bigr]
+M_{(+)\lambda,\mu}^{(2)k,j}
\bigl[Z^{(3)k}_{(\lambda,\nu)}
+Z^{(4)k}_{(\lambda,\nu)}\bigr]\Big]\cos(\mu\theta)
\\
&+2\!\sum_{\lambda,\mu,\nu\geq1}\Big[M_{(-)\lambda,\mu}^{(2)k,j}
\bigl[Z^{(1)k}_{(\lambda,\nu)}+Z^{(2)k}_{(\lambda,\nu)}\bigr]-M_{(-)\lambda,\mu}^{(1)k,j}
\bigl[Z^{(3)k}_{(\lambda,\nu)}
-Z^{(4)k}_{(\lambda,\nu)}\bigr]
\Big]\sin(\mu\theta)\,,
\ffin
where we used the definitions
\eemp
    &X_\lambda^k=\textrm{Re}\bigl(\eta_\lambda^{k}\bigr)\textrm{Re}\bigl(w_\lambda^{k}\bigr)+\textrm{Im}\bigl(\eta_\lambda^{k}\bigr)\textrm{Im}\bigl(w_\lambda^{k}\bigr)\,\,\,,\,\,\,
    Y_\lambda^{k,j}=\textrm{Re}\bigl(M_{\lambda,0}^{k,j}\bigr)\textrm{Re}\bigl(w_\lambda^{k}\bigr)
    +\textrm{Im}\bigl(M_{\lambda,0}^{k,j}\bigr)\textrm{Im}\bigl(w_\lambda^{k}\bigr)\,;
    \\
    &M_{(\pm)\lambda,\mu}^{(1)k,j}=\textrm{Re}\bigl(M_{\lambda,\mu}^{k,j}\bigr)\pm \textrm{Re}\bigl(M_{\lambda,-\mu}^{k,j}\bigr)\,\,\,,\,\,\,
    M_{(\pm)\lambda,\mu}^{(2)k,j}=\textrm{Re}\bigl(M_{\lambda,\mu}^{k,j}\bigr)\pm \textrm{Re}\bigl(M_{\lambda,-\mu}^{k,j}\bigr)\,;
    \\
    &w_{\pm(\lambda,\nu)}^{(1)k}=\textrm{Re}\bigl(w_{\lambda+\nu}^{\,k}\bigr)\pm\textrm{Re}(w_{\lambda-\nu}^{\,k}\bigr)\,\,\,,\,\,\,
    w_{\pm(\lambda,\nu)}^{(2)k}=\textrm{Im}\bigl(w_{\lambda+\nu}^{\,k}\bigr)\pm\textrm{Im}(w_{\lambda-\nu}^{\,k}\bigr)\,;
    \\
    &Z^{(1)k}_{(\lambda,\nu)}=\textrm{Re}\bigl(\eta_\nu^{k}\bigr)w_{+(\lambda,\nu)}^{(1)k}\,\,\,,\,\,\,
    Z^{(2)k}_{(\lambda,\nu)}=\textrm{Im}\bigl(\eta_\nu^{k}\bigr)w_{-(\lambda,\nu)}^{(2)k}\,\,\,,\,\,\,
    Z^{(3)k}_{(\lambda,\nu)}=\textrm{Re}\bigl(\eta_\nu^{k}\bigr)w_{+(\lambda,\nu)}^{(2)k}\,\,\,,\,\,\,
    Z^{(4)k}_{(\lambda,\nu)}=\textrm{Im}\bigl(\eta_\nu^{k}\bigr)w_{-(\lambda,\nu)}^{(1)k}\,.
\ffin

Similarly, \cref{eq:xiExp1} is now written as
\eemp\label{eq:xi1}
&\xi(k,M^{j,k},\eta^ k,\theta)=\,\textrm{Re}\bigl(M_{0,0}^{j,k}\bigr)\textrm{Re}\bigl(\eta_0^{k}\bigr)
+2\sum_{\lambda\geq1}\Big[R_\lambda^{j,k}+\textrm{Re}\bigl(\eta_0^{k}\bigr)
\bigl[\textrm{Re}\bigl(M_{\lambda,0}^{j,k}\bigr)\cos(\lambda\theta)-\textrm{Im}\bigl(M_{\lambda,0}^{j,k}\bigr)\sin(\lambda\theta)\bigr]\Big]
\\
&+2\sum_{\lambda,\,\mu\geq1}\bigl[M_{(+)\lambda,\mu}^{(1)j,k}\textrm{Re}\bigl(\eta_\mu^{k}\bigr)-M_{(-)\lambda,\mu}^{(2)j,k}\textrm{Im}\bigl(\eta_\mu^{k}\bigr)\bigr]\cos(\lambda\theta)
-2\sum_{\lambda,\,\mu\geq1}\bigl[M_{(+)\lambda,\mu}^{(2)j,k}\textrm{Re}\bigl(\eta_\mu^{k}\bigr)+M_{(-)\lambda,\mu}^{(1)k,j}\textrm{Im}\bigl(\eta_\mu^{k}\bigr)\bigr]\sin(\lambda\theta)\,,
\ffin
with $R_\lambda^{j,k}=\textrm{Re}\big(M_{0,\lambda}^{j,k}\big)\textrm{Re}\big(\eta_\lambda^{k}\big)-\textrm{Im}\big(M_{0,\lambda}^{j,k}\big)\textrm{Im}\big(\eta_\lambda^{\mathbb I\,k}\big)$.

\begin{center}{\bf The one dimensional case}\end{center}

The function $\psi$, satisfying the equation $\mathcal L_x\psi(x)=\phi(x)$, where $\mathcal L_x$ is defined according to \cref{eq:DE-gen}, can be found by means of the relations below. For DBC with either weight function or not
\begin{subequations}
\begin{align}\label{eq:psi1D-1}
    \psi^{j}&=-h^2\Big[\sum_{k=1}^{N-1}w^{k,j}A^{k,j}\phi^{k}+\frac{1}{2}w^{0,j}A^{0,j}\phi^{0}+\frac{1}{2}w^{N,j}A^{N,j}\phi^{N}\Big]+\frac{P^Nw^{N,j}\psi^NA^{N-1,j}}{1+\frac{h}{2}\frac{1}{Q^N}}
    +\frac{P^0w^{0,j}\psi^0A^{1,j}}{1-\frac{h}{2}\frac{1}{Q^0}}\,;\\\label{eq:psi1D-2}
    &=-h^2\Big[\sum_{k=1}^{N-1}A^{j,k}\phi^{k}+\frac{1}{2}A^{j,0}\phi^{0}+\frac{1}{2}A^{j,N}\phi^{N}\Big]+\frac{q^N\psi^NA^{j,N-1}}{1+\frac{h}{2}\frac{1}{Q^N}}
    +\frac{P^0\psi^0A^{j,1}}{1-\frac{h}{2}\frac{1}{Q^0}}\,.
\end{align}
\end{subequations}
    For NBC
    \begin{subequations}
    \begin{align}\label{eq:psi1D-3}
    \psi^{j}&=-h^2\Big[\sum_{k=1}^{N-1}w^{k,j}A^{k,j}\phi^{k}+\frac{1}{2}w^{0,j}A^{0,j}\phi^{0}+\frac{1}{2}w^{N,j}A^{Nj}\phi^{N}\Big]+hP^kw^{k,j}\psi\,'^{\,k}A^{k,j}\Big\vert_{k=0}^{N}\,;
    \\\label{eq:psi1D-4}
    &=-h^2\Big[\sum_{k=1}^{N-1}A^{j,k}\phi^{k}+\frac{1}{2}A^{j,0}\phi^{0}+\frac{1}{2}A^{j,N}\phi^{N}\Big]+hP^kA^{j,k}\bigl[\psi\,'^{\,k}+\psi^kf^k\bigr]\Big\vert_{k=0}^{N}\,.
\end{align}
\end{subequations}

\end{document}